\documentclass[aps,pre,twocolumn,groupedaddress]{revtex4}

\usepackage{graphicx}
\usepackage{amsfonts}
\usepackage{amsmath}
\usepackage{amssymb}
\usepackage{amsthm}
\usepackage{lscape}

\newcommand{\rb}{\mbox{\bf r}}
\newcommand{\ub}{\mbox{\bf u}}
\newcommand{\Db}{\mbox{\bf D}}
\newcommand{\ds}{\hat{\mbox{\bf d}}_s}
\newcommand{\del}{\mbox{\boldmath{$\nabla$}}}

\begin{document}

\title{A stochastic flow rule for granular materials}
\author{Ken Kamrin and Martin Z. Bazant}
\affiliation{Department of Mathematics, Massachusetts Institute
        of Technology, Cambridge, MA 01239}

\date{\today}
\begin{abstract}
  There have been many attempts to derive continuum models for dense
  granular flow, but a general theory is still lacking. Here, we start
  with Mohr-Coulomb plasticity for quasi-2D granular materials to
  calculate (average) stresses and slip planes, but we propose a
  ``stochastic flow rule'' (SFR) to replace the principle of
  coaxiality in classical plasticity. The SFR takes into account two
  crucial features of granular materials - discreteness and randomness
  - via diffusing ``spots'' of local fluidization, which act as
  carriers of plasticity.  We postulate that spots perform random
  walks biased along slip-lines with a drift direction determined by
  the stress imbalance upon a local switch from static to dynamic
  friction.  In the continuum limit (based on a Fokker-Planck equation
  for the spot concentration), this simple model is able to predict a
  variety of granular flow profiles in flat-bottom silos, annular
  Couette cells, flowing heaps, and plate-dragging experiments -- with
  essentially no fitting parameters -- although it is only expected to
  function where material is at incipient failure and slip-lines are
  inadmissible. For special cases of admissible slip-lines, such as
  plate dragging under a heavy load or flow down an inclined plane, we
  postulate a transition to rate-dependent Bagnold rheology, where
  flow occurs by sliding shear planes. With different yield criteria,
  the SFR provides a general framework for multiscale modeling of
  plasticity in amorphous materials, cycling between continuum
  limit-state stress calculations, meso-scale spot random walks, and
  microscopic particle relaxation.
\end{abstract}

\pacs{PACS number(s): }

\maketitle

\section{ Introduction }

For centuries, engineers have described granular materials using
continuum solid mechanics~\cite{nedderman,Sok,hill}. Dense granular
materials behave like rigid solids at rest, and yet are easily set
into liquid-like, quasi-steady motion by gravity or moving boundaries,
so the classical theory is Mohr-Coulomb plasticity (MCP), which
assumes a frictional yield criterion. The simplest model is the
two-dimensional ``Ideal Coulomb Material'' at limit-state, where the
maximum ratio of shear to normal stress is everywhere equal to a
constant (the internal friction coefficient), whether or not flow is
occurring. This model is believed to describe stresses well in static
or flowing granular materials, but, as we explain below, it fails to
predict flow profiles, when combined with the usual Coaxial Flow Rule
of continuum plasticity. Indeed, it seems continuum mechanics has not
yet produced a simple and robust model for granular flow.

In recent years, the sense that there is new physics to be discovered
has attracted a growing community of physicists to the study of
granular
materials~\cite{jaeger96,degennes99,kadanoff99,edwards01,halsey02,aranson06}. 
Unlike the engineers, their interest is mostly at the discrete particle
level, motivated by the breakdown of classical statistical mechanics
and hydrodynamics due to strong dissipation and long-lasting,
frictional contact networks. Dense granular materials exhibit many
interesting collective phenomena, such as force chains, slow
structural relaxation, and jamming. Similar non-equilibrium phenomena
occur in glasses, foams, and emulsions, as in granular materials, so
it is hoped that a general new statistical theory may
emerge. Presumably from such a microscopic basis, continuum models of
glassy relaxation and dense granular flow could be systematically
derived, just as dissipative hydrodynamics for granular gases can be
derived from kinetic theory with inelastic
collisions~\cite{campbell90}.

This dream has not yet been achieved, but many empirical continuum
models have been proposed~\cite{jaeger96,aranson06,alltheories}. The
difficulty in describing dense granular flow is evidenced by the
remarkable diversity of physical postulates, which include: coupled
static and rolling
phases~\cite{bouchaud94,bouchaud95a,boutreux98,boutreux99}, Bagnold
rheology~\cite{bagnold54} based on ``granular eddies''~\cite{ertas02},
granular temperature-dependent viscosity~\cite{savage98},
density-dependent viscosity~\cite{losert00,bocquet02}, non-local
stress propagation along arches~\cite{mills99}, self-activated shear
events due to non-local stress
fluctuations~\cite{pouliquen96,pouliquen01}, free-volume diffusion
opposing
gravity~\cite{lit58,mullins72,nedderman79,bazant06,rycroft06a},
``shear transformation zones'' coupled to free-volume
kinetics~\cite{lemaitre02,lemaitre02c}, and partial fluidization
governed by a Landau-like order parameter~\cite{aranson01,aranson02}.
Each of these theories can fit a subset of the experimental
data~\cite{midi04}, usually only for a specific geometry for which it
was designed, such as a flowing surface
layer~\cite{bouchaud94,bouchaud95a,boutreux98,boutreux99,aranson01},
inclined plane~\cite{bagnold54,ertas02}, Couette
cell~\cite{losert00,bocquet02}, inclined
chute~\cite{pouliquen96,pouliquen01}, or wide
silo~\cite{lit58,mullins72,nedderman79,bazant06,rycroft06a}, and none
seems to have very broad applicability. For example, we are not aware
of a single model, from physics or engineering, which can predict
velocity profiles in both draining silos and annular Couette cells,
even qualitatively.

The theory of partial fluidization of Aranson and Tsimring
has arguably had the most success in describing multiple flows within
a single theoretical framework~\cite{aranson01,aranson02}. Although
setting boundary conditions for the order parameter usually requires
additional {\it ad hoc} assertions, the model is nonetheless able to
reproduce known flow behavior in inclined chutes, avalanches, rotating
drums, and simple shear cells without many fitting parameters. It also
describes some unsteady flows.  However, the theory lacks
any clear microscopic foundation and is not directly coupled to a
constitutive stress model for static materials.  As such, it has only
been applied to problems with very simple solid stress fields,
limiting its current applicability to flows that depend on only one
spatial variable.

In an attempt to describe arbitrary geometries, such as silos and
Couette cells, we take the view that the engineers may already have a
reasonable continuum description of the mean stresses, so we start
with Mohr-Coulomb plasticity.  However, discreteness and randomness
clearly need to be taken into account in a granular material. For
static stresses, quenched randomness in material properties is known
to lead to statistical slip-line blurring in ``stochastic
plasticity''~\cite{ostoja05}, but this says nothing about how plastic
yielding actually occurs.

To describe yielding {\it dynamics}, we propose a ``stochastic flow
rule'' (SFR) where local fluidization (stick-slip transition)
propagates randomly along blurred slip-lines. We build on the recently
proposed Spot Model for random-packing dynamics~\cite{bazant06} by
viewing ``spots'' of free volume as carriers of plasticity in granular
materials, analogous to dislocations in crystals. Multiscale spot
simulations can reproduce quite realistic flowing packings in silo
drainage~\cite{rycroft06a}; here, we introduce a mechanical basis for
spot motion from MCP, which leads to a theory of considerable
generality for bulk granular flows.

The paper is organized as follows. Since plasticity is unfamiliar to
most physicists, we begin by reviewing key concepts from MCP in
section \ref{concepts}, both for stresses and for dense flows. In
section~\ref{short}, we highlight various shortcomings of the
classical theory, many of which we attribute to the Coaxial
Flow Rule. We then introduce the general spot-based SFR and a specific
simplification to be used for granular flow in
section~\ref{spotsec}. Next we apply the theory to four prototypical
examples: silo, Couette, heap, and plate-dragging flows in
section~\ref{expt}. Then in section~\ref{bag}, we explain how the last
two examples indicate a smooth transition from the SFR to Bagnold
rheology, when slip-lines become admissible, and we present a simple
composite theory, which extends the applicability of the model to
various shear flows. In section \ref{con}, we conclude by further
clarifying the range of applicability of the SFR and possible
extensions to other granular flows and different materials.

\section{ Concepts from continuum mechanics } 
\label{concepts}

\subsection{ Mohr-Coulomb plasticity: stresses }

In the eighteenth century, it was Coulomb, as a military engineer
designing earthen fortresses, who introduced the classical model of a
granular material, which persists to the present day: a continuous
medium with a frictional yield criterion. His ideas were expressed in
general continuum-mechanical terms by Mohr a century later, and a
modern mathematical formulation of ``Mohr-Coulomb plasticity'' (MCP),
which we also use below, is due to Sokolovskii\cite{Sok}. Although
other mechanical models exist, such as Drucker-Prager
plasticity~\cite{drucker}, MCP is perhaps the simplest and most widely
used for granular materials in engineering~\cite{nedderman}.  As such,
we choose to build our model of dense granular flow on the MCP
description of stresses, as a reasonable and time-tested first
approximation.
\begin{figure}
\centering{\includegraphics[width=3.2in, clip]{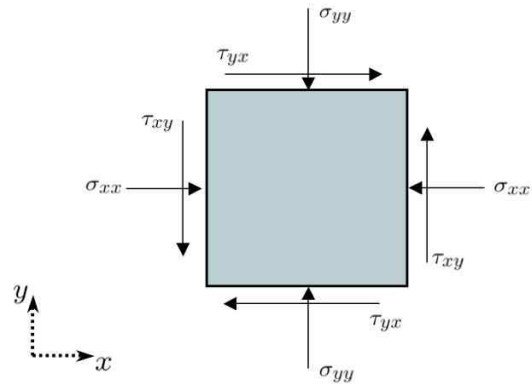}
\caption{Stresses on a material element. All vectors are pointing in the positive direction as per our sign convention.}\label{element}}
\end{figure}

\begin{figure}
\centering{\includegraphics[width=2.5in, clip]{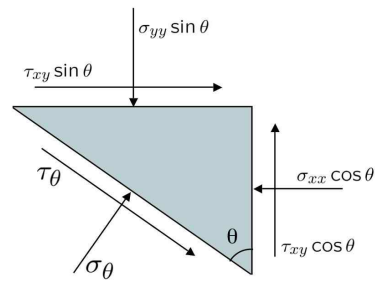}
\caption{Force diagram for a wedge. Hypotenuse length assigned to unity.}\label{wedge}}
\end{figure}

We begin in this section by reviewing relevant concepts from MCP,
e.g. following Nedderman~\cite{nedderman}. The
fundamental assumption is that a granular material can be treated as
an ``Ideal Coulomb Material'' (ICM), i.e. a rigid-plastic continuous media
which yields according to a Coulomb yield criterion
\begin{equation}
|\tau/\sigma| = \mu \equiv \tan \phi
\end{equation}
where $\tau$ is the shear stress, $\sigma$ is normal stress, and
$\phi$ the internal friction angle, akin to a standard friction law
with no cohesion.  Throughout, we accept the common tensorial
conventions for stresses with the key exception that normal stresses
are deemed positive in compression. This is a standard modification in
the study of non-cohesive granular materials since granular
assemblies cannot support tension. We will also focus entirely on
quasi-2D geometries.

Consider a small material element in static equilibrium and with no
body forces present (see Figure \ref{element}). The normal stresses
$\sigma_{xx}$ and $\sigma_{yy}$ can differ and the shear stresses
$\tau_{xy}$ and $\tau_{yx}$ must be equal in order to balance moments.
Likewise the variable $\tau_{yx}$ is redundant and will not be used
again in this paper.  To determine the stresses along any angle within
this element, we place a new boundary within the material at some
desired angle $\theta$ and observe force balance on the wedge that
remains (see Figure \ref{wedge}). After algebraic simplification, this
gives
\begin{eqnarray*}
\sigma_{\theta} &=&\frac{1}{2}(\sigma_{xx}+\sigma_{yy})+\frac{1}{2}(\sigma_{xx}-\sigma_{yy})\cos 2\theta-\tau_{xy}\sin 2\theta
\\
\tau_{\theta}&=&\frac{1}{2}(\sigma_{xx}-\sigma_{yy})\sin 2\theta +\tau_{xy}\cos 2\theta
\end{eqnarray*}
Now define
\begin{eqnarray*}
p&=&\frac{1}{2}(\sigma_{xx}+\sigma_{yy})
\\
\tan 2\psi &=& \frac{-2\tau_{xy}}{\sigma_{xx}-\sigma_{yy}}
\\
R&=&\sqrt{\left(\frac{\sigma_{xx}-\sigma_{yy}}{2}\right)^2 +\tau_{xy}^2}
\end{eqnarray*}
which allows us to write
\begin{eqnarray}
\sigma_{\theta}&=&p+R\cos(2\theta-2\psi)\label{stheta}
\\
\label{theta2}\tau_{\theta}&=&R\sin(2\theta-2\psi)\label{ttheta}
\end{eqnarray}
This implies that for all angles $\theta$, the locus of traction stresses
$(\sigma_{\theta},\tau_{\theta})$ is a circle centered at
$(p,0)$ with radius $R$. This circle is referred to as ``Mohr's
Circle''.

We have just derived Mohr's Circle without accounting for the possible
effects of body forces acting on the material element and gradients in
the stress field. Adjusting for these effects, however, would change
the results only negligibly as the element gets small in size.  If we
were to apply the same force-balancing analysis to a differentially
small material element with a body force and stress gradients, we
would find that the stress differences on the walls and the inclusion
of the differentially small body force within only add differentially
small terms to the equations for $\sigma_{\theta}$ and $\tau_{\theta}$.
Thus we can always use Mohr's Circle to obtain traction stresses along
a desired angle.

\begin{figure*}
\centering{\includegraphics[width=5in, clip]{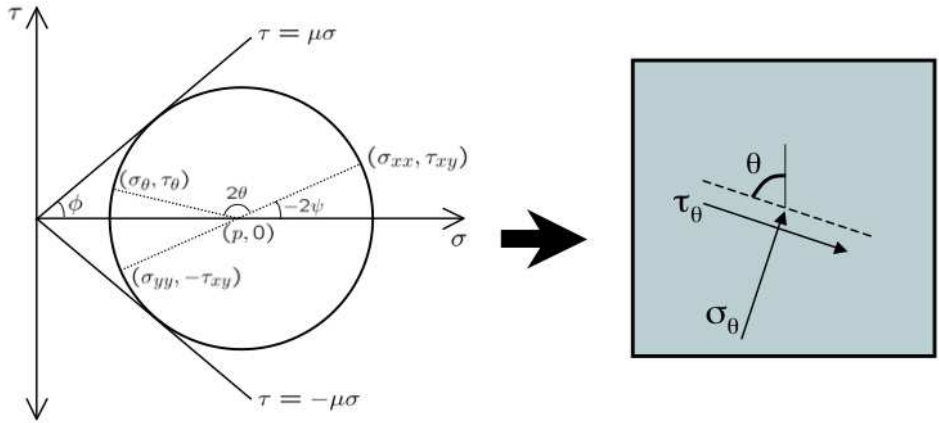}
\caption{Using Mohr's circle jointly with the Coulomb internal yield locus ($\tau=\pm\mu\sigma$) to determine the traction stresses along any plane within a material element.}\label{mohr}}
\end{figure*}

To ultimately define a full stress state for the material element, we
need one more equation--- we have 3 stress variables and only 2 force
balance equations:
\begin{eqnarray}
\frac{\partial\sigma_{xx} }{\partial x} - \frac{\partial\tau_{xy} }{\partial y }=F_{body}^x \label{xbalance}
\\
\frac{\partial \sigma_{yy}}{\partial y}-\frac{\partial \tau_{xy} }{\partial x}=F_{body}^y \label{ybalance}
\end{eqnarray}
We say a material element is at \emph{incipient failure} if the yield
criterion is fulfilled along some direction and $|\tau/\sigma|\leq\mu$
along all others. A material in which incipient failure occurs
everwhere is said to be at a \emph{limit-state}.  In a limit-state,
the Mohr's Circle at every point in the material must be tangent to
the locus $|\tau/\sigma|=\mu$.  As can be seen by applying
trigonometry in Figure \ref{mohr}, this requirement means that
$R=p\sin \phi$, enabling us to parameterize the stresses in terms of
$p$ and $\psi$ only, thereby closing the equations.  For this reason,
we restrain our anlysis to limit-state materials and refer to $p$ and
$\psi$ as the stress parameters or Sokolovskii variables. (The
limit-state stress treatment described here is also known as
``Slip-Line Theory''; to avoid possible confusion, we specify this is
not equivalent to Limit Analysis Plasticity concerned with upper and
lower collapse limits.)
 
Solving for the original stress
variables in terms of the stress parameters gives:
\begin{eqnarray}
  \sigma_{xx}& =& p(1+\sin\phi\cos 2\psi)\\
  \sigma_{yy} &=& p(1-\sin\phi\cos 2\psi)\\
  \tau_{xy}&=&-p\sin\phi\sin 2\psi
\end{eqnarray}
Using these expressions, we re-write equations (\ref{xbalance}) and (\ref{ybalance}):
\begin{align*}
 (1+&\sin\phi\cos2\psi)p_{x}-2p\sin\phi\sin2\psi\
 \psi_{x}  +\sin\phi\sin2\psi \ p_{y} \\&+2p\sin\phi\cos2\psi\ \psi_{y} = F_{body}^x \\
 \sin&\phi\sin2\psi\ p_{x}+2p\sin\phi\cos2\psi\ \psi_{x} +
 (1-\sin\phi\cos2\psi)p_{y} \\&+2p\sin\phi\sin2\psi\ \psi_{y}= F_{body}^y
 \end{align*}
 These will be referred to as the ``stress balance equations''. They
 form a hyperbolic system and thus can be solved using the method of
 characteristics.  The system reduces to the following two
 characteristic equations:
 \begin{align}
 dp\mp2p& \mu\ d\psi = F_{body}^y(dy\mp\mu\ dx) +F_{body}^x(dx\pm\mu\ dy) \nonumber
\\
 &\text{along curves fulfilling}\ \ \frac{dy}{dx} = \tan(\psi\mp\epsilon). 
 \end{align}
 To solve the stress balance equations, mesh the two families of
 characteristic curves in the bulk, then march from the boundaries in,
 progressively applying the two differential relationships above to
 approximate the  stress parameters at each intersection point
 in the mesh. More on this can be found in \cite{horne}. Other ways to
 solve the stress balance equations include the Two-Step Lax-Wendroff
 Method \cite{pitman} and the Galerkin Method \cite{gremaud}.

We return now to Mohr's Circle for a discussion of the stress properties
within a differential material element. Equations (\ref{stheta}) and
(\ref{ttheta}) show that Mohr's Circle can be used as a slide-rule to
determine the stresses along any angle $\theta$: One arrives at the
point $(\sigma_{\theta},\tau_{\theta})$ by starting at
$(\sigma_{xx},\tau_{xy})$ and traveling anti-clockwise around Mohr's
Circle for $2\theta$ radians (see Figure \ref{mohr}). Also note on the
diagram that the stresses along the $x$ and $y$ directions lie along a
diameter of Mohr's Circle; any two material directions differing by an
angle of $\pi/2$ lie along a diameter of the corresponding Mohr's
Circle diagram. Utilizing this property in reverse is perhaps the
easiest way to draw Mohr's Circle in the first place; draw the unique
circle for which $(\sigma_{xx},\tau_{xy})$ and
$(\sigma_{yy},-\tau_{xy})$ are endpoints of a diameter.

Let $(\sigma_1,0)$ and $(\sigma_3,0)$ be the points of intersection
between Mohr's Circle and the $\sigma$-axis, where
$\sigma_1>\sigma_3$. These points correspond to the two lines within a
material element along which the shear stress vanishes and the normal
stress is maximal or minimal.  $\sigma_1$ ($\sigma_3$) is called the
major (minor) principal stress and the line on which it acts is called
the major (minor) principal plane.

Mohr's Circle shows that the major principal plane occurs at an angle
$\psi$ anti-clockwise from the vertical (see Figure \ref{mohr}). Thus
the major principal stress points along an angle $\psi$ anti-clockwise
from the horizontal. This is the standard physical interpretation of
$\psi$. One might think of $\psi$ as the angle from the horizontal
along which a force chain would be predicted to lie.

By right-triangle geometry, a line segment connecting the center of
Mohr's Circle to a point of tangency with the internal yield locus
would make an angle of $\pi/2-\phi$ with the $\sigma$-axis. Each point
of tangency represents a direction along which the yield criterion is
met, i.e. a slip-line.  Mohr's Circle indicates that the slip-lines
are angled $(\pi/2-\phi)/2$ up and down from the minor principal
plane. But since the major and minor principal planes are orthogonal,
the major principal stress points along the minor principal
plane. Defining $\epsilon=\pi/4-\phi/2$, we deduce that slip-lines
occur along the angles $\psi\pm\epsilon$ measured anti-clockwise from
the from the horizontal.  Looking back at the characteristic
equations, we see that the slip-lines and the characteristic curves
coincide.  This means that information from the boundary conditions
propagates along the slip-lines to form a full solution to the stress
balance equations.

\begin{figure}
\centering{\includegraphics[width=1.7in, clip]{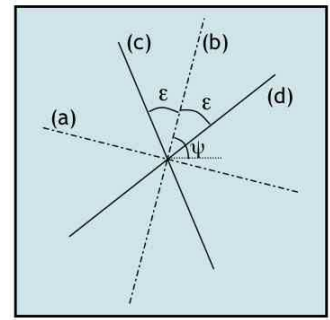}
\caption{Important lines intersecting each material point: (a) Major principal plane / Minor principal stress direction; (b) Minor principal plane / Major principal stress direction; (c)-(d) Slip-lines.}\label{frames}}
\end{figure}

It is worth noting that the stress balance equations are written for
static materials and do not appear to account for dynamic behavior
like dilatancy and convection stresses. The theory of critical state
soil mechanics \cite{wroth} was the first to rigorously approach the
issue of dilatancy (see appendix).  It concludes that when material
attains a flow state in which the density field stops changing in
time, all points in the flow lie along a \emph{critical state line} of
the form $|\tau/\sigma|=\delta$ for $\delta$ constant.  Since this
exactly mirrors the Coulomb yield criterion, we can keep the stress
balance equations and utilize $\delta= \mu$ (as in \cite{Jen}). As for
convection, adding the $\rho \ub\cdot\nabla\ub$ term into the stress
equations couples the stresses to the velocity and makes the problem
very difficult to solve. The practice of ignoring convection is
justified by our slow-flow requirement and is commonly used and
validated in basic solid mechanics literature \cite{hill,Sok,
  nedderman}. So we conclude that dynamic effects in flowing materials
do not preclude the use of the stress balance equations in slow,
steady flows.

\subsection{ Mohr-Coulomb plasticity: flow rules }

\begin{figure}
\centering{\includegraphics[width=3.2in, clip]{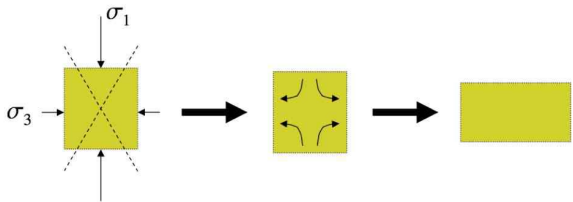}
\caption{Sketch of the Coaxial Flow Rule.}\label{coax}}
\end{figure}

To calculate flow, we assert incompressibility and a flow rule--- the
flow rule is a constitutive law chosen to reflect the general behavior
of the material at hand. The continuous nature of the ICM assumption
suggests that symmetry should be kept with respect to the principal
stress planes.  Based on this, Jenike proposed adopting the Coaxial
Flow Rule. The principle of coaxiality claims that material should
flow by extending along the minor principal stress direction and
contracting along the major principal stress direction; the principal
planes of stress are aligned with the principal planes of
strain-rate. The intuition for this constitutive rule is shown in
Figure \ref{coax}. Mathematically, this means that in a reference
frame where the minor and major principal stress directions are the
basis, the strain-rate tensor should have no off-diagonal components,
i.e.
\begin{equation}\label{rotate}
\mathbf{R_{\psi}}\mathbf{\dot{E}}\mathbf{R_{\psi}}^T \ \text{is diagonal,}
\end{equation}
where $\mathbf{R_\psi}$ rotates anti-clockwise by $\psi$ and $\mathbf{\dot{E}}$ is the strain-rate tensor
\begin{equation}
\mathbf{\dot{E}}=\frac{1}{2}\left(\nabla\ub+(\nabla\ub)^T\right).
\end{equation}
where $\ub=(u,v)$ is the velocity. Calculating the (1,2) component of
the matrix in equation (\ref{rotate}) and setting it to zero gives the
equation of coaxiality,
\begin{equation}\label{coaxial}
\frac{\partial u}{\partial y} +
\frac{\partial v}{\partial x}= \left(
\frac{\partial u}{\partial x}-\frac{\partial v}{\partial y}\right)\tan2\psi.
\end{equation}
This flow rule has played a dominant role in the development of
continuum plasticity theory and will be closely analyzed in this
work.

Coaxiality with incompressibility comprises another hyperbolic system
of equations enabling the velocity field to be solved via characteristics:
\begin{align}
 du +& \tan(\psi\mp\pi/4)\ dv = 0 \nonumber
\\
 &\text{along curves fulfilling}\ \
\frac{dy}{dx} = \tan(\psi\mp\pi/4).
 \end{align}
 So, given $\psi(x,y)$ from the stress balance equations, information
 about the flow travels from the boundaries into the bulk along curves
 rotated $\pi/4$ off from the principal stress planes--- using Mohr's Circle,
 we observe that these are the lines for which the shear stress is
 maximal (and the normal stresses are equal).

 Other flow rules have been suggested instead of coaxilaity. Of
 specific note, A.J.M. Spencer \cite{spencer64} has proposed the
 double-shearing flow rule.  Unlike coaxiality which can be understood
 as a simultaneous equal shearing along both slip-line families,
 double-shearing allows the shearing motion to be unequally
 distributed between the two families in such a way that the flow
 remains isochoric. For steady flows, the double-shearing flow rule is
\begin{eqnarray}\label{double}
& & \sin 2\psi\left(\frac{\partial u}{\partial x}-
\frac{\partial v}{\partial y}\right)-
\cos 2 \psi\left(\frac{\partial v}{\partial x}+ \frac{\partial u}{\partial y}\right) 
\nonumber \\
&=&\sin\phi\left(\frac{\partial v}{\partial x}
-\frac{\partial u}{\partial y}-2\ub\cdot\nabla\psi\right)
\end{eqnarray}
It can be seen that when the material neighboring a particle rotates
in sync with the rotation of the principal planes (i.e. as tracked by
the rate of change of $\psi$), the right side goes to zero and the
rule matches coaxiality. Under double-shearing, the characteristics of
stress and velocity align, easing many aspects of the numerics. Some
recent implementations of granular plasticity have utilized principles
of double-shearing \cite{anand00}. Though in this paper we deal
primarily with the comparison of coaxiality to our new theory, this
equation will be mentioned again in a later section.

\subsection{ The rate-independence concept }

We now more fully address the conceptual basis for the flow theory
just introduced.  The theory is fundamentally different from
traditional fluids where force-balance (including convection and
viscous stresses in the case of Newtonian fluids) can be used
alongside incompressibility to fully determine the fluid velocity and
pressure fields.  Unlike a fluid, granular materials can support a
static shear stress and thus force-balance plus incompressibility
alone is an underconstrained system. Rather, the stress-strain relationship
for granular material is presumed to be \emph{rate-independent} in
the slow, quasi-static regime we study.

This concept is best understood tensorially.  We can rewrite the
equations of coaxiality and incompressibility equivalently as: 
\begin{equation}\label{mises}
\mathbf{\dot{E}}=\lambda\mathbf{T_0}
\end{equation}
 where
\begin{eqnarray}
\mathbf{T}=\text{Stress tensor}=\left(
\begin{array}{cc}
-\sigma_{xx} &  \tau_{xy} 
\\
 \tau_{xy} &  -\sigma_{yy}
\end{array}\right)
\\
\mathbf{T_0}=\mathbf{T}-\frac{1}{2}(\text{tr}\mathbf{T})\mathbf{I}=\text{Deviatoric stress tensor,}
\end{eqnarray}
and $\lambda$ is a multiplier which can vary in space. Equation
(\ref{coaxial}) is merely the ratio of the (1,2) component and the
difference of the (2,2) and (1,1) components of equation
(\ref{mises}), thus cancelling $\lambda$, and incompressibility is
automatic since we relate to the deviatoric stress tensor. Equation
(\ref{mises}) gives a simple and highly general form for plastic
material deformation applicable to a broad range of deformable
materials and so it is ideal for illustrating the role of
rate-dependency. In MCP, we solve for $\mathbf{T_0}$ a priori from the
stress balance equations.  $\lambda$ adds the extra degree of freedom
necessary to make sure the strain-rate field is actually compatible
with a real velocity field--- $\lambda$ is not any specific function
of the stress or strain-rate variables and it adjusts to fit different
velocity boundary conditions.  Thus the stress alone does not imply the
strain-rate and vice versa.

Supposing on the other hand that we were dealing with a
\emph{rate-dependent} (i.e. visco-plastic) material like Newtonian
fluid, the above tensorial equation would still apply but we cannot
claim to know $\mathbf{T_0}$ in advance since material motion changes
the stresses. Instead we prescribe a functional form for $\lambda$,
like $\lambda=\text{{viscosity}}^{-1}=\text{constant}$, and write the
force balance equations in terms of $\mathbf{\dot{E}}$. Thus
$\mathbf{\dot{E}}$ is computed very differently for the two cases: in
the rate-independent case, (\ref{mises}) is solved using a known form
for $\mathbf{T_0}$ and in the rate-dependent case, (\ref{mises}) is
solved using a known form for $\lambda$.

The physical intuition for rate-independent \emph{flow} can be easily
understood with an example.  Suppose we slide two frictional blocks
against each other at two different non-zero sliding rates. In most
rudimentry dry friction laws, the shear stress required to slide one
block against another is proportional to the normal stress--- there is
no mention whatsoever of the rate of sliding.  Thus the two sliding
rates are modeled to be attainable with the same shear stress and
likewise the stress-strain relationship is deemed rate-independent.
For slow granular flows with long-lasting interparticle contacts,
comparisons with this example are especially instructive.



\section{ Shortcomings of Mohr-Coulomb plasticity }
\label{short}

The use of the stress balance equations with incompressibility and the
Coaxial Flow Rule will be referred hitherto as Mohr-Coulomb Plasticity
(MCP).  The theory has the benefit of being founded on mechanical
principles, but does have some marked drawbacks. We point out a few:

\begin{itemize}
\item The theory frequently predicts highly discontinuous velocity fields.
\item The Coaxial Flow Rule is conceptually troubling in some simple geometries.
\item The assumption of limit-state stresses is overreaching.
\item MCP is a continuum theory and thus cannot model discreteness and randomness.
\end{itemize} 

We will now discuss these four points in detail.

\subsection{ Discontinuous ``shocks'' in stress and velocity }
\begin{figure}
\centering{\includegraphics[width=3.1in, clip]{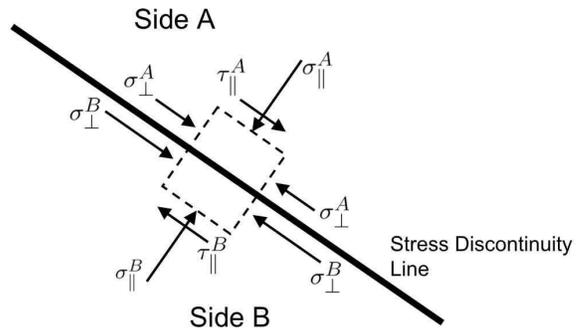}
\caption{Stresses on a control volume intersected by a discontinuity.  Note how a jump in  $\sigma_{\perp}$ places no net force on the control volume.}\label{discont}}
\end{figure}

The two stress PDEs and two flow PDEs are each fully hyperbolic
systems meaning that continuous solutions do not necessarily exist for
arbitrary choices of the boundary conditions.  Instead, discontinuous
solutions are constructed utilizing intuitive jump conditions. For
stresses, a jump in the stress parameters across a discontinuity line
is only allowable if such a jump places no net forces on a small
control volume surrounding the line thereby ensuring particle
stability.  This means the normal and shear stresses along the
direction of the discontinuity must be the same on both sides of the
discontinuity.  However, the normal stress along the perpendicular
direction can have a jump upon crossing the discontinuity as this
places no net force on the control volume (see Figure \ref{discont}).

In terms of the stress parameters, this means that $p$ and $\psi$ can
jump as long as
\begin{equation}
\frac{(1+\sin\phi\cos(2\Theta-2\psi^B))}{(1+\sin\phi\cos(2\Theta-2\psi^A))}=\frac{\sin(2\Theta-2\psi^B)}{\sin(2\Theta -2\psi^A)}=\frac{p^A}{p^B}
\end{equation}
where $\Theta$ is the angle from the vertical along which the
stress discontinuity lies. 

As for velocity, incompressibility forces us to impose a simpler jump
rule in that the component tangent to the velocity discontinuity is
the only one allowed to jump. We note that whenever a stress shock
exists, the jump in the stress parameters will usually place a jump in
the flow rule and may cause a velocity discontinuity to form
coincident with the stress shock. A velocity discontinuity can form
even when the stress field is smooth since the velocity PDEs are
themselves hyperbolic. (It is worth noting that when shocks are
allowed in the solution, multiple solutions sometimes arise to the
same problem; introduction of the so-called ``entropy condition'' can
be used to choose the best of the possible
solutions~\cite{Dre,nedderman}.)  Overall, the MCP equations are
mathematically very poorly behaved, and have also been shown to give
violent instabilities and finite-time
singularities~\cite{pitman87,schaeffer87}.

\begin{figure}
\begin{center}
\includegraphics[width=3in, clip]{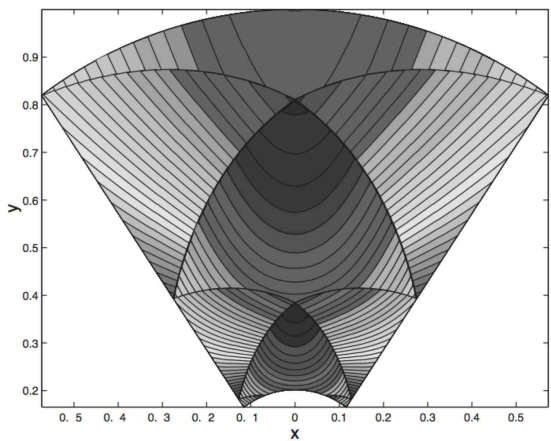}
  \caption{Numerical solution to MCP in a wedge hopper with non-radial
    stresses on the top boundary.  Normal stress in the radial
    direction displayed. (Courtesy of the authors of
    Ref.~\cite{gremaud}.) \label{disconthopper}}
\end{center}
\end{figure}

Aside from its mathematical difficulties, MCP theory also does not
match experiments or our everyday experience of granular flows. In
particular, MCP commonly predicts complicated patterns of velocity
discontinuities in situations where experiments indicate smooth flow
in steady-state.  In Figure \ref{disconthopper},  the numerically
determined stress field for a wedge hopper with only slightly
non-radial boundary conditions on the top surface exhibits a fan-like
array of shocks. The associated velocity field (not shown) will at
best exhibit a similar pattern of discontinuities and at worse add
even more discontinuities.  Such a broken velocity field is clearly
unphysical.  As the grain size becomes very small (sands),
discontinuous velocity fields with no relationship to the stress field
have been observed, but these are only temporary; the shocks commonly
blur away immediately after the onset of flow, which has been
attributed to some instability mechanism \cite{drescher2}. Literature
on the topic \cite{nedderman} is quick to concede that infinitessimally
sharp velocity jumps are physically nonsensical and should be understood
as being spread over at least a few particle widths.  Below, we will
see that our model naturally provides a mechanism for the blurring of
velocity shocks even in the presence of a stress shock, with large
velocity variations occurring only at the scale of several particle
diameters.

Typically, to avoid the task of having to track/capture shocks in the
stress/velocity field, approximations to MCP are invoked which give
continuous solutions either by altering the boundary conditions or
simplifying the PDE's. Smooth stress approximations are especially
useful when attempting to solve for the velocity field--- tracking
flow shocks coming from both a discontinuous stress field \emph{and}
hyperbolicity in the velocity equations is an enormous job.  To our
knowledge, a full solution to MCP has never been obtained
either numerically or analytically in cases where the underlying
stress field has shocks. Instead, shock-free approximate solutions
have mostly been pursued.

Arguably, the two most successful and commonly used results of
MCP are actually approximations, not full solutions.  The
Jenike Radial Solution \cite{Jen, jenike61} for wedge hopper flow
solves the MCP equations exactly, and with no
discontinuities, but does not allow for a traction free top
surface. It is a similarity solution of the form
\begin{eqnarray}
p=r f(\theta)
\\
\psi=g(\theta)
\\
\vec{v}=-\frac{h(\theta)}{r}\hat{r}
\end{eqnarray}
which reduces the entire system to 3 ODE's with $(r,\theta)$ the
position ($r=\text{distance from the hopper apex}$ and $\theta$ is
measured anticlockwise from the vertical). Though this solution
enables the material to obey a wall yield criterion along the hopper
walls, the stress parameters at the top surface have very little
freedom.  This is why most claim the Radial Solution to only hold near
the orifice, considerably away from the actual top surface. 

Another commonly used simplification is called the Method of
Differential Slices, although it only applies to stresses and not flow
(our focus here).  Originally proposed by Janssen in 1895 and
significantly enhanced since then, it is used to determine wall
stresses in bins and containers. The method makes some very
far-reaching assumptions about the internal stresses: $p$ is presumed
to only depend on height and the $\psi$ field is assumed to be
identically $\pi/2$ or $0$. These assumptions reduce the stress
balance equations to one ODE and ultimately give the famous result
that wall stresses increase up to a certain depth and then saturate to
a constant value. (This saturation behavior is not a byproduct of the
approximation; the discontinuous, full solution to the stress balance
equations in a bin also gives similar stress saturation behavior.)
While this effect has been verified extensively in experiment, the
underlying assumptions clearly cannot hold since, for example, the
walls exert an upward shear stress on the material which contradicts
the assumption about $\psi$.~\cite{nedderman}.

In summary, the equations of MCP theory have very limited
applicability to granular flows. There are very few, if any, solutions
available (either numerical or analytical) for many important
geometries such as planar or annular Couette cells, vertical
chutes, inclined places, etc. In the case of silos, MCP has been used
extensively to describe stresses, although the equations are difficult
to solve and poorly behaved from a mathematical point of view, as
noted above. There have also been some attempts to use MCP to describe
granular drainage from silos, in conjunction with the coaxial flow
rule, but this approach has met with little success, as we now
elaborate.

\subsection{Physical difficulties with coaxiality }

It is instructive to review the existing picture of silo drainage in
MCP theory, to highlight what we will view as a major concern in the
use of coaxiality for granular flows. Suppose we have a flat-bottomed
quasi-2D silo with smooth side-walls. Under standard filling procedures,
the walls provide only enough pressure to keep particles from sliding
farther out. These wall conditions, known as the ``active case'', give
the following solution to the stress balance equations as found by
marching down characteristics starting from the flat, pressure free,
top surface:

 \begin{eqnarray}
\psi(x,y)&=&\pi/2 \label{silostress1}
\\
p(x,y)&=&\frac{f_g y}{1+\sin\phi} \label{silostress2}
\end{eqnarray}
where $f_g\equiv\rho g$ is the weight density of the material and $y$
is positive downward. Since the $\psi$ field is identically
$\pi/2$ everywhere, the slip-lines are thus perfectly straight lines
angled at $\pm \epsilon$ from the vertical.
 
\begin{figure}
\begin{center}
(a)\includegraphics[width=1.5in, clip]{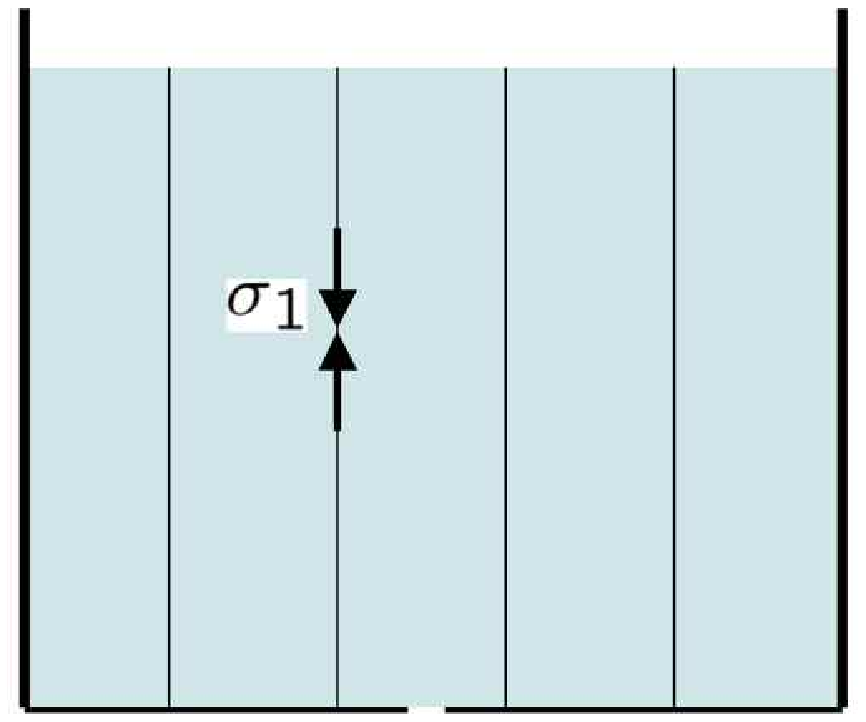}
(b)\includegraphics[width=1.5in, clip]{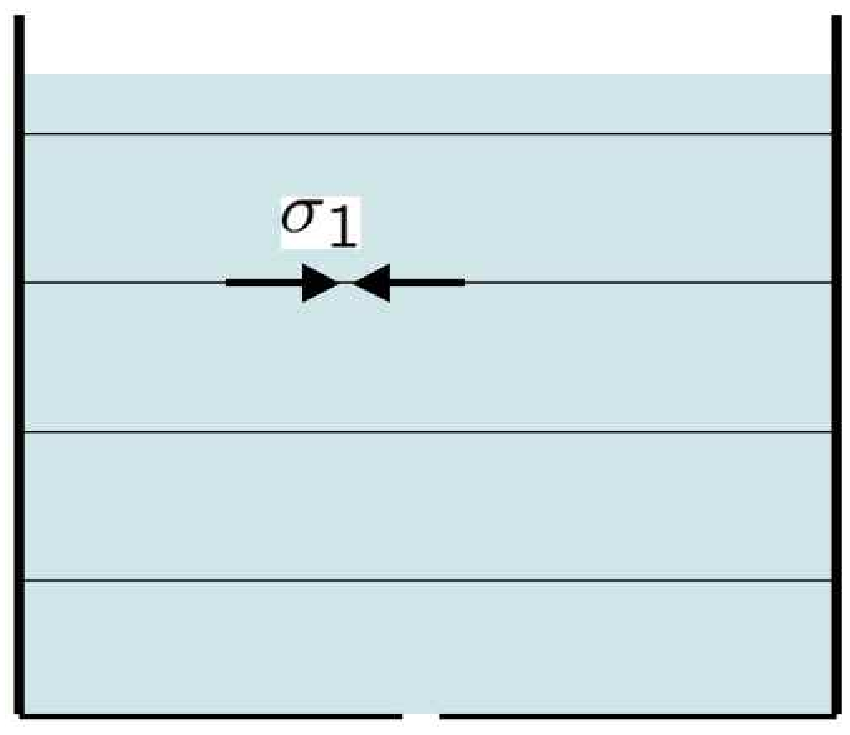}
\caption{Major principal stress chains in a quasi-2D silo for the (a) active case, and (b) passive case.}\label{activepassive}
\end{center}
\end{figure}

Refer again to Figure \ref{coax}.  The material deforms based solely
on principal plane alignment. For a slow, dense flow in the silo
geometry, coaxiality is troubling. Since the major principal stress is
everywhere vertical, coaxiality requires material to stretch
horizontally making it geometrically impossible for the material to
converge and exit through the silo orifice.  This issue has been
dodged historically by asserting that a drastic change in the wall
stresses occurs once the orifice opens, such that the walls drive the
flow, not gravity \cite{nedderman}.  The silo supposedly enters a ``passive
state'' where the walls are pushing so hard on the material that the
material is literally squeezed through the orifice by the walls.  Even
with this heavy-handed presumption, the solution predicted by equation
(\ref{coaxial}) is absurd; it predicts the only non-stagnant
regions in the silo are two narrow, straight channels which converge
on the silo opening and are angled at $\pm 45^{o}$ from the vertical.

Coaxiality also suffers thermodynamically. The equation itself only
ensures there is no shear strain-rate in the principal stress
reference frame and actually does not directly enforce that of the two
principal strain-rate axes, the axis of maximal compression (i.e. the
major principal strain-rate direction) must align with the major
principal stress direction, as was the physical intuition shown in
Figure \ref{coax}. Coaxiality just as easily admits solutions for
which the minor principal stresses align with the major principal
strain-rate. When this happens, the plastic power dissipated per unit
volume can be written

\begin{eqnarray*}
P&=&\mathbf{T}:\mathbf{\dot{E}}
\\
&=&\left(
\begin{array}{cc}
-\sigma_{1} & 0 
\\
 0 &  -\sigma_{3}
\end{array}\right):\left(
\begin{array}{cc}
|\dot{\gamma}| &  0 
\\
 0 &  -|\dot{\gamma}|
\end{array}\right)
\\
&=&|\dot{\gamma}|(\sigma_3-
\sigma_1)
\\
&<&0
\end{eqnarray*}
where $\mathbf{A}:\mathbf{B}=\Sigma A_{ij}B_{ij}$ and
$\pm|\dot{\gamma}|$ are the principal strain-rate values. This type of
behavior violates the second law of thermodynamics as it implies that
material deformation \emph{does} work on the system and likewise is
non-dissipative. In more advanced plasticity theories, the
thermodynamic inequality is upheld by requiring $\lambda$ in equation
(\ref{mises}) to be non-negative, but in the basic limit-state
framework we discuss, this contraint cannot be directly enforced.

We should quickly mention that in constructing the limit-state stress
field for the discharging silo, we have used as a boundary condition
that flow ensues when the pressure $p$ above the hole drops
differentially from the value it takes when the hole is closed.  This
claim allows us to preserve the continuous stress field described in equations 
(\ref{silostress1}) and (\ref{silostress2}) for slow, quasi-static
flow.

\subsection{ Incipient yield everywhere }

The constraint of assuming a limit-state stress field of incipient
yield everywhere is also questionable. Granular flows can contain
regions below the yield criterion within which the plastic strain-rate
must drop to zero.  For example in drainage from a wide silo with a
small orifice, the lower regions far from the orifice are completely
stagnant~\cite{samadani99,choi05}, and thus can hardly be considered
to be at incipient yield. In fact, discrete-element simulations show
that grains in this region essentially do not move from their initial,
static packing~\cite{rycroft06a}. Simulations also reveal that high
above the orifice, where the shear rate is reduced, the packing again
becomes nearly rigid~\cite{rycroft06b}, suggesting that the yield
criterion is not met there either. As the silo example illustrates, a
more general description of stresses coming from an elasto-plasticity
theory may be necessary to properly account for sub-yield material
\cite{hill,bazantsr}.

Elasto-plasticity also alleviates another major concern with MCP which
is that it is only well-defined in two dimensions. Three-dimensional
stress tensors have six free variables, too many to be uniquely
described by just force balance and incipient failure (altogether only
4 equations). We mention in passing that extensions of MCP to
axisymmetric three-dimensional situations have been developed. For
example, the Har Von Karman hypothesis, which assumes that the
intermediate principal stress $\sigma_2$, where
$\sigma_1\leq\sigma_2\leq\sigma_3$, is identical to either $\sigma_1$
or $\sigma_3$, is frequently utilized in solving for conical hopper
flow. However, elasto-plasticity can handle three dimensions without
this hypothesis, while also allowing for stress states below the yield
criterion in different regions.

\subsection{ Neglect of discreteness and randomness }

\begin{figure}
\centering{\includegraphics[width=1.5in, clip]{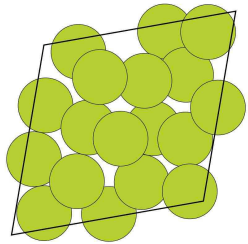}
  \caption{A meso-scale object containing a small number of randomly
    packed, discrete grains, which controls the dynamics of dense
    granular flow, analogous to the ``representative volume element''
    in classical continuum mechanics. }\label{balls}}
\end{figure}

Beyond its practical limitations and mathematical difficulties, the
most basic shortcomings of MCP are in its assumptions. Above all, a
granular material is not continuous. The microscopic grains composing
it are usually visible to the naked eye, and significant variations in
the velocity often occur across a distance of only several particle
lengths, e.g. in shear bands and boundary layers.  Of course, the
general theory of deterministic continuum mechanics is only expected
to apply accurately when the system can be broken into
``Representative Volume Elements'' (RVE's) of size $L$ fulfilling $d
\ll L \ll L_{macro}$ for $d$ the microscale (particle size) and
$L_{macro}$ the size of the system \cite{hill2}, which is clearly
violated in many granular flows. Therefore, the discrete, random
nature of the particle packing must play an important role in the
deformation process.  To incorporate this notion coherently, it may be
useful to seek out a dominant meso-scale object as a substitute for
the RVE, upon which mechanical flow ideas apply, but in a
non-deterministic, stochastic fashion (see Figure \ref{balls}).  This
concept is somewhat comparable to the ``Stochastic Volume Element'' in
the theory of plasticity of heterogeneous
materials~\cite{ostoja05}. In that setting, it is known that (what
physicists would call ``quenched'') randomness in material properties
leads to blurring of the slip-lines, but, to our knowledge, this
effect has not been considered in MCP for granular materials.

More importantly, however, since the meso-scale should only be a few
grains in width, there must also be randomness in the {\it dynamics}
of yielding to an applied stress or body force, since the theoretical
concept of a continuous slip-line is incompatible with the reality of
a discrete, random packing.  A stochastic response also seems
fundamentally more consistent with the assumption of inicipient yield:
If the material is just barely in equilibrium, it must be very
sensitive to small, random fluctuations, causing localized yielding.

We conclude that the shortcomings of MCP may be related to
the deterministic Coaxial Flow Rule, so we now proceed to replace it
with a more physically appropriate Stochastic Flow Rule. The
Mohr-Coulomb description of stresses is more clearly grounded in
principles of solid mechanics and is widely used in silo engineering,
so we will assume that it still holds, on average, in the presence of
slow flows, as a first approximation.

\section{ The stochastic flow rule}
\label{spotsec}
\subsection{ Diffusing ``spots'' of plastic deformation }

It has been noted in a variety of experiments that dense granular
flows can have velocity profiles which ressemble solutions to a
diffusion equation. By far the best example is drainage from a wide
silo, which has a well-known Gaussian profile near the orifice,
spreading vertically as the square root of the height (with parabolic
streamlines) in a range of
experiments~\cite{nedderman79,tuzun79,samadani99,medina98a,choi05}.
Recently, experiments in the split-bottom Couette geometry have
demonstrated precise error-function profiles of the velocity spreading
upward from the shearing circle, albeit with more complicated
scaling~\cite{fenistein03}. Shear bands, when they exist, tend to be
exponentially localized near moving rough walls, but we note that
these too can be viewed as solutions to a steady drift-diffusion
equation with drift directed toward the wall.

It seems, therefore, that a successful flow rule for dense granular
materials could be based on a stochastic process of deformation,
consistent with our general arguments above based on discreteness and
randomness. This begs the question: What is the diffusing carrier of
plastic deformation? In crystals, plasticity is carried by
dislocations, but it is not clear that any such defects might exist in
an amorphous material. The Gaussian velocity profile of granular
drainage was first explained independently by
Litwiniszyn~\cite{lit58,lit63b} and Mullins~\cite{mullins72} as being
due to the diffusion of voids upward from the orifice, exchanging with
particles to cause downward motion. However, this model cannot be
taken literally, since granular flows have nearly uniform density with
essentially no voids and with far less cage-breaking than the model
would predict.

\begin{figure}
\begin{center}
(a)
\\
\includegraphics[width=1.5in, clip]{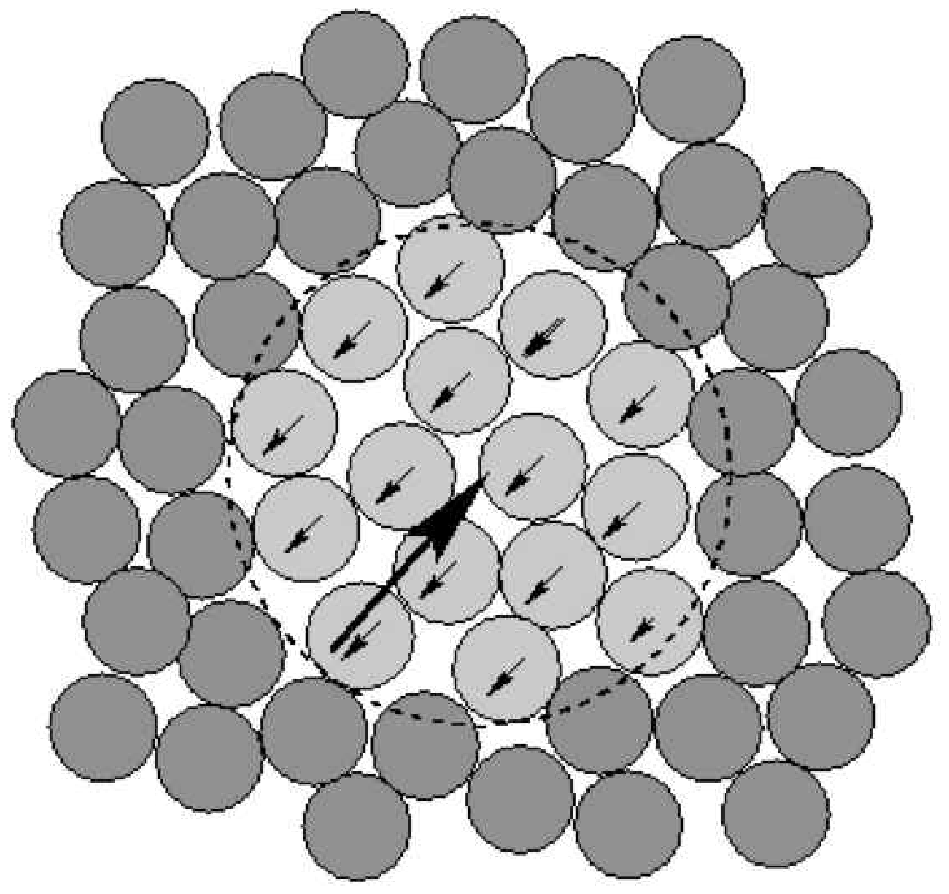}
\\
(b)
\\
\includegraphics[width=2.1in, clip]{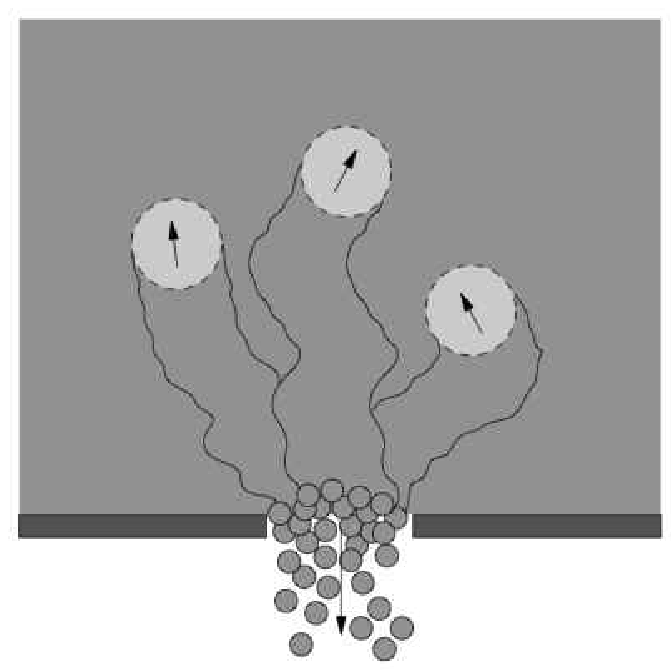}
\\
(c)
\\
\includegraphics[width=2.1in, clip]{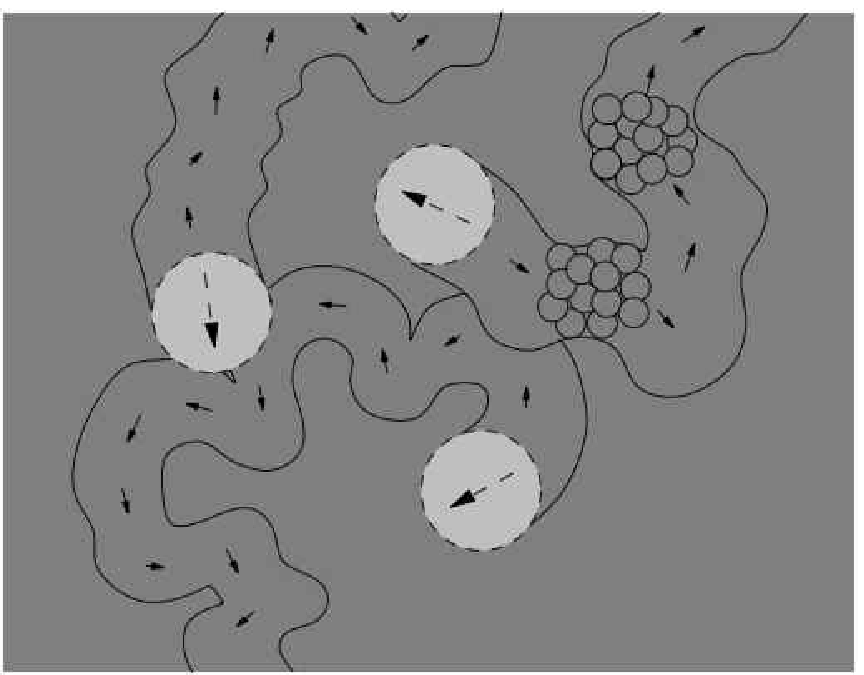} 
\caption{ Spots as carriers of plastic deformation in amorphous
  materials. (a) Cartoon of basic spot motion. A spot of local
  fluidization, carrying some free volume, moves to the upper right
  causes a cooperative displacement of particles, on average to the
  lower left, opposing the spot displacement. (b) In silo drainage,
  spots are injected at the orifice and perform random walks biased
  upward by gravity, causing downward motion of particles. (c) In
  other situations, we conjecture that spots are created during
  initial shear dilation and perform random walks biased by local
  stress imbalances and body forces during steady flow.}\label{spot}
\end{center}
\end{figure}



\begin{figure}
  \centering{\includegraphics[width=3in, clip]{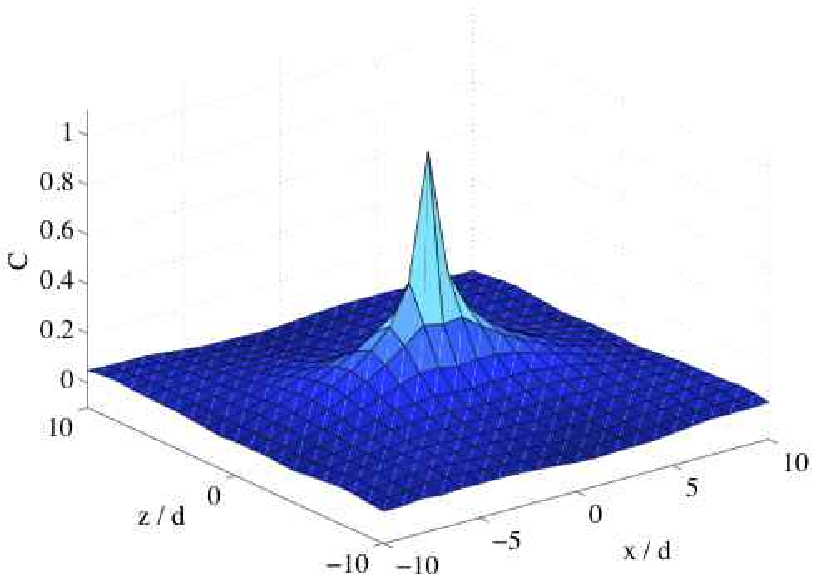}
\includegraphics[width=3in, clip]{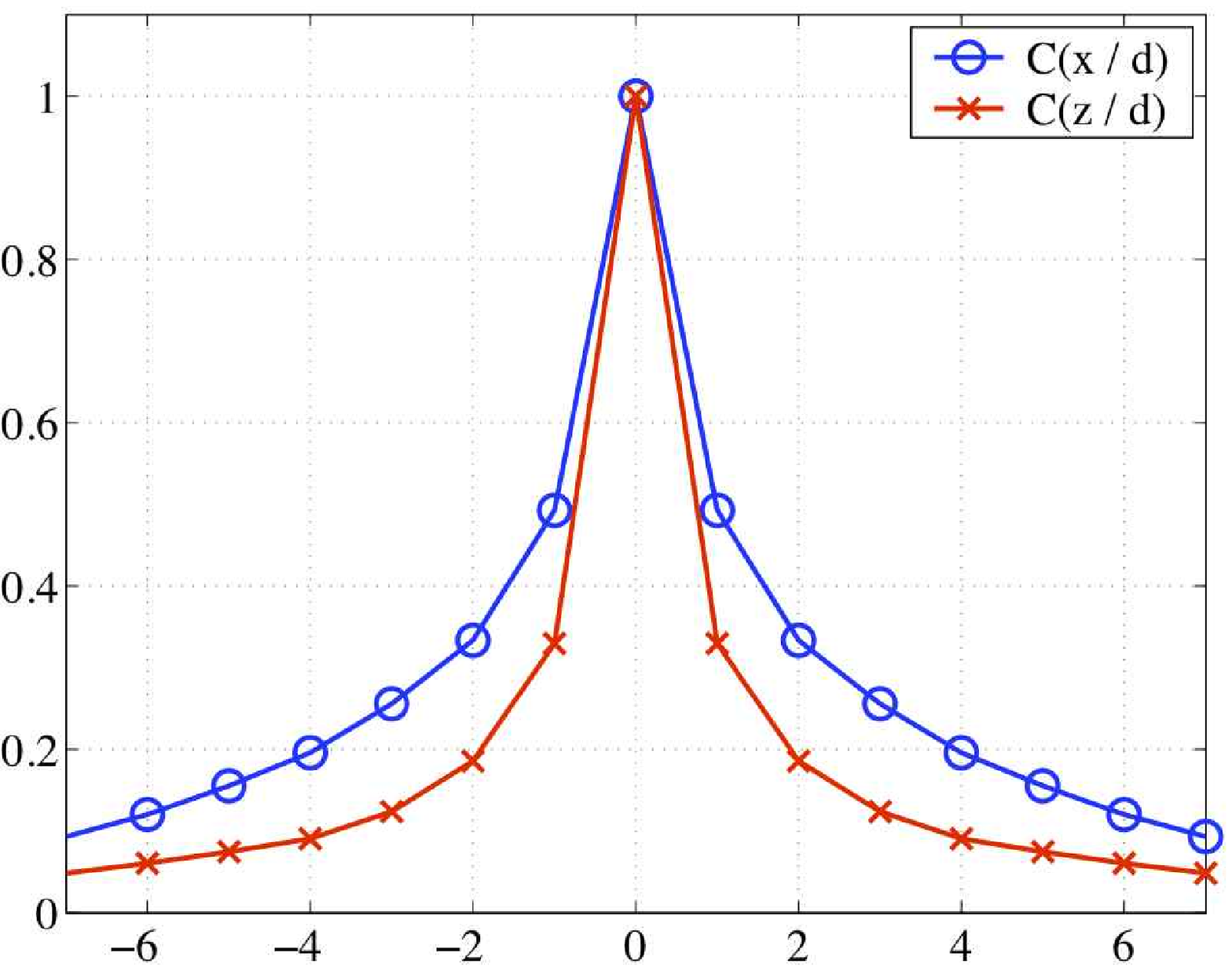}
    \caption{Spatial velocity correlations in silo drainage
      experiments as in Refs.~\cite{choi04,choi05} with glass beads
      ($d=3$mm) obtained high-speed digital-video particle tracking.
      The correlation coefficient of instantaneous displacements of a
      pair different particles is plotted as a function of their
      separation, averaged over all pairs and all times in the
      video. (Courtesy of the authors of
      Ref.~\cite{choi04}.)}\label{corr}}
\end{figure}

Instead, the starting point for our theory lies in the work of
Bazant~\cite{bazant06}, who proposed a general model for the flow of
amorphous materials (dense random packings) based on diffusing
``spots'' of cooperative relaxation, as illustrated in Figure
\ref{spot}. The basic idea is that each random spot displacement
causes a small block-like displacement of particles in the opposite
direction.  This flow mechanism takes into account the tendency of
each particle to move together with its cage of first neighbors, so
the size of a spot is typically three to five particle diameters, $L_s
\approx 3-5 d$. This expectation has been confirmed in dense silo
drainage as the length scale for spatial velocity correlations in the
experiments of Choi et al.~\cite{choi04,choi05} (data shown in
Figure~\ref{corr}) as well as the discrete-element simulations of
Rycroft et al.~ \cite{rycroft06a}.  In the continuum mechanics
terminology, this suggest that the spot may be an appropriate meso-scale
replacement for the RVE.

A major motivation for our work comes from the recent demonstration
by Rycroft et al. that the Spot Model can be used as a basis for
realistic multiscale simulations of dense granular drainage in a wide
silo, assuming that spots perform upward random walks, biased
uniformly by gravity~\cite{rycroft06a}. Using only five fitting
parameters (the size, volume, diffusivity, drift speed, and creation
rate of spots), the spot simulations were able to accurately reproduce
the statistical dynamics of several hundred thousand frictional,
visco-elastic spheres in discrete-element simulations of drainage from
a wide silo. This suggests that a general understanding of
dense granular flows may come from a mechanical theory of spot
dynamics. 

\subsection{ General form of the flow rule }

In this work, we propose such a mechanical theory, based on the
assumption that MCP provides a reasonable description of the mean
stresses in slow dense granular flows. The key idea
is to replace the Coaxial Flow Rule with a ``Stochastic Flow Rule''
based on mechanically biased spot diffusion. In the continuum
approximation, the general form of the flow rule thus consists of two
steps~\cite{bazant06}: (i) a Fokker-Planck (drift-diffusion) equation
is solved for the probability density (or concentration) of spots,
$\rho_s(\rb,t)$:
\begin{equation}
\frac{\partial \rho_s}{\partial t} + \frac{\partial}{\partial
  x^\alpha}\left(D_1^\alpha \rho_s\right) = \frac{\partial}{\partial
  x^\alpha}\frac{\partial}{\partial
  x^\beta} \left(D_2^{\alpha\beta} \rho_s\right)
\label{eq:spotfp}
\end{equation}
where $\{D_1^\alpha\}$ is the drift vector and $\{D_2^{\alpha\beta}\}$
the diffusivity tensor of spots, determined by the stress field
(below); and (ii) the mean drift velocity of particles $\ub = \{
u^\alpha\}$ is constructed to oppose the local flux of spots:
\begin{eqnarray}
u^\alpha &=& - \int d\rb^\prime \, w(\rb,\rb^\prime) 
\left[D_1^\alpha(\rb^\prime,t) \rho_s(\rb^\prime,t)  \right. \nonumber \\
& & \left. - \frac{\partial}{\partial x^\beta}\left(
  D_2^{\alpha\beta}(\rb^\prime,t)\rho_s(\rb^\prime,t)
\right)\right] \label{eq:up}  
\end{eqnarray}
where $w(\rb,\rb^\prime)$ is the (dimensionless) spot ``influence
function'' specifying how much a particle at $\rb$ moves in response
to a spot displacement at $\rb^\prime$.  Without making the continuum
approximation, the same physical picture can also be the basis for a
multiscale model, alternating between macroscopic continuum stress
computation and discrete spot-driven random-packing
dynamics~\cite{rycroft06a}.

The mean flow profile (\ref{eq:up}) is derived from a nonlocal
stochastic partial differential equation for spot-driven particle
dynamics, in the approximation that spots do not interact with each
other~\cite{bazant06}. Here, we have assumed the centered Stratonovich
definition of stochastic differentials~\cite{risken}, which means
physically that the spot influence is centered on its displacement. In
contrast, Bazant used the one-sided reverse-\^{I}to definition, where
the spot influence is centered on the end of its displacement, which
leads to an extra factor of two in the last term~\cite{bazant06}. This
choice is mathematically unrestricted (the ``stochastic
dilemna''~\cite{risken}), but we find the centered definition to be a
somewhat more reasonable physical hypothesis. Rycroft et al. have also
found that the centered definition produces more realistic flowing
random packings in multiscale spot simulations of granular drainage,
when compared to full discrete-element simulations
\cite{rycroft06a}. If we use the simple approximation $w\approx
\delta(|\rb-\rb'|)$ in the integral for particle velocity, the
Stratonovich interpretation has the benefit of automatically upholding
volume conservation.

Without even specifying how local stresses determine spot dynamics,
the general form of the flow rule (\ref{eq:up}) predicts continuous
velocity fields, even when the mean stresses are discontinuous. For
example, shocks in the MCP stress field may lead to
discontinuities in the spot drift vector, ${\bf D}_1$, and thus the
spot flux. However, due to the nonlocal nature of the spot model, the
particle flux is a convolution of the spot flux with the spot
influence function, thus preserving a continuous velocity field, which
varies at scales larger than the spot size, $L_s$. This is a direct
consequence of the geometry of dense random packings: The strong
tendency for particles to move with their nearest neighbors smears
velocity changes over at least one correlation length.

In the simplest approximation, the spot influence is translationally
invariant, $w = w(\rb - \rb^\prime)$, so that spots everywhere in the
system have the same size and shape. The spot influence decays off for
$r>L_s$, as a Gaussian among other possibilities.


In (\ref{eq:up}) we allow for the
likelihood that the spot influence may not be translationally and
rotationally invariant, e.g. since the local stress state always
breaks symmetry. This is actually clear in the experimental
measurements of Figure~\ref{corr}, where velocity correlations are
more short-ranged, without roughly half the decay length, in the
vertical direction (parallel to gravity) compared to the horizontal
(transverse) direction. This suggests that spots are generally
non-spherical and may be more elongated in the directions transverse
to their drift (or the body force). If anisotropy in the spot
influence were taken into account, it would also be natural to allow
for an anisotropic spot diffusivity tensor, which depends on the local
stress state. However, such effects seem to be small in the granular
flows we consider below, which are well described by a much simpler
model.

Another likely possibility is that spots come in a range of shapes and
sizes. There could be a statistical distribution of regions of local
fluidization or plastic yield related to the local packing and stress
state. It is thus more reasonable to think of the spot influence
function $w(\rb,\rb^\prime)$ as averaging over this distribution, just
as a spatial velocity correlation measurement averages over a large
number of collective relaxations. One advantage of taking the
continuum limit of a Fokker-Planck equation (\ref{eq:spotfp}) in
applying the SFR is that such details are buried in the coefficients,
which could in in principle be derived systematically from any
microscopic statistical model, or simply viewed as a starting point
for further physical hypotheses (as we do below).

Finally, we mention that there are also surely some nontrivial
interactions between spots, which would make the SFR nonlinear and
could lead to interesting new phenomena, such as spontaneous pattern
formation. For example, spots may have a medium range attraction,
since it is more difficult to propagate particle rearrangements and
plastic deformation into less dense, less mobile regions; there could
also be a short range repulsion if the spot density gets too high,
since grains will collapse into overly dilated regions.  Such effects
may be responsible for intermittent density waves in draining hoppers
with narrow orifices~\cite{behringer89,poschel94}, and perhaps even
shear banding in other amorphous materials, such as metallic glasses,
with a different plastic yield criterion. However, we will see that
the hypothesis of non-interacting spots already works rather well in
cases of steady, dense granular flows.

\subsection{ A simple model for steady flows }

Due to efficient dissipation by friction, granular materials subjected
to steady forcing typically relax very quickly to a steady flowing
state.  For example, when a silo's orifice is opened, a wave of
reduced density (spots) progates upward, leaving in its wake a nearly
steady particle flow, which we associate with a steady flow of
spots. This initial density wave can be seen very clearly in
discrete-element simulations of various hopper-silo
geometries~\cite{rycroft06b}.  For a narrow orifice, we have noted
that intermittent flows with density waves can be
observed~\cite{behringer89,poschel94}, but typical drainage flows are
rather steady in time~\cite{choi04,choi05}. Similarly, when a rough
inner cylinder is set into uniform rotation in a Couette cell, shear
dilation propagates outward, raising the level of the packing, until a
steady flow profile is reached. We interpret this initial dilation as
signaling the creation of spots on the cylinder, which quickly reach a
steady distribution in the bulk.

Hereafter, we focus on describing steady flow profiles, with
equilibrium spot densities. For simplicity, let us assume isotropic
spot diffusion, $D^{\alpha\beta} = D_2 \delta_{\alpha\beta}$, since
fluctuations are dominated by the (largely isotropic) geometry of
dense random packings. Using the spot size $L_s$ as the natural length
scale, we can express the spot drift speed, $|\Db_1| = L_s/\Delta t_1$,
and diffusivity, $D_2 = L_s^2/2\Delta t_2$, in terms of the times, $\Delta t_1$
and $\Delta t_2$, for drift and diffusion to reach this length.

The flow profile of a draining silo, normalized by the outflow speed,
is approximately constant over a wide range of flow speeds, as has recently
been verified to great precision in the experiments of Choi et
al. ~\cite{choi04}. Not only is the mean velocity profile independent
of flow rate (over an order of magnitude in mean velocity), but
fluctuations about the mean, such as vertical and horizontal diffusion
and measures of ``cage breaking'', also depend only on the distance
dropped, and not explicitly on time (or some measure of ``granular
temperature''). In statistical terms, changing the flow rate is like
watching the same movie at a different speed, so that the random
packing goes through a similar sequence of geometrical configurations
regardless of the velocity. Similar features have also been observed
in shearing experiments in Couette cells \cite{mueth03} and numerical
simulations of planar shear \cite{lois05}.

The experimental and simulational evidence, therefore, prompts the
crucial approximation that $\Delta t_1/\Delta t_2=$ constant so as to
uphold statistical invariance of the particle trajectories under
changing the overall flow rate. This can be justified if spots perform
random walks with displacements selected from a fixed distribution,
set by the geometry of the random packing~\cite{bazant06}. Here, we
make the stronger assumption that the characteristic length of these
random walks is the spot size $L_s$, so that $\Delta t_1=\Delta
t_2\equiv\Delta t$. Our physical picture is that a spot represents a
``cell'' of localized fluidization (or plastic yield) of typical size
$L_s$, which triggers further fluidization ahead of it and randomly
propagates to a neighboring cell of similar extent. This picture is
also consistent with the interpretation of $L_s$ as a velocity
correlation length above. 

With these hypotheses, the Fokker-Planck equation (\ref{eq:spotfp})
takes the simple time-independent form,
\begin{equation}
\nabla\cdot(\ds \, \rho_s) = \frac{L_s}{2} \nabla^2 \rho_s \label{rhofield}
\end{equation}
where $\ds(\rb) = \Db_1/|\Db_1|$ is the spot drift direction,
determined by the mechanics of plastic yielding (below). The flow
field is then
\begin{equation}
\ub = - \frac{L_s}{\Delta t} \int d\rb^\prime \, w(\rb,\rb^\prime) 
\left(\ds(\rb^\prime) \rho_s(\rb^\prime) - \frac{L_s}{2}
  \nabla \rho_s(\rb^\prime) \right) \label{eq:u}
\end{equation}
Equations (\ref{rhofield}) and (\ref{eq:u}) define a simplified
Stochastic Flow Rule, with only one parameter, $L_s$, which need not
be fitted to any flow profile. Instead, it can be measured
independently as the velocity correlation length, which may be viewed
as a dynamical material property.

\subsection{ A mechanical theory of spot drift }

The main contribution of this paper is a simple theory connecting the
spot drift direction to the stresses in MCP. The basic idea is to view
the displacement of a spot as being due to a local event of material
failure or fluidization. To make a quantitative prediction, we first
define a cell of the material as the roughly diamond shaped region
encompassed by two intersecting pairs of slip-lines, separated by
$L_s$. When a spot passes through this cell, it fluidizes the material
and thus locally changes the friction coefficient from the static
value $\mu$ to the kinetic value $\mu_k$.  This upsets the force
balance on the cell and may cause a perturbative net force to
occur.

\begin{figure}
\begin{center}

(a)\includegraphics[width=2.8in, clip]{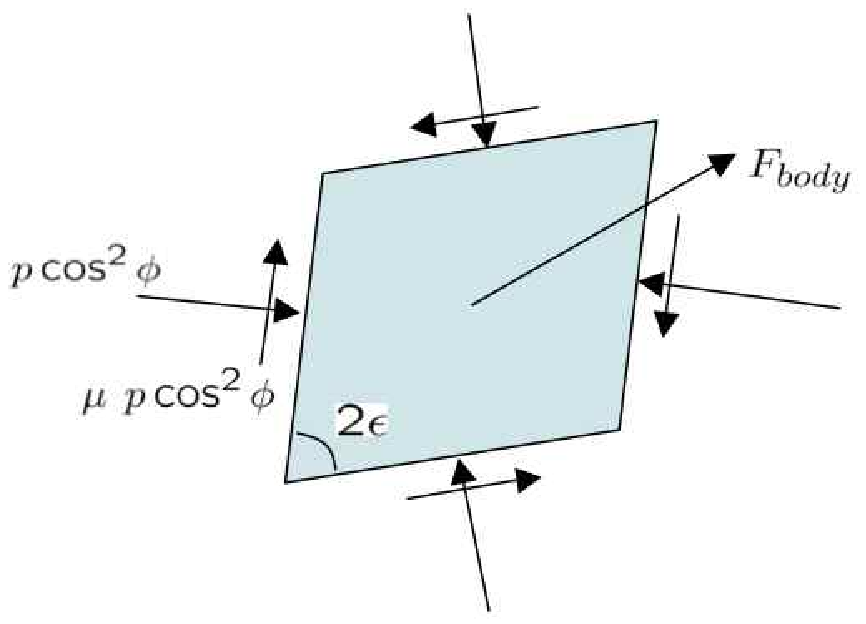}
\\
(b) \includegraphics[width=2.8in, clip]{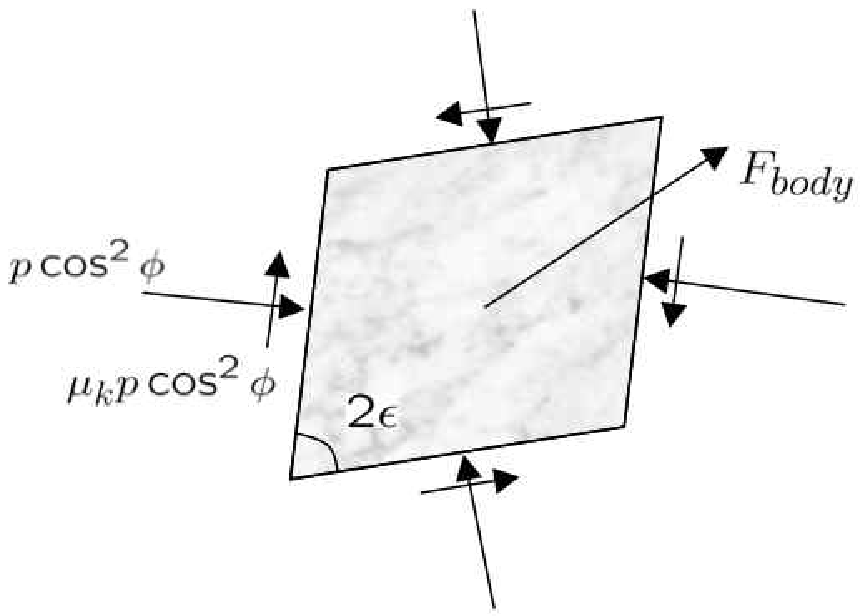}
\\
(c) \includegraphics[width=3.2in, clip]{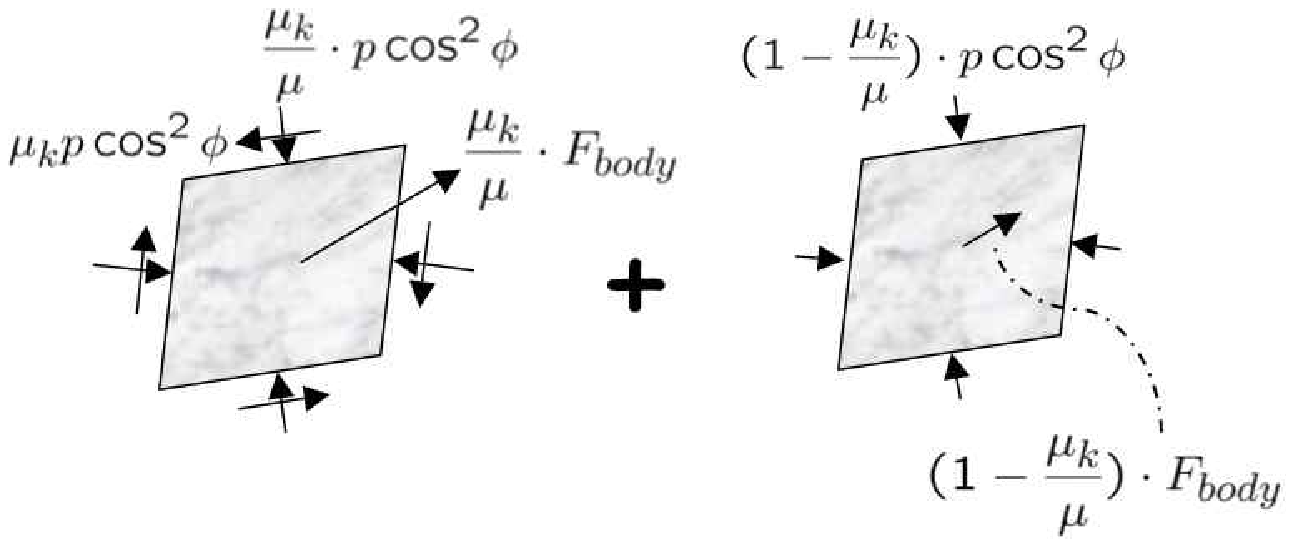}
 \caption{(a) Material cell in static equilibrium.(b) A spot enters the
   cell fluidizing the material.  In the force diagram, this means
   $\mu$ decreases to $\mu_k$. (c) The force diagram for the fluidized
   material cell is best analyzed by breaking it into the sum of two
   diagrams.}\label{fluidize}
\end{center}
\end{figure}

The force diagram for a material cell occupied by a spot can be broken
into the sum of two diagrams (Figure \ref{fluidize}), one which is the
static diagram multiplied by $\mu_{k}/\mu$ and one which contains only
normal contact force contributions and a body force term.  MCP requires the static diagram to be balanced, thus the latter
is the only cause for a net force.  A well-known corollary of the
divergence theorem enables us to express the surface integral of
normal contact stresses in terms of a gradient of $p$ giving
\begin{equation}\label{fnet}
\vec{F}_{net}=\left(1-\frac{\mu_{k}}{\mu}\right)
\left(\vec{F}_{body}-\cos^2\phi\vec{\nabla}p\right)
\end{equation}
as an effective force which pulls on a cell as it is fluidized by a
passing spot, causing the spot to preferentially drift in the opposite
direction.

A spot cannot move in an arbitrary direction, however, since the
material is at incipient yield only along the two slip
lines. Therefore, the net force is constrained to pull the material
cell along one of the slip-lines.  The spot drift direction is then
obtained by projecting (minus) the force, $-\vec{F}_{net}$, onto the
slip-lines and averaging these two projection vectors with equal
weight:
\begin{eqnarray}
 \vec{\xi}^{(\pm)} &=& - (\vec{F}_{net}\cdot \hat{n}_{\psi\pm\epsilon})\hat{n}_{\psi\pm\epsilon}
\\
  \ds &=&\frac{\vec{\xi}^{(+)}+\vec{\xi}^{(-)}}{\left|\vec{\xi}^{(+)}+\vec{\xi}^{(-)}\right|} \label{bias}
\end{eqnarray}
where $\hat{n}_{\theta}=(\cos\theta,\sin\theta)$.  With a formula for
$\ds$ now determined, the SFR as stated in equations (\ref{rhofield}) and
(\ref{eq:u}) is now fully defined and ready for use.

This continuum mechanical theory of spot drift also helps us
understand the sources of spot diffusion. As noted above and sketched
in Figure~\ref{balls}, a material cell is a small fragment of a random
packing, which is unlikely to be able to accomodate shear strain
precisely along the slip-lines of the {\it mean} continuum stress
field. Instead, the instantaneous slip-lines are effectively blurred
by the discrete random packing.  Still, we preserve the picture of
spots moving along slip-lines in constructing $\ds$, but represent the
additonal bluriness in the slip-line field by enforcing isotropic spot
diffusivity.

\subsection{ Frame indifference }

Finally, we must check that our flow rule satisfies frame
indifference; solving for flow in different rigidly moving reference
frames cannot give different answers in a fixed reference frame.
Since the SFR is a 2D steady-state flow rule, the only flows we need
to check for indifference are those with rotational/translational
symmetry.  In these cases, the particle velocity is a function of only
one spatial variable and equation (\ref{rhofield}) for $\rho_s$ becomes
a second-order ODE.  In solving the boundary value problem, we must
ensure that grains along the walls have a velocity vector tangent to
the walls. This constrains one of the two degrees of freedom in the
set of possible solutions for $\rho_s$.  Since (\ref{rhofield}) is
homogeneous, the other degree of freedom must come out as a
multiplicative undetermined constant. Thus the velocity profile is
unique up to a multiplicative constant.

With only one constant, we cannot match boundary conditions for
particle speed along more than one wall in general.  So to solve for a
flow between two walls, we must add rigid-body motions to the
reference frame of the observer until we have the unique frame for
which a solution exists matching both boundary conditions.  This is an
unexpected and welcome bonus of the SFR.  Most flow rules in continuum
mechanics enforce material frame indifference directly, i.e. the flow
rule itself is derived to be automatically satisfied by any rigid-body
motion, ensuring the same results independent of reference frame.
Coaxiality achieves this by relating stress information only to
strain-rate variables for instance. The SFR, however, upholds frame
indifference indirectly in that the solution does not exist
\emph{unless} the problem is solved in exactly one ``correct'' frame
of reference.

We have thus integrated the spot concept with the theory of plastic
stresses treating spots as the ``carriers of plasticity''. We note
that up to our granular-specific determination of the drift direction,
the SFR principle can be applied to any amorphous isotropic material
with a small characteristic length scale (dominant randomness) and a
yield criterion.

\section{Applications to Granular Flow} 
\label{expt}

The Stochastic Flow Rule is quite general and in principle can be
applied to any limit-state plasticity model of stresses, with
different choices of the yield function to describe different
materials. In this section, we apply the simplest SFR
(\ref{rhofield})--(\ref{eq:u}) to granular materials with MCP stresses
and compare its predictions to a wide range of existing experimental
data for steady dense flows. In calculating stresses, we assume a
typical friction angle of $\phi=30^o$.  It is known that for spherical
grains, the friction angle usually lies in a somewhat narrow range of
about $20^o-30^o$ and can be as large as $50^o$ for some anisotropic,
highly angular materials \cite{nedderman}.  In the examples we
consider, however, varying the $\phi$ value in this range has very
little macroscopic effect in our model.

The spot size $L_s$ has a much larger effect, so we focus on its role
in a variety of dense flows. We emphasize that we do not fit $L_s$ to
any flow profile below. Instead, we simply use the range $L_s=3-5d$
for dense flowing sphere packings inferred independently from spatial
velocity correlations in silo drainage experiments~\cite{choi04} and
simulations~\cite{rycroft06a} (see Fig.~\ref{corr}). This is
consistent with our view of the correlation length, $L_s$, as a
fundamental geometrical property of a flowing granular material.

Without any fitting parameters, we will apply the simple SFR to
several prototypical flows. Each has different forcing and symmetries
and, to our knowledge, they cannot be simultaneously described by any
existing model.  Our first example is granular drainage to a small
orifice in a wide flat-bottomed silo, driven entirely by gravity. Our
second example is shear flow in an annular Couette cell driven by a
moving rough inner cylinder, where gravity plays no role. Our third
example is the dragging of a loaded plate over a semi-infinite
material at rest, which combines gravitational forces and boundary
forcing. Lastly we apply the SFR to a canonical free-surface flow, the
continuous avalanching of a granular heap.  The transition to a 
rapidly flowing surface shear layer on a heap will also lead us into a
discussion of how rate-dependent effects, such as Bagnold rheology, may
naturally extend into our model.

Throughout our treatment of the various examples, the first step will
be to solve (\ref{rhofield}) to obtain the ``unconvolved'' velocity
field
\begin{equation}
\ub^*=-L_s \ds \rho_s + \frac{L_s^2}{2}\nabla \rho_s
\end{equation}
which corresponds to the SFR velocity (\ref{eq:u}) for a point-like
influence function $w=\delta(|\rb-\rb'|)$. For the most part, $\ub^*$
is the ``skeleton'' for the full solution $\ub$ because convolving
$\ub^*$ with a general spot influence merely blurs out the sharper
features of $\ub^*$. In some situations with wide shear zones, such as
silo flow, the convolution has only a minor effect, but in others
with narrow shear bands, at the scale of the spot size, the
convolution step is essential for self-consistency and accuracy.

\subsection{ Silos }

\begin{figure*}
\begin{center}
\includegraphics[width=6in, clip]{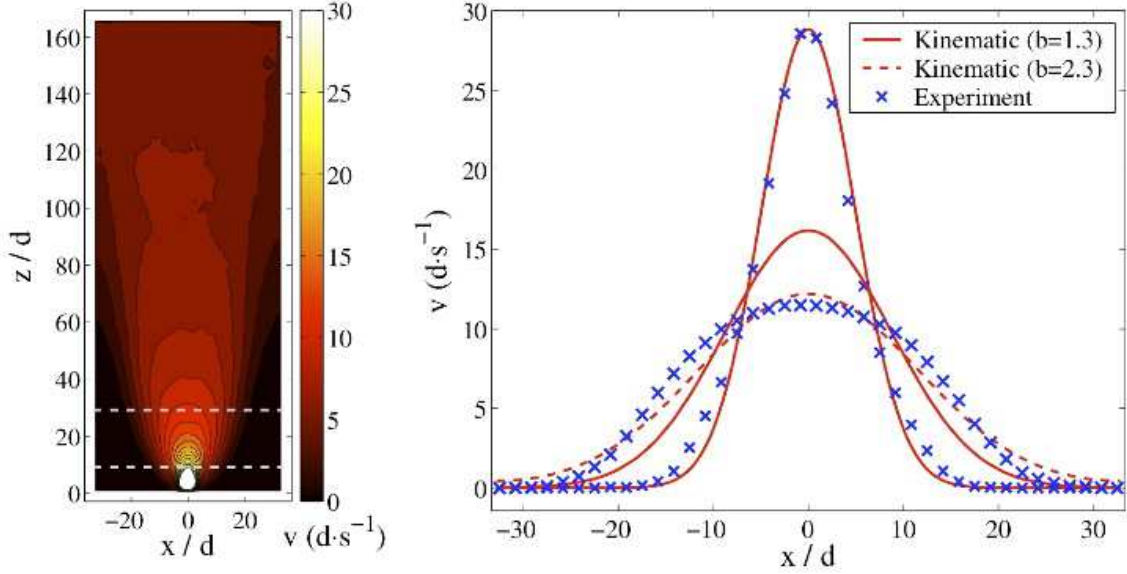}
\caption{ The mean velocity profile in a wide quasi-2d silo of 3mm
  glass beads from Ref.~\cite{choi05}. Horizontal slices of the
  downward velocity component near the orifice, indicated in the
  complete flow profile on the left, are shown on the right, and
  compared to the predictions of the Kinematic Model with two choices
  of the parameter $b$. The Stochastic Flow Rule for MCP for a wide
  silo (without side walls) gives a similar velocity field with
  $b\approx 1.5-2.5d$.}
\label{spread}
\end{center}
\end{figure*}

The flow profile in a flat-bottom silo geometry is well-known for its
striking similarity to the fundamental solution of the diffusion
equation. As noted above, early models of silo flow explained this
based on the upward diffusion of voids from the
orifice~\cite{lit58,lit63b,mullins72}. Without reference to a specific
microscopic mechanism, Nedderman and T\"{u}zun later derived the same
equations based on a continuum constitutive
law~\cite{nedderman79,nedderman}. They asserted that the horizontal
velocity component $u$ is proportional to the horizontal gradient of the
downward velocity component $v$,
\begin{equation}
u = b\, \frac{\partial v}{\partial x}
\end{equation}
since particles should drift from regions of slow, dense flow toward
regions of faster, less dense (more dilated) flow. Assuming small
density fluctuations, mass conservation applied to the 2D velocity field,
$\ub=(u,-v)$ then yields the diffusion equation for the downward velocity,
\begin{equation}
\frac{\partial v}{\partial z}=b\, \frac{\partial^2 v}{\partial x^2}
\end{equation}
where the vertical direction $z$ acts like ``time''. The diffusivity
$b$ is thus really a ``diffusion length'', to be determined
empirically. An advantage of the continuum formulation is that it
avoids the paradox (resolved by the Spot Model~\cite{bazant06}) that
the classical picture of void random walks requires $b \ll d$, while
experiments invariably show $b > d$.

Solving the Kinematic Model in the wide flat-bottomed silo geometry with a
point orifice gives the familiar Green function for the diffusion
equation,
\begin{equation}
v(x,z)=\frac{e^{-x^2/4 b z}}{\sqrt{4 \pi b z}}. \label{eq:kmg}
\end{equation}
This gives an excellent match to experimental data close to the
orifice (e.g. see Figure \ref{spread}), although the fit gradually
gets worse with increasing height, as the flow becomes somewhat more
plug-like. Nevertheless, Gaussian fits of experimental data have
provided similar estimates of $b = 1.3d$ \cite{choi04}, $1.3-2.3d$
\cite{choi05}, $2.3d$ \cite{nedderman79}, $3.5d$ \cite{samadani99},
and $2d-4d$ \cite{medina98a} for a variety of granular materials
composed of monodisperse spheres.

\begin{figure}
\begin{center}
\includegraphics[width=2.7in, clip]{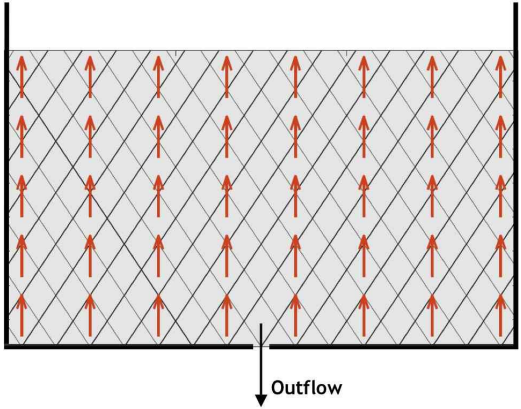}
\caption{ The flat-bottomed silo geometry.  The intersecting black
  lines represent the slip-line field as determined from solving the
  stress balance equations of MCP, and the red vector field is the
  spot drift direction, as determined from the SFR. In this highly
  symmetric geometry, the spot drift precisely opposes the
  gravitational body force, $\hat{d} = -\hat{g}$. }\label{silodrift}
\end{center}
\end{figure}

We now apply our theory to this flow geometry and see how it connects
to the Kinematic Model. Applying equation (\ref{bias}) using the
stress field described by equations (\ref{silostress1}) and
(\ref{silostress2}) gives uniform upward spot drift; $\vec{F}_{net}$
comes out as pointing uniformly downward and the slip-lines are
symmetric about the vertical (see Figure \ref{silodrift}). The SFR
(\ref{rhofield}) then reduces to 
\begin{equation}
  \frac{\partial \rho_s}{\partial
    z}=\frac{L_s}{2}\left(\frac{\partial^2\rho_s}{\partial x^2} +
      \frac{\partial^2\rho_s}{\partial z^2}\right) \label{eq:silorho}
\end{equation}
although we emphasize that this form applies only when the walls are
smooth or equivalently when the silo width is large. The last term, which
represents vertical diffusion of spots (relative to the mean upward
drift), makes this equation for the spot density differ somewhat from
the simple diffusion equation for the downward velocity of the
Kinematic Model. Consistent with our model, vertical diffusion, with a
similar (but not identical) diffusion length as horizontal diffusion,
has been observed in recent silo-drainage
experiments~\cite{choi04,choi05}.

The general solution of (\ref{eq:silorho}) can be expressed as a
Fourier integral,
\begin{equation}
 \rho_s=\frac{1}{2\pi}\int_{-\infty}^{\infty}e^{i k
   x}A(k)e^{\frac{z}{L_s}(1-\sqrt{1+L_s^2k^2})}dk
\end{equation}
where $A(k)$ is the Fourier transform of the spot density at the
bottom ($z=0$). The narrowest possible orifice allowing for flow is
the case of a point source of spots, $\rho_s(x,0) \propto \delta(x)$,
$A(k) \propto 1$ (which is also the Green function). Convolution with
a spot influence function of width $L_s$ produces a downward velocity
profile on the orifice of width $L_s$. Unlike the Kinematic Model (or
any other continuum model which does not account for the finite grain
size), our theory thus predicts that flow cannot occur unless the
orifice is at least as wide as one spot, $L_s = 3-5 d$.

The details of flow very close to the orifice, $z = O(L_s)$, are
controlled by the choice of boundary condition, reflecting the
dynamics of dilation, contact-breaking, and acceleration at the
orifice, which are not described by our bulk dense-flow model. Rather
than speculate on the form of this boundary condition, we focus on the
bulk region slightly farther from the orifice. For $z \gg L_s$ (and
$L_sk \ll 1$), the vertical diffusion term becomes unimportant, and
the Green function tends to a Gaussian
\begin{equation}
v(x,z) \sim \frac{e^{-x^2/2\sigma_v^2(z)}}{\sqrt{2\pi\sigma_v^2(z)}} \label{eq:vsfr}
\end{equation}
where the variance is 
\begin{equation}
\sigma_v^2(z) \sim L_s z + O(L_s^2).
\end{equation}
(The second term is an offset from convolution with the spot influence
function, which also depends on the choice of boundary conditions.)

There has been no prior theoretical prediction of the kinematic
parameter $b$, which we interpret as the spot diffusion
length~\cite{bazant06}. Comparing (\ref{eq:kmg}) and
(\ref{eq:vsfr}), we obtain $b = L_s/2 = 1.5-2.5 d$ without any
fitting, beyond the independent determination of $L_s$ from velocity
correlations. This prediction is in excellent agreement with the
experimental measurements listed above. However, the model does not
predict the apparent increase of $b$ with height, as the flow becomes
more plug like. This may be due to the breakdown of the assumption of
incipient yield higher in the silo, where the shear is greatly
reduced, and it may require modeling stresses more generally with
elasto-plasticity. 

In any case, we are not aware of any other model of silo flow with a
plausible basis in mechanics. It is noteworthy that we assume
\emph{active} silo stresses (driven by gravity), as typically assumed
in a quasi-static silo. As a result, we do not require a sudden switch
to passive stresses (driven by the side walls) upon flow initiation,
as in existing plasticity theories based on the Coaxial Flow
Rule~\cite{nedderman}. Our use of the standard MCP model for stresses
in quasi-static silos also suggests that the SFR may predict
reasonable dependences on the geometry of the silo or hopper, wall
roughness, and other mechanical parameters. In contrast, the Kinematic
Model fails to incorporate any mechanics, and, not surprisingly, fails
to describe flows in different silo/hopper geometries in
experiments~\cite{choi05}. Testing our model in the same way would be
an interesting direction for future work, since it has essentially no
adjustable parameters.

\subsection{ Couette cells }

The key benefit of our model is versatility; we will now take exactly
the same model, which is able to describe wide silo flows driven by
gravity, and apply it to Taylor-Couette shear flows in annular cells
driven by a moving boundary. The granular material is confined between
vertical rough-walled concentric cylinders and set into motion by
rotating the inner cylinder. The flow field has been characterized
extensively in experiments and simulations, and several theories have
been proposed for this particular
geometry~\cite{losert00,bocquet02,latzel03,midi04}. For example, a
good fit of experimental data for Couette cells can be obtained by
postulating a density and temperature dependent viscosity in a fluid
mechanical theory~\cite{bocquet02}, but it is not clear that the same
model can describe any other geometries, such as silos, hoppers, or
other shear flows.

\begin{figure}
\begin{center}
(a) \\
\includegraphics[width=3.1in, clip]{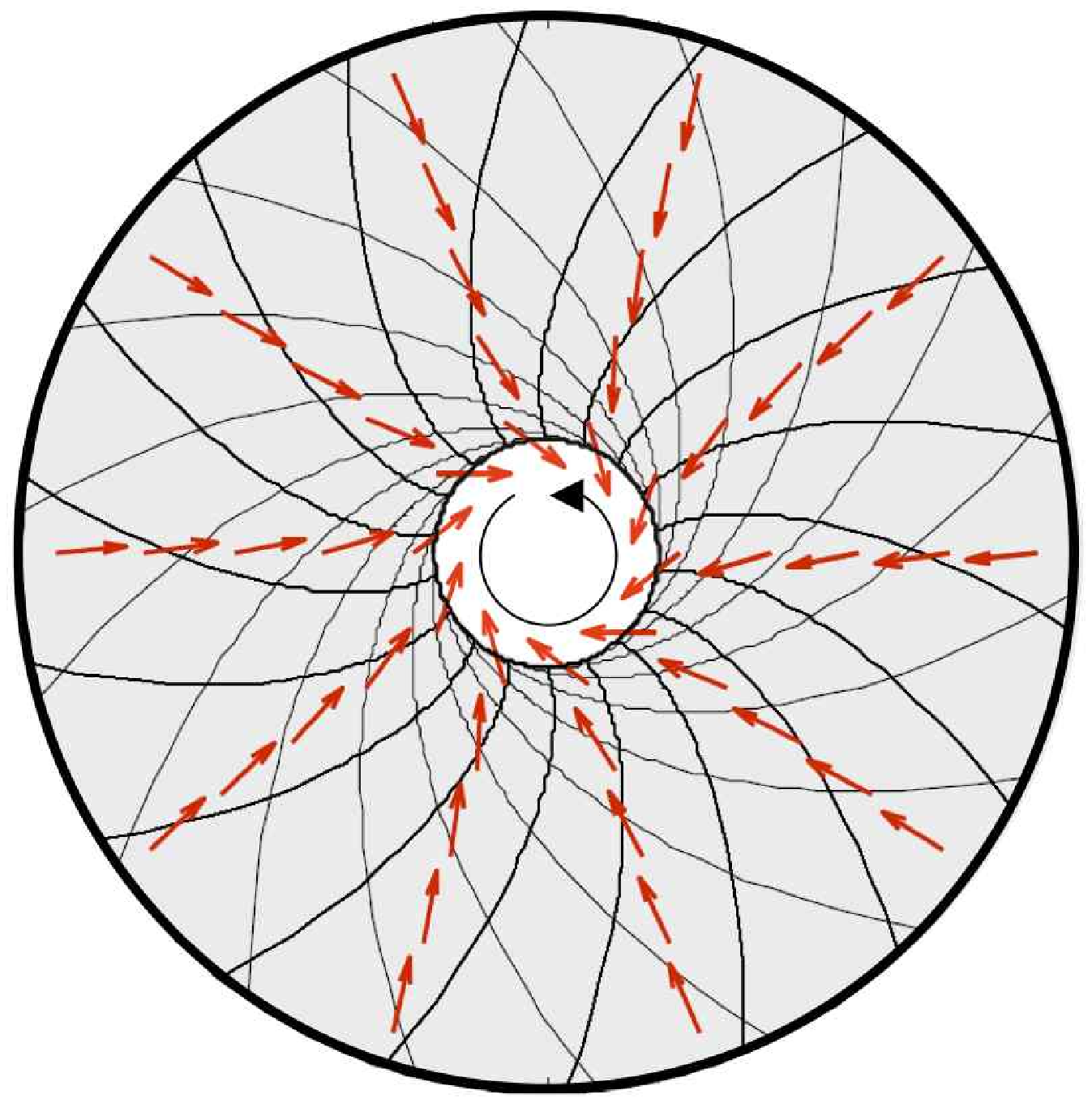}
\\
(b) 
\includegraphics[width=3.2in, clip]{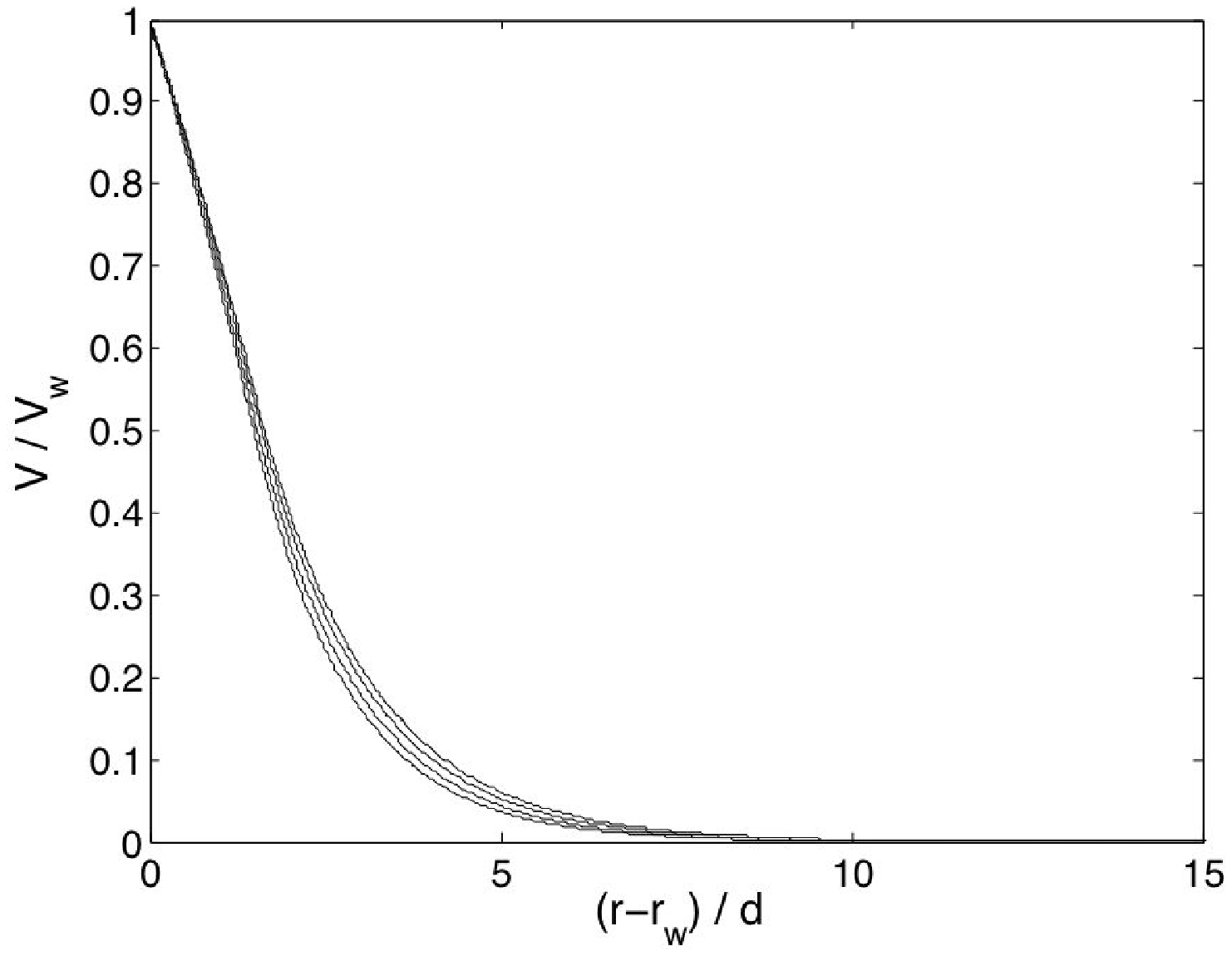}
\caption{(a) A plan view of the annular Couette cell geometry, where
  the granular material is confined between concentric vertical
  cylinders.  The rough wall is rotated anti-clockwise while the outer
  wall is held fixed. The crossing black lines within the material are
  the slip-lines as found from MCP, and the red vector field is the
  spot drift as determined by the SFR. (b) Normalized SFR velocity
  from as a function of distance from the inner wall with inner
  cylinder radius 15d, 25d, 50d, and 100d (from bottom to top curves,
  respectively). The friction angle is $\phi=30^\circ$, and the spot
  size is $L_s=3d$.  }
\label{couetteSFR}
\end{center}
\end{figure}

To solve for the MCP stresses in the annular Couette geometry, we
first convert the stress balance equations to polar coordinates
$(r,\theta)$ and require that $p$ and $\psi$ obey radial symmetry.
This gives the following pair of ODEs:
\begin{eqnarray}
\frac{\partial\psi^*}{\partial r}&=&-\frac{\sin 2\psi^*}{r(\cos 2\psi^* +\sin\phi)}
\\
\frac{\partial\eta}{\partial r}&=&-\frac{2\sin \phi}{r(\cos 2\psi^*+\sin \phi)}
\end{eqnarray}
where $\eta=\log p$ and $\psi^*=\psi+\frac{\pi}{2}-\theta$. Although
$\psi^*$ has an implicit analytical solution, $\eta$ does not, so we
solve these equations numerically using fully rough inner wall
boundary conditions $\psi^*(r_w)=\frac{\pi}{2}-\epsilon$ and any
arbitrary value for $\eta(r_w)$. The resulting slip-lines are shown in
Figure ~\ref{couetteSFR}(a). 

In the Couette geometry, the average normal stress, $p$, decreases with
radial distance, which implies that the fluidization force on
material, $\vec{F}_{net}$, is everywhere directed outward.  We then
apply equation (\ref{bias}) to calculate the drift direction $\ds(r)$
by projecting the vector $\vec{F}_{net}$ onto slip-lines, and implement the SFR,
exploiting symmetry which allows only a nonzero velocity in the
$\hat{\theta}$ direction. This implies
\begin{equation}
\ub^*\cdot\hat{r}=0=-L_s(\ds\cdot\hat{r})\rho_s+\frac{L_s^2}{2}\frac{\partial
  \rho_s}{\partial r}
\end{equation}
which yields a solution for $\rho_s$ up to a scalar factor. We then
use $\rho_s$ to calculate the $\theta$ component of the (unconvolved)
velocity once again using the SFR equation,
\begin{equation}
\ub^*\cdot\hat{\theta}=-L_s(\ds\cdot\hat{\theta})\rho_s.
\end{equation}
It turns out, as we may have expected, that $\ub^*$ has a shear band
at the inner wall with nearly exact exponential decay. The length
scale of this decay is the spot size, $L_s$, since this is
the velocity correlation length, beyond which the inadmissible shear
at the inner cylinder can be effectively dissipated by the
material. 

The thinness of the shear band requires that, for consistency, we must
take into account the finite spot size in reconstructing the velocity
field through the convolution integral (\ref{eq:u}).  For simplicity
we will use a uniform spot influence function, i.e.
\begin{equation}
w(\rb)=\frac{4}{\pi L_s^2}\, H\left(\frac{L_s}{2}-|\rb|\right)
\end{equation}
where $H(x)$ is the Heaviside step function. To evaluate the integral
(\ref{eq:u}), we also must make a hypothesis about how spots operate
when they overlap one of the walls. Random packing dynamics near walls
is different than in the bulk and sensitive to surface roughness, and
further detailed analysis of experiments and simulations will be
required to elucidate the collective mechanism(s). Here, the precise
shape of spots near the wall has little effect, except to flatten out
the spike in velocity that occurs near the wall in the unconvolved
velocity $\ub^*$. As simple first approximation, used hereafter in
this paper, we will view the space beyond each boundary as containing
a bath of non-diffusive spots at uniform concentration whose flux is
such that the particle velocity invoked ``inside'' the boundaries
directly mimics the rigid motion of the walls. This effectively
``folds'' part of the spot influence back into the granular material
when it overlaps with the wall. The resulting velocity field is shown
in Figure \ref{couetteSFR}(b), where normalized velocity is shown
versus distance from the inner cylinder wall for $L_s=3d$ for a wide
range of inner cylinder radii.

\begin{figure}
\begin{center}
(a) \\
\includegraphics[width=3.1in, clip]{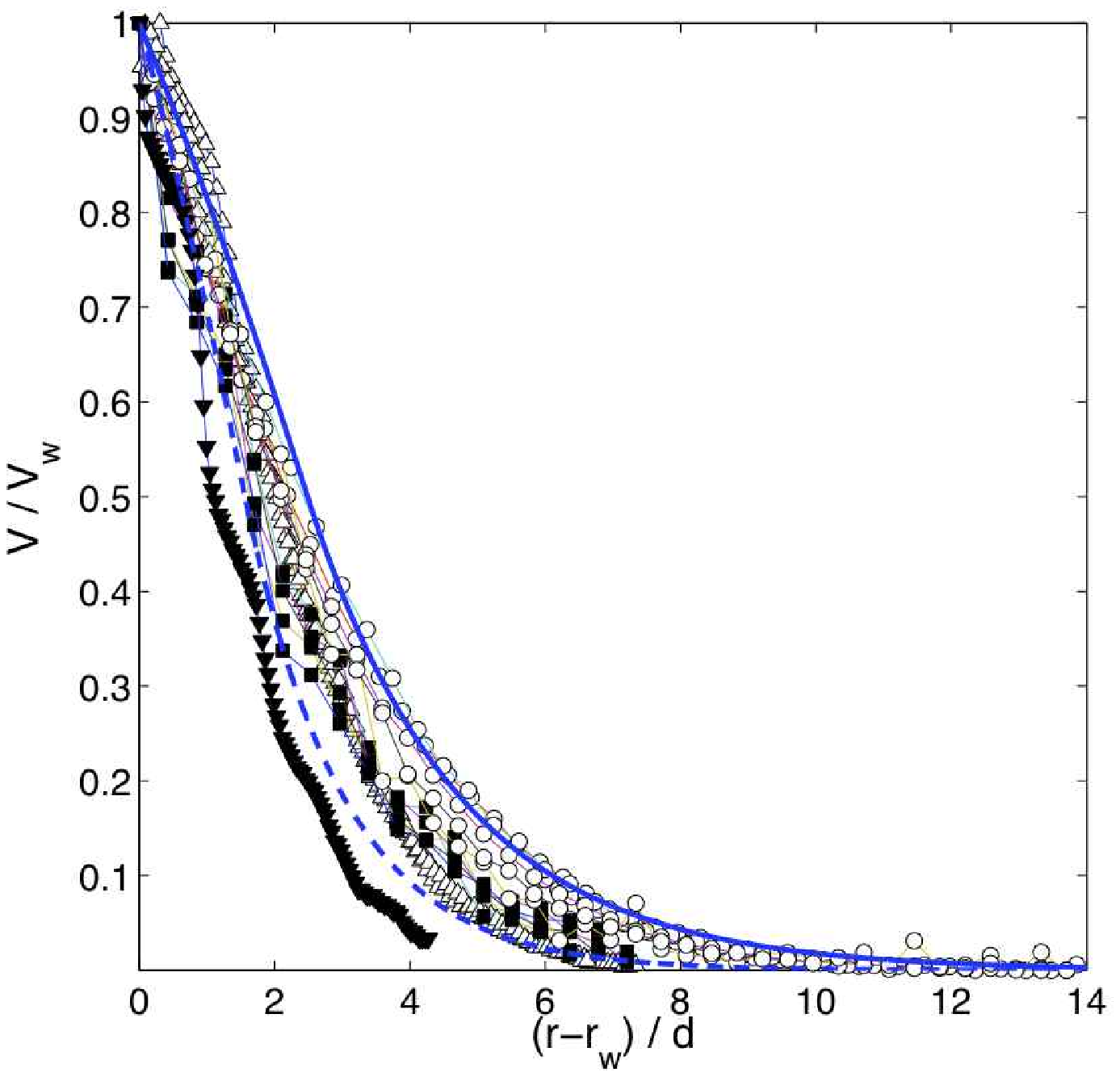}\\
(b) \includegraphics[width=3.3in, clip]{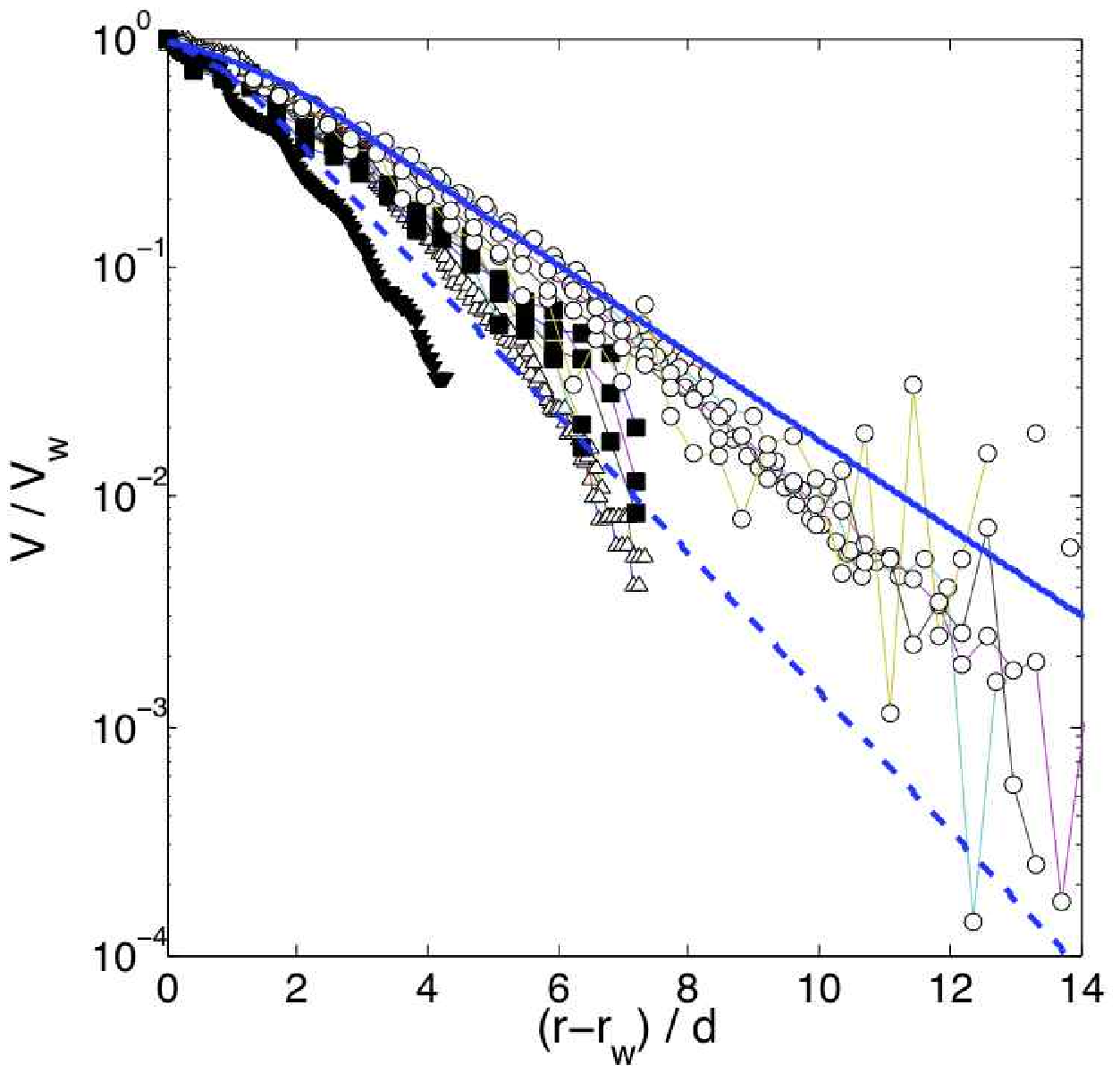}
\caption{ Theory versus experiment for the normalized velocity in
  annular Couette cells on (a) linear and (b) semilog plots. The
  dashed curve is the predicted SFR velocity field with $L_s=3d$,
  while the solid line is for $L_s=5d$; both curves are for an inner
  cylinder radius of $r_w=80d$ and $\phi=30^\circ$. Experimental
  measurements (points) for a wide range of inner and outer cylinder
  radii are shown from the compilation of data shown in Figure 3c of
  \cite{midi04}. (The experimental data is courtesy of GDR Midi and
  originates from the work of \cite{dacruz_phd}, \cite{bocquet01},
  \cite{mueth00}, and \cite{chambon03}.)  }\label{annular}
\end{center}
\end{figure}

The predicted flow field -- without any fitting -- is in striking
agreement with a large set of data from experimental and
discrete-element simulations for different cylinder radii and grain
sizes~\cite{losert00,bocquet01,latzel03,midi04}. As shown in
Figure~\ref{annular}, the experimental data compiled by GDR
Midi~\cite{midi04} falls almost entirely within the predicted SFR
velocity profiles, by setting the spot size to the same typical range
of correlation lengths, $L_s=3-5 d$, measured independently in a
quasi-2D silo (Figure~\ref{corr}). Viewing the data on a semilog plot
shows that the agreement extends all the way into the tail of the
velocity distribution. We emphasize that the same simple theory, with
the same range of $L_s$ values, also accurately predicts silo flows
above, as well as other situations below. Unifying all of this data in
a single simple theory without any empirical fitting constitutes a
stringent quantitative test.

It is interesting to note the behavior close to the wall, especially
in thin Couette cells. In experiments~\cite{midi04}, annular flow
profiles are known to have a Gaussian correction term when the
thickness of the cell becomes non-negligible in terms of particle
size. This slight flattening near the wall is apparent in our model as
well and is a byproduct of convolving with the spot influence. We thus
interpret this feature as another sign of the strongly correlated
motion of particles, primarily with the ``cage'' of nearest neighbors,
as approximately described by the spot mechanism. In this calculation,
we used a uniform spot influence, but have noticed relatively little
sensitivity of the predicted flow profile, for different strongly
localized influence functions, such as a Gaussian, $w \propto
e^{-2r^2/L_s^2}$. A detailed comparison of the model to experimental
data may provide fundamental insights into the spot influence, and
thus the collective dynamics of random packings, near a rough wall at
the discrete particle level.

The experimental results shown in Figure~\ref{annular} come from
apparati with inner wall radii ranging from $14d-100d$. The relatively
small variations in the data sets over such a large range of inner
radii clearly indicates that the inner wall radius is not a crucial
length scale in the flow. The plotted theoretical prediction uses an
inner radius of $\approx 80d$, but, as can be seen in Figure
\ref{couetteSFR}(b), our results depend only minimally on the inner
cylinder radius.  Indeed, the meso-scale correlation length of $L_s=
3-5d$ is the dominant length scale in our theory for this
geometry.

\begin{figure}
\begin{center}
  \includegraphics[width=3.2in, clip]{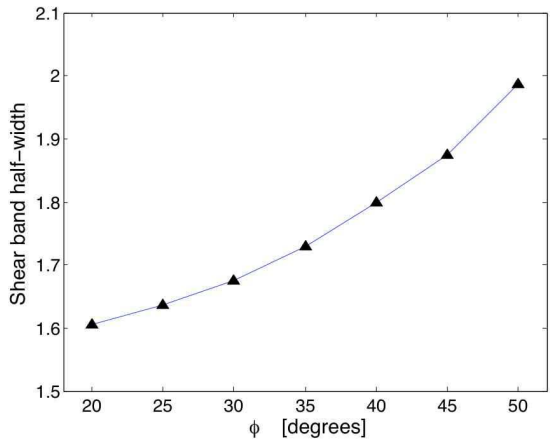}
  \caption{Predicted variation in the width of the shear band with SFR
    over the standard range of granular friction angles for the
    annular Couette geometry. ($L_s=3d$,
    $r_w=80d$)}\label{frictionband}
\end{center}
\end{figure}

To substantiate an earlier claim, we now consider how the friction
angle $\phi$ affects the flow properties (holding $L_s$ fixed)
according to our model. We can see this most clearly by observing how
the shear band half-width (i.e. the distance from the wall to the
location where velocity is half-maximum) varies over the $\phi$ range
for usual granular materials ($\approx 20^o-50^o$).  As shown in
Figure \ref{frictionband} the half-width changes by $<0.4d$ over the
entire range and by $<0.1d$ for the range of laboratory-style
spherical grains. This very weak influence of internal friction agrees
with simulations in the Couette geometry by Sch\"ollmann
\cite{schollmann98}.

\subsection{ Plate dragging }

We now examine perhaps the simplest situation where gravity affects
the shear band caused by a moving rough wall. Consider slowly dragging
a rough plate horizontally across the upper surface of a deep
(semi-infinite) granular material. The plate maintains full contact by
pressing down on the surface with pressure $p_0\cos^2\phi$. The profile of the
shear band which forms below the plate depends on the relative loading
pressure, $q_0 = p_0/f_g$, where $f_g$ is the weight (gravitational
body force) density of the material.

\begin{figure*}
\begin{center}
(a) \includegraphics[width=2.6in, clip]{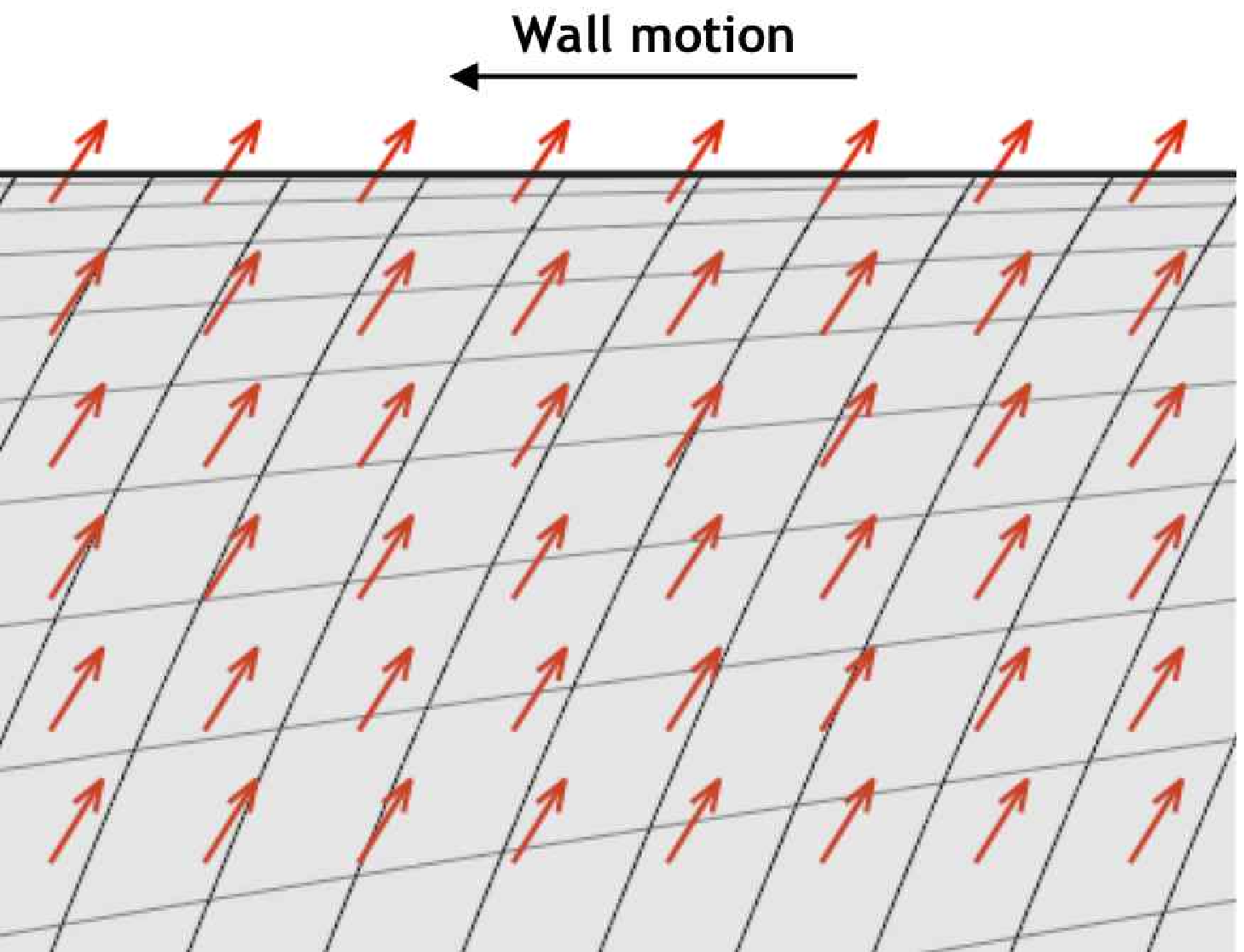}
\\
(b) \includegraphics[width=3.2in, clip]{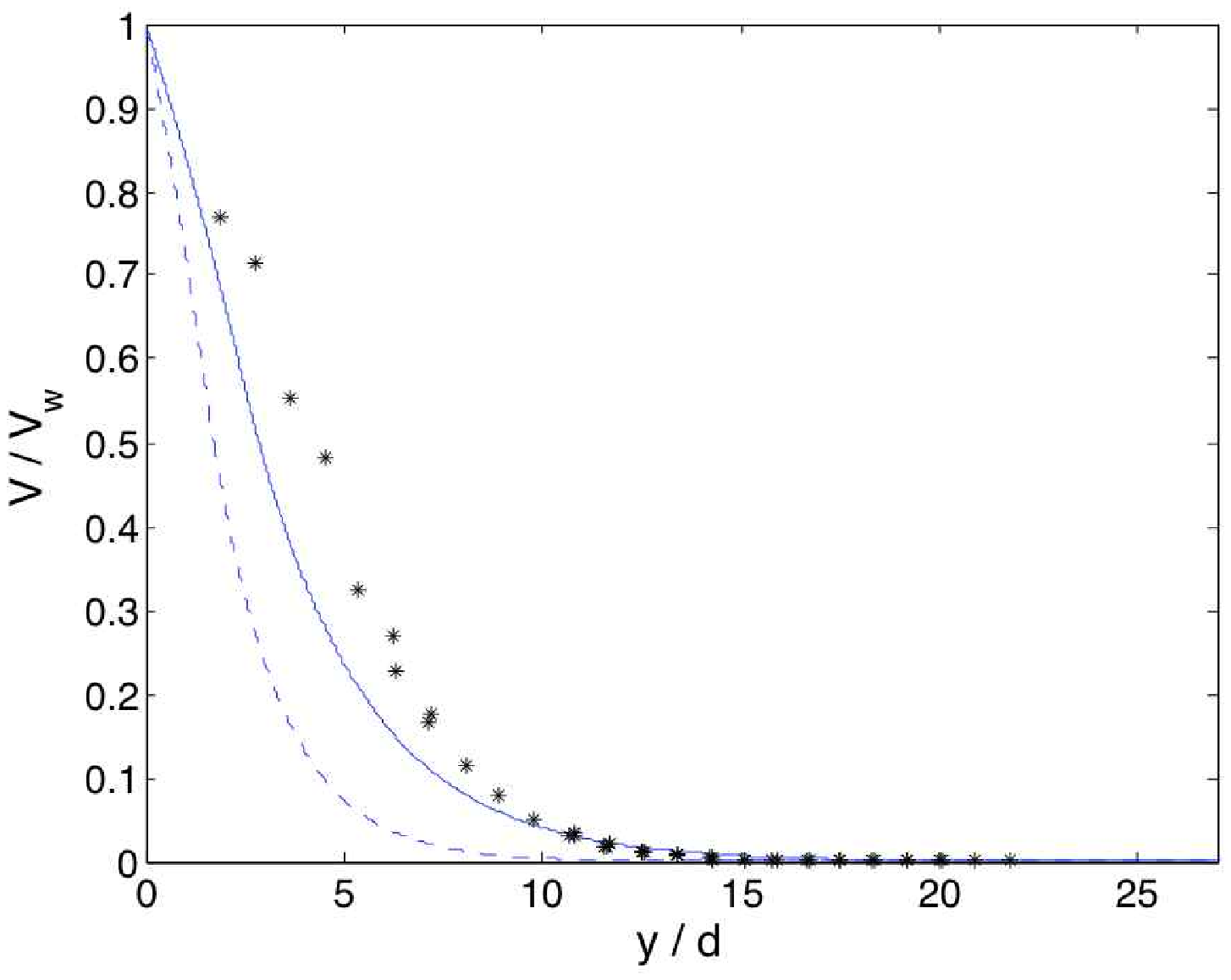}
\includegraphics[width=3.2in, clip]{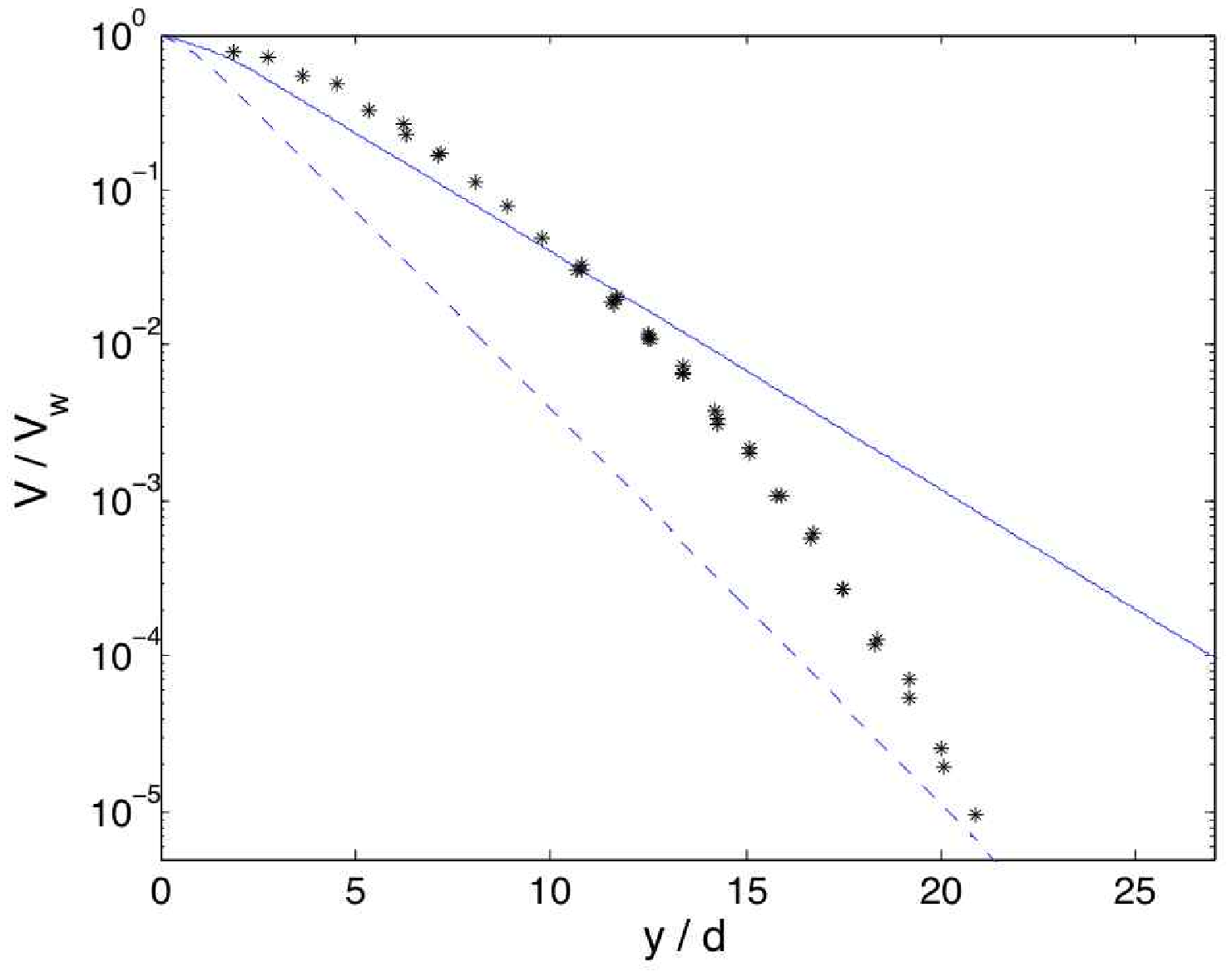}
\caption{(a) The plate-dragging geometry.  The top wall drags along
  the top of a bed of granular material.  The crossing black lines
  within the material are the slip-lines as found from MCP, and the
  vector field is the spot drift as determined by the SFR.  (b) Theory
  against experiment for the plate dragging geometry.  Theory:  L=3 (-
  - -), L=5 (---). Experiment ($*$) courtesy of the authors of
  \cite{tsai04} .} \label{drag} 
\end{center}
\end{figure*}

The plate-dragging flow field can be found using a procedure analagous
to the annular Couette cell, but enforcing horizontal instead of
radial symmetry. With $y$ measuring distance below the plate,
the stress balance equations give
\begin{eqnarray*}
\psi_y&=&\frac{- \sin 2\psi}{2q(\cos 2\psi -\sin \phi)}
\\
q_y&=&\frac{ \cos 2 \psi}{\cos 2\psi -\sin \phi}
\end{eqnarray*}
where $q(y)=p(y)/f_g$ is the average normal stress scaled to the
weight density. The fluidization force will push material downward and
spots upward resulting in a flow profile that decays close to
exponentially near the moving wall. 

Experiments ~\cite{tsai05, siavoshi06} and simulations~\cite{
  thompson91, volfson03, jalali02} offer differing assessments of the
details of the flow profile away from the shear band, but the dominant
exponential decay behavior is clearly observed in all.  The displayed
SFR prediction (Figure \ref{drag}) uses loading parameters from Tsai
and Gollub \cite{tsai04} in order to appropriately compare with their
results. Although the general properties of the flow appear to be
represented well by the model, we do notice that the predicted range
of typical flows is too small to fully encompass the experimental data.
There could be a number of reasons for this discrepancy, but it is
worth pointing out that the quasi-2D plate-dragging geometry is rather
difficult to realize in experiments. For example, this experiment was
executed by rotating a loaded washer-type object on top of an annular
channel, and it was observed that the sidewalls pushing in the third
dimension actually did play some role.

\begin{figure}
\begin{center}
\includegraphics[width=3in, clip]{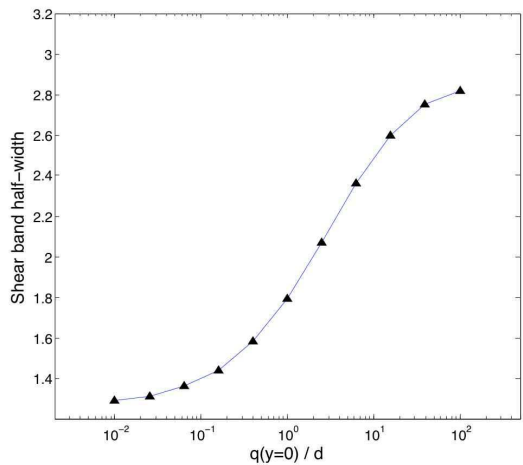}
\caption{Plot of theoretical shear band half-width vs relative loading
  pressure $q_0 = p(y=0)/f_g$ of the top plate (where $f_g$ is the
  weight density). The calculation assumes $L_s=5d$}\label{widthdrag}
\end{center}
\end{figure}

In Refs.~\cite{volfson03} and \cite{dacruz05b}, simulations of this
environment indicate that the shear band width increases with
increasing loading of the top plate.  As can be seen in Figure
\ref{widthdrag}, our theory captures this general trend of increasing
loading causing increasing shear band width.  However, the swing in
band size predicted by our theory is not large enough to match the
range of band sizes in simulations~\cite{volfson03} and
\cite{dacruz05b} in which the shear band half-width can be as large as
several tens of particle diameters for large enough $q_0$ and
diverges as $q_0\rightarrow\infty$ (i.e. zero gravity).  In cases such
as these where the value of $q_0$ becomes very large, as we will
discuss in more depth after the next section, we believe a new phenomenon
begins to dominate our meso-scale argument and that this phenomenon
may be attributed to a particular property of the slip-line field.
 
\subsection{Slow heap flows}

We now examine a prototypical free surface flow. Very close to the
respose angle, a granular heap which is slowly but consistently re-fed
grains undergoes a particular type of motion characterized by
avalanching at the top surface and a slower ``creeping'' motion
beneath.  This type of flow has been studied in experiments
\cite{lemieux00} and simulations \cite{silbert03}.  Though heap flows with
faster top shear layers have also been
studied~\cite{komatsu01,midi04} we will focus for now on the
slower regime, which more closely resembles a quasi-static flow
where the SFR might apply.

This kind of flow is stable, but indeed quite ``delicate'' in the
sense that relatively small changes to the system parameters
(i.e. flow rate, height of the flowing layer) can invoke large changes
to the qualitative flow profile especially in the top layers
\cite{silbert03}.  We will describe and attempt to explain this effect
more in the next section.

\begin{figure}
\begin{center}
(a)\includegraphics[width=3in, clip]{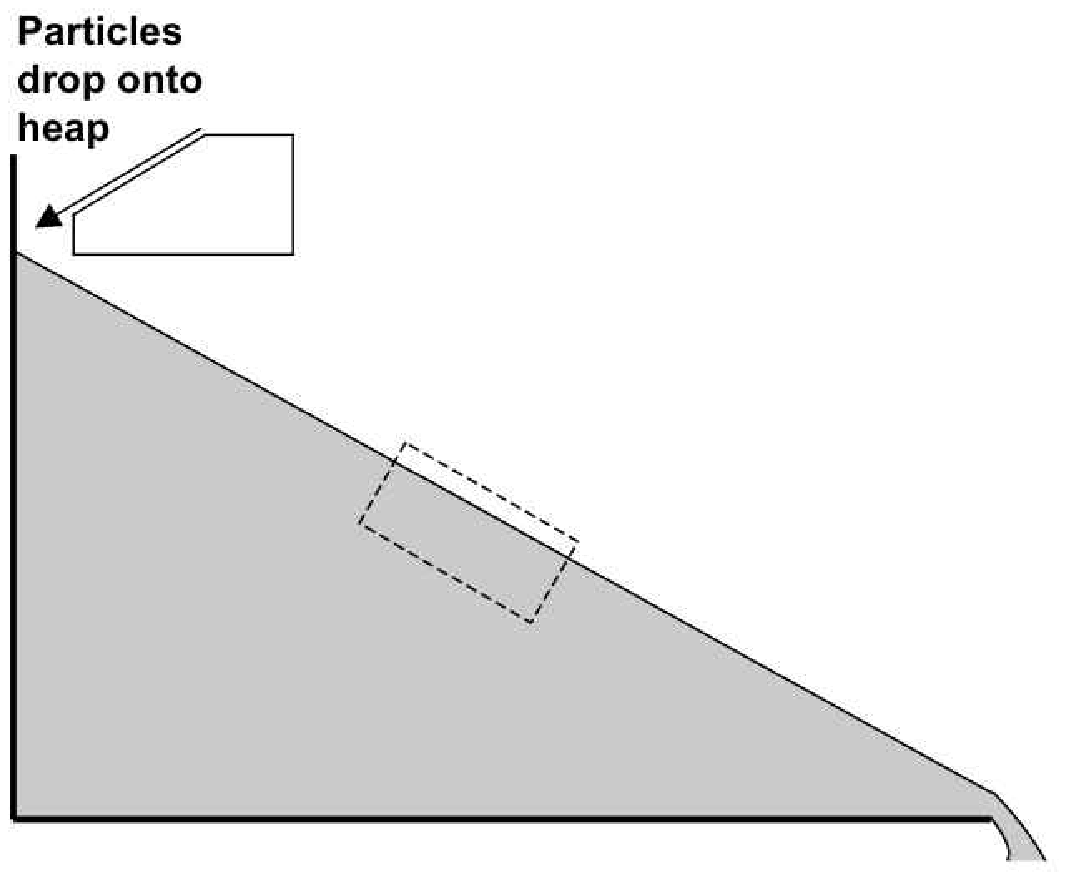}
(b)\includegraphics[width=3in, clip]{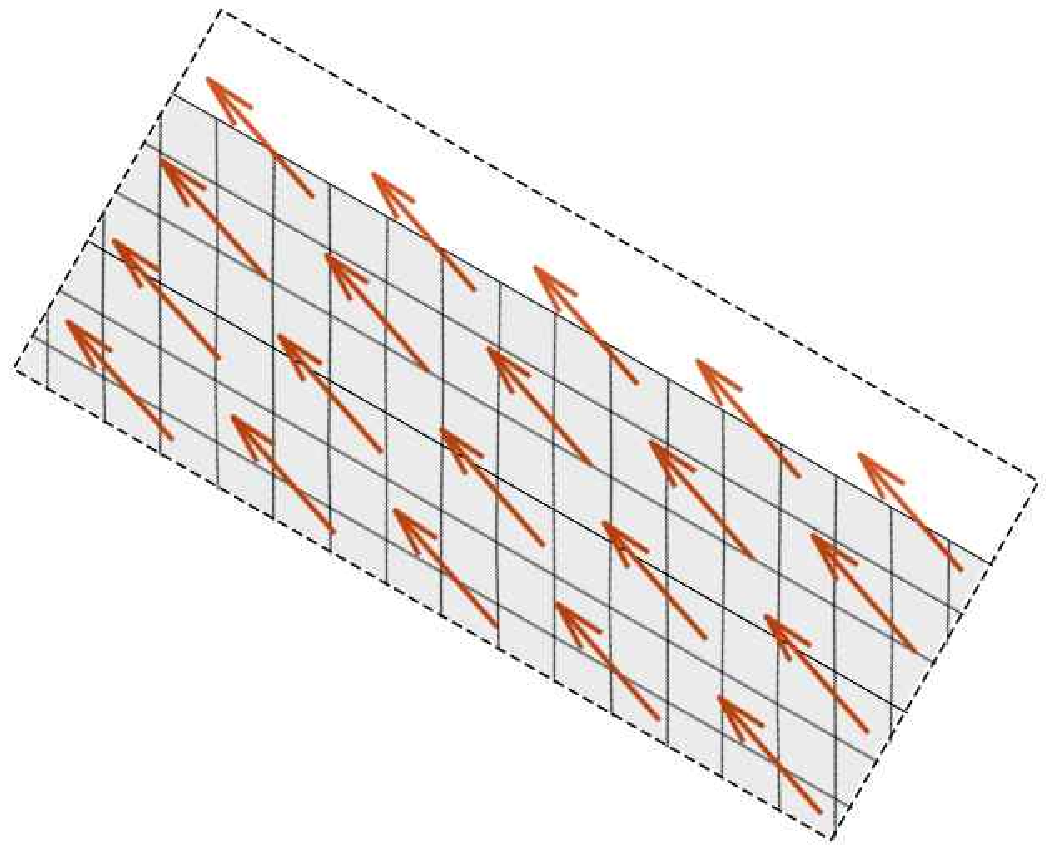}
\caption{(a) The heap flow setup. (b) The dashed rectangle in (a) is enlarged; the slip-line field from MCP is plotted along with the drift field from the SFR. }\label{heapdrift}
\end{center}
\end{figure}

The heap geometry is depicted in Figure \ref{heapdrift} along with the
corresponding spot drift field and slip-line field.  Any gravity
driven free surface flow problem for which the stresses and flow are
approximately invariant in the direction parallel to the
surface will have limit-state stresses that obey the following
relations:
\begin{eqnarray}
\psi&=&-\epsilon
\\
p&=&\frac{f_g y}{\cos\phi}
\end{eqnarray}
where $y$ is the depth measured orthogonally from the free
surface. Note that in limit-state theory, for self-consistency, the
static angle of repose is identical to the internal friction angle
$\phi$, which is a reasonable assertion but still debated in the
community. (By ``static repose angle'' we refer to the angle of
inclination below which a flowing system jams; in simulations of flow
down a rough inclined plane, it has been shown that this angle does
vary in a narrow range depending on the height of the flow
\cite{silbert03,silbert01}.)  Applying equation (\ref{bias}) to these
equations gives a simple expression for the spot drift vector:
\begin{equation}
\ds=\frac{(1+\sin^2\phi, \ -\sin\phi\cos\phi)}{|(1+\sin^2\phi, \
  -\sin\phi\cos\phi)|}
\end{equation}
We may then apply the SFR, which simplifies upon requiring that the
flow run parallel to the free surface (i.e. $\ub^*=(-u,0)$).

Since the drift field is uniform, we obtain an analytical solution for
the unconvolved velocity:
\begin{equation}\label{heap}
u=u(0)\exp\left(\frac{-y \sin 2\phi}{L_s}\sqrt{\frac{2}{5-3\cos2\phi}}\right)
\end{equation}
Thus our model predicts that the velocity decays exponentially off the
free surface.  In cases like these, where the boundary of the flow
makes no contact with a rigid wall, it is less clear how the spots
(and free volume) might behave near the flowing free surface. To avoid
addressing this issue in detail, we neglect convolution with the spot
influence function and simply assume $\ub\approx\ub^*$.

In their experiments on slow heap flow, Lemieux and Durian
\cite{lemieux00} have shown that the velocity profile in the
flowing top layers is indeed well approximated by an exponential
decay. Furthermore, they found the flow in this regime to be
continuous and stable. The decay law they obtained is
\[u/u(0)\approx\exp(-\frac{y}{4.5d})\]
which is very close to our predicted solution for $L_s=3d$:
\[u/u(0)=\exp(-\frac{y}{4.58d}) \] Silbert et. al \cite{silbert03}
report finding a similar decay profile at low flow rates in
simulations of flow down a rough inclined plane, although the
avalanching at the surface was intermittent. In conclusion, we have
demonstrated a fourth, qualitatively different situation where the
same simple MCP/SFR theory predicts the flow profile, without
adjusting any parameters.

\section{ Transition from the SFR to Bagnold rheology }
\label{bag}
\subsection{ Breakdown of the SFR }

In the last two examples, plate dragging and slow heap flow, there are
limits where the SFR fails to predict the experimental flow
profiles. In this section, we will explain why the breakdown of the
SFR is to be expected in these cases and others, whenever slip-lines
approach ``admissibility'' and coincide with shear planes. In this
singular limit of the SFR, we postulate a transition to Bagnold
rheology. The stochastic spot-based mechanism for plastic
yielding is thus replaced by a different physical mechanism, the free
sliding along shear planes.

For example, consider the case of plate dragging above.  At large
relative loading, the flow field resembles that of a zero-gravity
horizontal shear cell (between shearing flat plates), and it appears
that the SFR breaks down: With body forces and $\del p$ both going to
0, equation (\ref{bias}) gives $\vec{F}_{net}=\mathbf{0}$ implying
that spots have no drift and consequently the only SFR solution is
$\ub=\mathbf{0}$. 

Problems also occur with flow down a rough inclined plane: Slightly
increasing the flow rate (and consequently the flow height) or
inclination angle causes the velocity vs depth relationship to exit
the exponential decay regime detailed above and undergo significant
changes, passing first through a regime of linear dependence
\cite{ancey02, berton03} to a regime resembling a $3/2$
power-law of depth \cite{pouliquen99, silbert01, prochnow_thesis,
  azanza_thesis} opposite in concavity to the exponential decay
regime.

Why does the velocity profile for inclined plane flow undergo many
different qualitative regimes depending delicately on system
parameters, while others (e.g. silos, annular cells) appear to be only
weakly affected and almost always exhibit the same (normalized)
velocity profile? Tall inclined plane flows and zero-gravity planar
shear flows have been successfully described in multiple experiments
and simulations \cite{dacruz05,lois05, pouliquen99, prochnow_thesis,
  silbert01} by the empirical scaling law of
Bagnold~\cite{bagnold54}. In this section, we suggest a means to
reconcile and perhaps eventually combine these theories into a
coherent whole.

\subsection{ Bagnold rheology }

Let us briefly review Bagnold's classical theory of granular shear
flow. In its original form, ``Bagnold scaling'' expresses a particular
rate-dependency for granular flow whenever the solid fraction is
uniform throughout:
\begin{equation}\label{bagnold1}
\tau \propto \dot{\gamma}^2.
\end{equation}
where $\dot{\gamma}$ is the rate of simple shear. To account as well
for static stresses arising from the internal friction, a related
variant of this scaling law is commonly used \cite{duran_book}:
\begin{equation}\label{bagnold} 
\tau-\mu\sigma\propto\dot{\gamma}^2.
\end{equation}
It is in some sense a law for how the yield criterion can be exceeded
when non-negligible strain-rates can absorb the extra shear
stress. This constitutive law alone is an incomplete flow theory since
it provides no way of predicting whether or not the solid fraction
will be uniform during flow or how a non-uniform solid fraction
affects the above rheology. Bagnold originally explained the quadratic
relationship between stress and strain-rate in terms of binary
collisions as the joint effect of both the particle collision rate and
the momentum loss per collision being directly proportional to the
strain-rate \cite{bagnold54}.  Despite this collision-based argument,
however, Bagnold scaling has been observed to hold well into the dense
regime, whenever the solid fraction is approximately constant
throughout the system.  This seemingly contradictory observation can
be justified in the hard-sphere limit (without body forces) by a
Newtonian invariance argument \cite{lois05}, although it calls into
question the underlying physical mechanism.

Zero-gravity planar shear flow and thick inclined plane flow both
exhibit nearly uniform density and thus have been employed as test
cases for Bagnold scaling.  In the planar-shear environment, the shear
and normal stresses acting on the shear planes are spatially constant
throughout the flow.  Equation (\ref{bagnold}) therefore implies that
the strain-rate is uniform; as a result, the velocity varies linearly
from one wall to the other.  This result is known as Uniform Shear
Flow (USF) and is easily verified in simulations of Lees-Edwards
boundary conditions. For example, the rheology (\ref{bagnold}) has
been demonstrated in the simulations of \cite{dacruz05}.

Applying Bagnold scaling to the inclined plane geometry, slightly above
static repose, gives a shear stress excess which grows linearly with
depth and thus a shearing rate that grows as the square root of
depth. This implies a velocity profile of the form
\begin{equation}\label{powerlaw}
u\propto h^{3/2}-y^{3/2}
\end{equation} 
for $y$ the depth variable and $h$ the height of the flowing material
(with no-slip bottom boundary condition).  In this way, Bagnold
scaling successfully explains the $3/2$ power law dependence noted
above.

\subsection{ Slip-line admissibility }

The seemingly disparate flow mechanisms of the SFR and Bagnold
rheology can be reconciled very naturally by considering the geometry
of the slip-lines.  In plasticity theory, all flows can be classified
based on ``slip-line admissibility''. For admissible slip-lines,
boundary conditions are such that the flow can, and presumably does,
take place by continuous shearing along only one family of
slip-lines. In mathematical terms, the slip-lines are admissible for a
given flow, whenever the double-shearing continuum flow-rule
(\ref{double}) allows multiple solutions to the boundary value
problem.

\begin{figure}
\begin{center}
(a)\includegraphics[width=2.1in, clip]{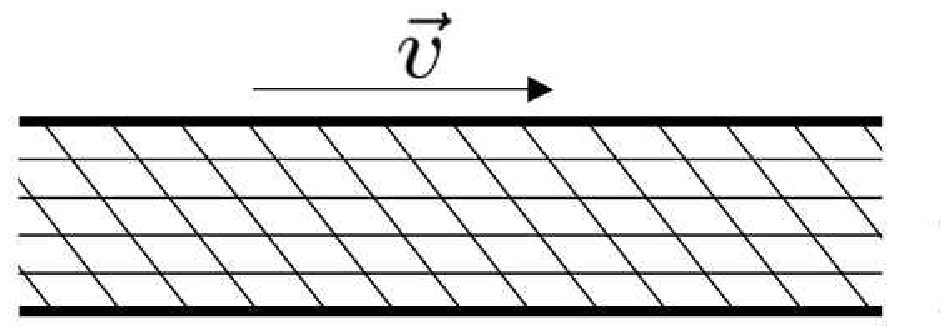}
(b)\includegraphics[width=2.1in, clip]{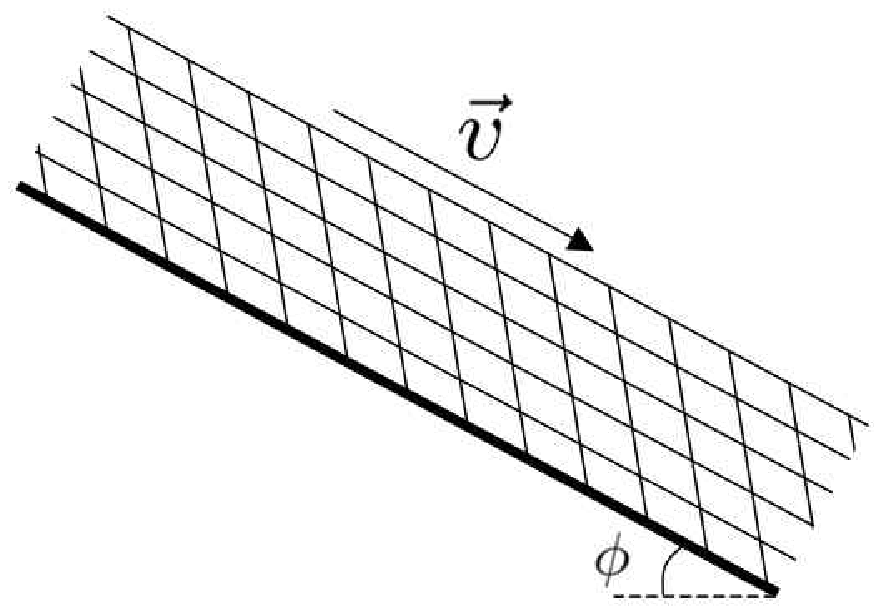}
\caption{(a) Slip-line field for gravity-free planar shear flow, (b) Slip-line field for inclined plane flow. Note that in both cases the shear planes will be aligned completely with one slip-line family.}\label{admissi}
\end{center}
\end{figure}

Slip-line admissibility is the exception, not the rule, since it is
highly unlikely that the prescribed velocity boundary conditions are
fulfilled by a continuous shear on either slip-line family. Be that as
it may, it so happens that planar shear flow and inclined plane flow
are both slip-line admissible. This special property is shared by no
other flow geometry studied in this paper, or, to our knowledge,
elsewhere in the granular materials community. (Contrast the slip
lines in Figure \ref{admissi} with those in Figures
 \ref{silodrift} and \ref{couetteSFR}.) 

There is also an interesting difference in the density
distributions. For admissible flows, the volume fraction is nearly
uniform, and Bagnold rheology has a reasonable physical
justification. For the more common case of inadmissible flows, as in
silos and Couette cells, the volume fraction is typically highly
nonuniform. In such cases, the SFR seems to provide an excellent
description of the flow, and Bagnold rheology clearly does not apply.

These observations motivate us to think of admissibility as a
criterion for two very different microscopic mechanisms for granular
flow: In admissible flows, material motion is a viscous dragging of
material ``slabs'' along one slip-line family (Bagnold dominated),
whereas in inadmissible flows there is no clear choice as to which
slip-line family should control the motion and thus material randomly
chooses between both slip-line families (SFR dominated). Perhaps
admissibility in the slip-line field \emph{causes} the solid fraction
to remain roughly uniform because the material in no sense has to
collide head-on into neighboring material for it to move. In flows
where non-uniform dilation does occur, experiments have shown the
motion is nearly independent of any local Bagnold rheology,
encouraging our strong distinction between the dynamics of flow
problems of differing admissibility status \cite{tardos98, bocquet01}.

These considerations all lead us to the fundamental conjecture:
\begin{quote} Slip-line admissibility is a geometrical and mechanical
  indicator as to the relative importance of rate-dependency (Bagnold
  rheology) over rate-independency (SFR) in a dense granular
  flow.
\end{quote}
This means that a flow which has an admissible
limit-state stress field will be dominated by rate-dependent effects
when the yield criterion is only slightly exceeded.

\subsection{ Redistribution of excess shear stress }

A more rigoruous physical justification of our conjecture can be made
utilizing limit-state stresses and observing the effect of pushing the
system above yield.  Bagnold rheology is a statement connecting the
shear stress excess (i.e. amount by which $\tau$ exceeds $\mu\sigma$)
along a shear plane to the rate of simple shear along the plane.  We
must emphasize that shear planes and slip-lines are \emph{not}
equivalent terms; slip-lines are defined by the quasi-static stresses
as lines along which $\tau-\mu\sigma=0$, whereas shear planes are
defined entirely by the velocity profile. In an admissible system, the
shear planes coincide with one slip-line family.  In inadmissible
systems, the shear planes almost everywhere do not coincide with
slip-lines.

With admissible slip-lines, excess shear stress tends to be uniformly
distributed throughout the system, resulting in global Bagnold
rheology. For example, consider a zero-gravity planar shear cell. If
we were to apply additional shear stress to the body in a manner
aligned with the admissible slip-line family, e.g. by increasing the
wall shear stress above yield by some amount $\Delta \tau$, that
additional shear stress would distribute itself within the material
precisely along the slip-lines. Every horizontal slip-line within the
bulk would thus receive a boost in shear stress of size $\Delta \tau$.
In limit-state theory the slip-lines have the highest possible
$\tau-\mu\sigma$ value a quasi-static material element can take --
zero. Adding $\Delta \tau$ additonal shear stress to the slip-lines
means that $\tau$ will maximally exceed $\mu\sigma$ precisely along
the slip-lines, and, by admissibility, every shearing plane. As a
result, there will be a Bagnold contribution everywhere.  Similarly,
if we took a limit-state inclined plane geometry and increased the
tilt angle some amount, an analagous boost in shear stress along the
admissible slip-line family would occur causing $\tau$ to exceed
$\mu\sigma$ precisely along all the shear planes.

In contrast, with inadmissible slip-lines, excess shear stress tends
to remain localized where it is applied, and the SFR dominates the rest of
the flow. For example, consider annular Couette flow. As can be seen
in Figure \ref{couetteSFR}, the slip-lines only coincide with the
shear planes (which in this case are concentric circles) along the
inner wall of the cell.  If the inner wall were given an increase
$\Delta \tau$ in applied shear stress, torque balancing requires the
shear stress along any concentric circle within the material to
receive a boost of $\Delta \tau \cdot r_{w}/r$.  Suppose the shear
cell has inner wall radius $40d$ and the boost in wall shear is
significant, say $\Delta \tau= \tau_w/10$.  Solving for
$\tau-\mu\sigma$ along the shear planes in this situation gives a very
different result than in the previous case--- here $\tau$ will only
exceed $\mu\sigma$ along the shear planes that are less than $1.4d$
off the inner wall.  So, regardless of whether the density is or is
not uniform, Bagnold scaling would at best only apply in an almost
negligibly thin region near the wall.  If the wall friction were less
than fully rough, this region would further decrease.

\subsection{ A simple composite theory }

The preceding discussion indicates that, in general, one can use the
admissibility status of the system to choose whether or not the flow
should obey the SFR or Bagnold rheology, or perhaps some combination
of the two. Indeed, it seems reasonable that when slip-lines are
\emph{approaching} admissibility (e.g. plate-dragging with high $q_0$)
or when an admissible system is only slightly pushed above yield
(e.g. inclined plane flow near static repose) we must account for
contributions from both effects simultaneously. It is beyond the scope
of this paper to postulate the precise microscopic dynamics (and
derive corresponding continuum equations) for this regime, but we can
at least give a sense of how the more general theory might look. 

In general, we envision a smooth transition from rate-independent SFR
dynamics to rate-dependent Bagnold dynamics controlled by the
distribution of shear stress excess. This implies the coexistence of
(at least) two different microscopic mechanisms: SFR and admissible
shear. The SFR contribution would derive from the usual spot-based,
quasi-static stochastic dynamics; the Bagnold contribution would come
from a rate-dependent shearing motion along the appropriate slip-line
family whenever there is a small excess shear stress (beyond the limit
state) applied on a boundary which causes shear stress excess along
the shear planes within.

The two mechanisms should have different statistical signatures. For
shear deformation along admissible slip-lines, we would expect
anisotropic velocity correlations. In the direction perpendicular to
the shear plane, the correlation length should be somewhat shorter
than the typical spot size, since slip-line admissibility allows flow
to occur with less drastic local rearrangements, farther from
jamming. In the directions parallel to the shear plane, however the
correlation length could be longer, since material slabs sliding along
shear planes may develop more rigid, planar regions. It would
certainly be interesting to study velocity correlations in heap flows
at different inclinations and plate-dragging experiments under
different loads to shed more light on the microscopic mechanisms
involved in the SFR to Bagnold transition.

For the remainder of this section, we make a first attempt at a
composite model, simply a linear superposition of SFR and Bagnold
velocity fields:
\begin{equation}\label{composite}
\ub=\alpha\ub_{_{\text{SFR}}}+\beta\ub_{_{\text{Bag}}}. 
\end{equation}
which could have its microscopic basis in a random competition between
the two mechanisms, when slip-lines are near admissibility.  Here,
$\ub_{_{SFR}}$ is an SFR solution for the flow, using the limit-state
stress field everywhere, and $\ub_{_{Bag}}$ is obtained from the
excess shear stress on a boundary by integrating the Bagnold strain
rate $\dot{\gamma}=\sqrt{\tau-\mu\sigma}$ over those shear planes for
which $\tau-\mu\sigma$ has been boosted above zero.  (Note that we
ignore the condition of uniform density for Bagnold rheology since we
conjecture that uniform density is actually a geometric consequence
of slip-line admissibility and will arise naturally whenever Bagnold
rheology dominates the flow.)

A reasonable first approximation is that the SFR and Bagnold solutions
individually fulfill the necessary boundary conditions for the
velocity profile since, under the right circumstances, either can be
made to dominate the other.  The constant $\beta$ is the Bagnold
proportionality constant which may depend on the density of the flow
among other parameters \cite{dacruz05}. Since the SFR is a
rate-independent flow model, $\ub_{_{SFR}}$ can always be multiplied
by a positive constant (observe that equation (\ref{rhofield}) is
homogeneous in $\rho_s$), and thus we allow the scalar multiple
$\alpha$. Given some determinable form for $\beta$, $\alpha$ is chosen
such that $\ub$ fits the velocity boundary conditions. This seems
reasonable for moving walls (as in plate dragging), but not for free
surfaces, whose boundary velocity should also be predicted by the
theory. In such cases, where $\alpha$ is not clearly defined in this
simple model, one could use other empirical relations, such as the
Pouliquen Flow Rule for inclined plane flows~\cite{pouliquen99}, to
deduce the free boundary velocity, and thus $\alpha$.



\subsection{ Some applications of the composite theory }

Using our very simple composite theory, we will now revisit a few
geometries that were troublesome for the SFR alone. Extending the
theory with a smooth transition to Bagnold scaling controlled by
slip-line admissibility seems to resolve the experimental puzzles and
capture the basic physics of granular shear flows.  In the cases we
consider below, we do not change the value of $\alpha$ as we increase
the shear stress excess; this way the relative importance of Bagnold
effects are easier to isolate.

\subsubsection{ Planar shear cell }

In a zero-gravity planar shear cell, $\ub_{_{SFR}}=0$, \emph{but} the Bagnold
solution for any amount of shear stress excess is of the form
$\ub_{_{Bag}}=k y$, and thus the composite solution, regardless of the
values $\alpha$ and $\beta$, is a homogeneous flow between the two
rough plates.  The lack of a ``background'' SFR solution in this case
may explain why Bagnold rheology is almost exactly observed in
simulations of this geometry over a wide range of strain rates
\cite{lois05}.

\subsubsection{ Rough inclined plane }

\begin{figure}
\includegraphics[width=3.2in, clip]{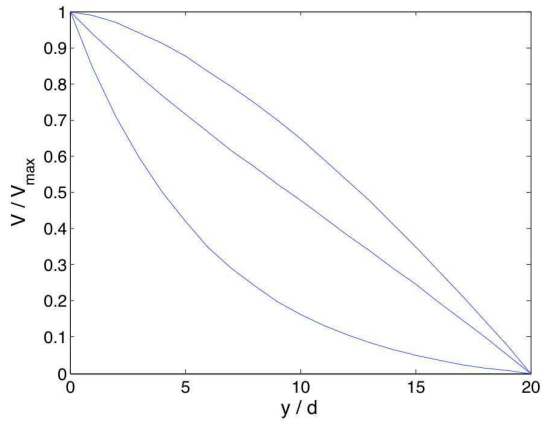}
\caption{Some predicted velocity profiles for flow down a rough
  inclined plane as a function of depth. Note that all three known
  flow behaviors; exp decay, linear, 3/2 power law; appear in proper
  relationship to the inclination (see
  Ref. \cite{silbert03}). (Bottom) Incline near static repose, fully
  SFR dominated ($L_s=4d$); (Middle) Increased inclination angle,
  Bagnold to SFR ratio of 3:1; (Top) Further increase to inclination;
  fully Bagnold dominated.}\label{sfrbag}
\end{figure}

For flow down a rough inclined plane, the SFR solution is an
exponential decay (\ref{heap}), and the Bagnold solution is a 3/2
power law (\ref{powerlaw}). When the material is only slightly above
static repose, a shear stress excess along the shear planes will exist
but will be very small; it goes as $\sqrt{\Delta \theta}$ for an
incline $\Delta\theta$ above static repose \cite{duran_book}.  As a
result, $\ub_{_{Bag}}$ will be small in magnitude, and the SFR
solution will show through as the ``creeping flow'' with exponential
decay. As $\Delta \theta$ increases, the increased shear stress excess
will cause the Bagnold contribution to increase, and the flow will
eventually morph into the 3/2 power law dependence of pure Bagnold
scaling.  In between, where both contributions are of similar
magnitude, the superposition of the two flow fields gives a predicted
profile that appears approximately linear, since the SFR and Bagnold
solutions are of opposite concavity. Thus, the composite SFR-Bagnold
formulation appears to be able to explain the various flow regimes in
inclined plane flow, which have been observed in experiments and
simulations (see Figure \ref{sfrbag}). 

Recent experimental work of Pouliquen \cite{pouliquen04} seems to
support this analysis; it is found that inclined plane flow occurring
at lower inclination angles exhibits spatial velocity correlations
near the typical spot size (as the SFR would imply), but as
inclination increases, the correlation length appears to decrease, an
effect we might attribute to an increased dominance of Bagnold scaling
(a phenomenon not goverened by a correlation length) over the SFR.


\subsubsection{ Rapid heap flows }

\begin{figure}
\includegraphics[width=3in, clip]{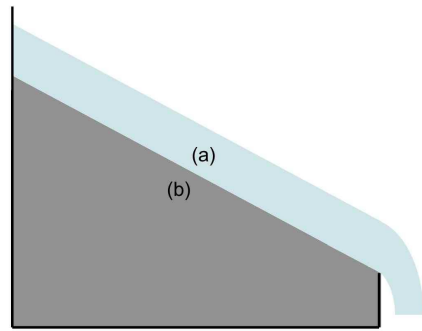}
\caption{Standard heap flows enable one to see both the SFR and
  Bagnold contributions spearately in one flow geometry. (a) The
  surface region is dominated by Bagnold scaling. (b) The creep-region
  beneath adheres to the SFR.}\label{admissiheap}
\end{figure}

The composite theory also seems consistent with rapid heap flows.
When the flow rate down the heap increases, the region near the
surface resembles inclined plane flow in any one of its various flow
regimes, whereas the region beneath the surface flow undergoes creep
motion which decays close to exponentially \cite{komatsu01, midi04}
(see Figure \ref{admissiheap}).  We can justify this in terms of
slip-line admissibility: The slip-lines throughout the system (see
Figure \ref{admissi}) have the same form in both regions. In the
surface region, the slip-lines are admissible because there is nothing
blocking the motion from being a simple shearing along the
slip-lines. In the creep region, however, the gate (or the ground)
prevents global shearing along the slip-lines and thus the slip-lines
are inadmissible and the SFR dominates.

We can equivalently explain heap flow in terms of shear stress
excess. The excess incurred by increasing the heap angle will
distribute itself differently in the two regions.  In the surface
region, the excess can only be absorbed along the shear planes by
inducing a strong Bagnold dependence.  However, the gate at the bottom
of the heap will support any shear stress excess on the creep
region. (Note that the slip-lines in the creep region all hit the
gate, or the ground.)  Thus the full flow will be the sum of an
exponentially decaying SFR solution superposed with a significant
Bagnold-type solution which starts at the surface and cuts off at the
interface with the creep zone.

\subsubsection{ Plate dragging under a heavy load }

\begin{figure}
\begin{center}
\includegraphics[width=3in, clip]{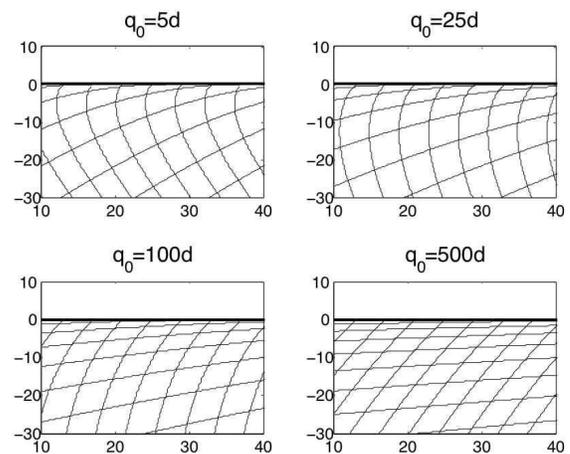}
\caption{Plate-dragging slip-lines approaching admissibility as $q_0$ increases.}\label{slipload}
\end{center}
\end{figure}

We will now explain the comment made at the end of the plate-dragging
section.  The slip-lines of a plate-dragging geometry can be pushed
drastically close to full admissibility by simply increasing the
relative loading of the top plate, $q_0$, above a certain
non-excessive amount (see Figure \ref{slipload}). To see the effects
of approaching admissibility more carefully, say we take a limit-state
plate-dragging setup and pull the plate slightly harder, inducing a
super-yield shear stress boost of $\Delta \tau$ under the plate. Bagnold
effects should appear wherever, as a consequence of stress balancing,
a shear stress excess results along a shear plane.  The shear planes
are horizontal lines in this case, and at limit-state, the stresses
along any horizontal obey
\begin{equation}
\tau-\mu\sigma=-\mu f_g y=-\mu \frac{p_0}{q_0} y.
\end{equation}
When the extra shear stress is applied, a shear stress excess of
$\Delta \tau-\mu p_0 y/q_0$ will form for $0\leq y \leq \frac{\Delta
  \tau}{\mu p_0} q_0$. Accordingly the Bagnold flow contribution will have
a shear zone whose depth extends into the granular bed as an
increasing linear function of $q_0$ for a fixed shear stress boost
$\Delta \tau$ and fixed downward plate-pressure (i.e. we decrease the
material weight density to increase $q_0$). Integrating the Bagnold
shear rate gives that the relative size of the Bagnold contribution
should also increase with increasing $q_0$. With $q_0$ large enough,
therefore, the SFR contribution will be dwarfed by a Bagnold term with
a larger shear band. As $q_0\rightarrow \infty$ the slip-lines become
completely admissible and the shear band width diverges as we would expect.

The mismatch in shear band size between the data of Tsai and Gollub
and the SFR could be due, among other possible reasons, to the fact
that $q_0\approx 380d$ was large enough to make the Bagnold
contributution sizeable.  This could also explain the large shear
bands found in \cite{volfson03}.  A detailed comparison of experiments
and simulations with different versions of a composite SFR/Bagnold
theory would be an interested direction for future work.



\section{ Conclusion } 
\label{con}
\subsection{ Highlights of the present work }

We have proposed a stochastic flow rule (SFR) for granular materials,
assuming limit-state stresses from Mohr-Coulomb plasticity (MCP). In
the usual case where slip-lines are inadmissible (inconsistent with
boundary conditions), we postulate that flow occurs in response to
diffusing ``spots'' of local fluidization, which perform random walks
along slip-lines, biased by stress imbalances.  The spot-based SFR
corrects many shortcomings of classical MCP and allows some of the
first reasonable flow profiles to be derived from limit-state
stresses, which engineers have used for centuries to described the
statics of granular materials.

Our theory notably differs from all prior continuum theories (cited in
the introduction) in that it is derived systematically from a
microscopic statistical model~\cite{bazant06}. The Spot Model is
already known to produce realistic flowing random
packings~\cite{rycroft06a}, and what we have done is to provide a
general mechanical theory of spot dynamics. Evidence for spots has
been consistently found in spatial velocity correlations in
simulations~\cite{rycroft06a} and experiments~\cite{choi04}
(Fig.~\ref{corr}) on silo drainage.

Beyond its fundamental physical appeal, the SFR seems to have
unprecedented versatility in describing different granular flows. It
has only two parameters, the friction angle $\phi$ and correlation
length (spot size) $L_s$, which are not fitted; they are considered
properties of the material which can be measured independently from
flow profiles. For monodisperse frictional spheres, the SFR can
predict a variety of different flows using the same spatial velocity
correlation length, $L_s\approx 3-5 d$, measured in experiments and
simulations. This is perhaps the most compelling evidence in favor of
the spot mechanism which underlies the SFR.

We have shown that the SFR can describe a rather diverse set of
experimental data on granular flows. Some flows are driven by body
forces (silo and heap flows); others have body forces, but are driven
by applied shear (plate-dragging); still others are driven by applied
shear without body forces (annular shear flow).  Some geometries have
straight boundaries (silos, heaps, plate-dragging), and yet the theory
works equally well for highly curved boundaries (annular shear
flow). Some of the flows exhibit shear localization (annular shear
flow, plate-dragging, heap flow), and yet the theory correctly
predicts wide shear zones in silo flow. It is noteworthy that the
same, simple model, correctly predicts and places shear bands in
geometries where they arise for very different reasons--- gravity
causes the shear band in plate-dragging, and yet the geometry (through
the $\del p$ term in the drift) causes the shear band in annular
Couette flow. We are not aware of any other theory (including
classical MCP) which can quantitatively describe more than one of
these flows, let alone without empirically fitting the velocity
profiles.

\subsection{ Comparison with partial fluidization }

It is interesting to compare our approach to the continuum theory of
partial fluidization of Aranson and
Tsimring~\cite{aranson01,aranson02}.  Although it lacks any
microscopic basis, their theory also introduces a
diffusing scalar field to control the dynamics, as opposed to a
classical stress/strain-rate relation. The analog of our spot density
is the ``order parameter'' $\rho$, which measures the degree of
``fluidization'' of the continuum by mixing two different types of
stresses, corresponding to distinct ``liquid'' ($\rho=0$) and
``solid'' ($\rho=1$) phases.  Given the stress tensor for the material
in a static solid state, $\sigma_{ij}^0$, the stresses in a flowing
granular material are modeled by adding some degree of viscous
stresses, as in a Newtonian liquid:
\begin{equation}\label{opstress}
\sigma_{ij}=(\rho+(1-\rho)\delta_{ij})\sigma_{ij}^0+\eta\dot{E}_{ij}
\end{equation}
where $\eta$ is the viscosity. The order parameter controlling the
balance of these two stress tensors is postulated to obey a
reaction-diffusion equation,
\begin{equation}\label{op}
  (\Delta t)\frac{\partial \rho}{\partial t}=l^2\del^2\rho+\rho(1-\rho)(\rho-\delta)
\end{equation}
for collision time $\Delta t$, grain length scale $l$, and a function
$\delta$ of the stress state, which is greater than 1 where the
material is above the static yield criterion, less than 0 where below
the dynamic yield criterion, and between 0 and 1 otherwise.  One
benefit of this model is that it can be used for unsteady flows. In
principle, the SFR may also describe time-dependence through the spot
Fokker-Planck equation (\ref{eq:spotfp}), but we have only developed
and tested the theory so far for steady flows, starting from
(\ref{rhofield}).

For the sake of comparison, consider a steady flow modeled by partial
fluidization and the SFR.  The difference is that the spot equation
(\ref{rhofield}) couples diffusion to a drift depending on frictional
yielding, whereas the order parameter equation (\ref{op}) balances
diffusion with a nonlinear source term, ressembling a chemical
reaction rate, which indirectly mimicks the effect of a Coulomb yield
criterion.  Interestingly, if the SFR could be extended to an
elasto-plastic model without making the incipient failure assumption
(see below), a similar non-linear source term may have to be added to
the spot equation to account for the need to destroy spots when they
enter zones below yield.  It is also notable that our argument for why
a spot drifts, i.e. a localized stick-slip type of shear stress
decrease along the spot boundary, is reminiscent of equation
(\ref{opstress}) wherein the shear stress goes down in the presence of
fluidization. In this sense, a higher spot concentration in our model
is similar to a higher degree of partial fluidization.

The similarities, however, seem to end there. One difficulty with the
partial fluidization approach is that it cannot easily describe
rate-independent effects since the motion stems from a viscous form in
the stress tensor. Also in sharp contrast to our approach based on
plasticity, partial fluidization does not provide a clear theory of
the static solid stresses in the limit of no flow, opting instead to
deal with environments for which the open components of this tensor
are not needed (simple shear flows). This could perhaps be modified,
but if the theory is to be fully general, equation (\ref{opstress})
would need a frame-independent form where foreknowledge of the shear
planes is not necessary to properly apply fluidization.  These
considerations as well as selecting boundary conditions on the order
parameter, seem to be the primary limitations in testing partial
fluidization in more general situations, such as those considered here
with the SFR.

\subsection{ Future directions  }

In spite of some successes, we still do not have a complete theory of
dense granular flow. There are at least three basic limitations of the
SFR, about which we can only offer some preliminary ideas to guide
future work.

\subsubsection{ Slip-line inadmissibility }

Although most slip-line fields are inadmissible, the SFR breaks down
as slip-lines approach admissibility.  We have already begun to extend
the model into this regime by conjecturing that slip-line
admissibility is associated with Bagnold rheology, as excess shear
stress (above the limit-state) drives a local shear rate along the
most admissible slip-lines. We have shown that a simple linear
superposition of Bagnold and SFR flow fields with appropriate boundary
conditions can describe a variety of composite flows, exhibiting both
Bagnold and SFR behavior in different limits or segregated into
different regions. These include planar zero-gravity shear, various
inclined-plane and heap flows, and plate-dragging at large
loading. However, more work is needed to develop and test a composite
SFR/Bagnold theory, both at the continuum level and in terms of the
two microscopic mechanisms.

\subsubsection{ 2D symmetry }

\begin{figure}
\centering{\includegraphics[width=3.4in, clip]{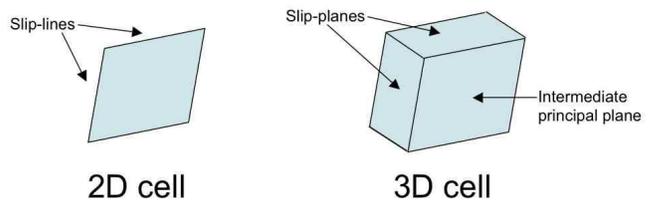} \label{cells}}
\caption{ The difference between 2D and 3D continuum cells which could
  be used to construct the SFR. }
\end{figure}

Through MCP, the SFR is currently used only in quasi-2D geometries. In
efforts to extend the theory to 3D, a good test case would be the
split-bottom Couette cell, which displays a wide, diffusive shear
band~\cite{fenistein03}, reminscent of a draining silo.  The 2D
limitation may not be so difficult to overcome, although any
plasticity theory is more complicated in three dimensions, than in
two. As usual, constructing a 3D limit-state stress field requires an
additional hypothesis to close the stress equations. In 3D, a general
material point at incipient failure with distinct principal stresses
$\sigma_1>\sigma_2>\sigma_3$ is intersected by a pair of slip planes
angled $2\epsilon$ apart. We cannot therefore encase a 3D cell of
material within slip planes as we are able to do to a 2D cell with
slip-lines. However, the principal plane on which $\sigma_2$ acts, the
intermediate principal plane, can be used along with the slip-planes
to encase a 3D material cell.  This is legitimate because, if such a
cell underwent slip-plane fluidization, the net material force would
be guaranteed to point parallel to the intermediate principal plane;
since the intermediate principal plane offers no shear resistance, the
material can slide along this plane, while simultaneously sliding
along a slip-plane.

  To apply the SFR then, the drift vector should still be calculated
  from equations (\ref{fnet}) and (\ref{bias}), but all vectors must
  be projected first into the $\sigma_1\sigma_3$-plane since the
  $\sigma_2$ direction is not involved in slip-plane fluidization. The
  shape of a spot and its diffusivity would likely be anisotropic,
  with different values in the intermediate direction, since the main
  source of diffusion is slip-plane fluidization.  

  If ever the intermediate principal stress equals either the major or
  minor principal stress, as in the Har Von Karman hypothesis,
  incipient failure is upheld on a cone instead of intersecting
  slip-planes. When this degenerate case occurs, the material cell can
  be encased serveral different ways depending on the surrounding
  stress states. This must be determined before we can rigorously
  define how to apply fluidization and the SFR.

\subsubsection{ Incipient yield everywhere }

The SFR assumes stresses near a limit state. While incipient yield is
believed to be a good hypothesis in many situations, even during dense
flow (which should be checked further in DEM simulations), it clearly
breaks down in some cases, at least in certain regions. This is
perhaps the most difficult limitation to overcome, since the
limit-state assumption is needed to fully determine the stress
tensor. Without it, the material effectively enters a different state
most likely governed by some non-linear elastic stress strain law
which is far more difficult to apply.

We have already argued that such a transition away from incipient
yield must exist in some granular flows, due to the strong tendency of
granular materials to compactify into a rigid solid state when shaken
(e.g. by nearby flowing regions) but not sufficiently sheared.  A good
example is a tall narrow silo with smooth side walls, where the SFR holds
near the orifice, but breaks down in the upper region, ressembling a
vertical chute. The broad shear band localizes on the side walls, as a
rigid central plug develops, which likely falls below incipient yield.

A more robust elasto-plastic theory for the stress state would relax
our limit-state constraints and allow for material to fall below the
yield criterion where it is described by elasticity. The SFR
could then be applied only where the material is at yield and
everywhere else the material does not deform plastically.
Elasto-plasticity theory operates just as well in 3D as in 2D which is
a key benefit over limit-state plasticity. However, our model as we
have already presented it is \emph{far} simpler than elasto-plasticity
and yet still manages accurate results when applied to limit-state
stress fields.

\vskip 18pt

As the SFR matures as a theory of granular flow, it would also be
interesting to apply it to other amorphous materials, such as metallic
glasses, and to develop new simulation methods. The basic idea is very
general and applies to any material with a yield criterion. It has
already been suggested that the Spot Model could have relevance for
glassy relaxation~\cite{bazant06}, and the SFR provides a general
means to drive spot dynamics, based on solid mechanical
principles. The Spot Model also provides a multiscale algorithm for
random-packing dynamics, which works well for silo
drainage~\cite{rycroft06a}, so the SFR could enable a general
framework for multiscale modeling of amorphous materials. The idea
would be to cycle between continuum stress calculations, meso-scale
spot random walks, and microscopic particle dynamics.

\begin{acknowledgements}

  The authors are gratefully indebted to the many researchers noted
  within who made their experimental data available for comparison
  with the present theory. Special thanks go to  the members of GDR
  Midi (especially Olivier Pouliquen) and to J.-C. Tsai for their help
  in the transferring of data. The authors also thank Zden{\v e}k
  P. Ba{\v z}ant for helpful comments on the manuscript, and Pierre
  Gremaud for providing Fig. \ref{disconthopper}. 

\end{acknowledgements}

\appendix
\section{Critical State Soil Mechanics}

A common precept in plasticity is the notion of \emph{normality} or
\emph{associatedness}.  Flows that obey normality have a flow rule
defined in terms of the yield function $Y$ as follows:

\begin{equation}
\mathbf{\dot{E}}=\lambda\frac{\partial Y}{\partial \mathbf{T}}
\end{equation}
where $\lambda$ is a positive multiplier. For a 3D flow, this means
that if the yield function were plotted in 6-space as a function of
all 6 independent entries in the 3D stress tensor, the strain-rate
matrix would be a `vector' pointing normal to the yield surface
oriented toward greater values of $Y$.

One of the first gripes about the use of friction-based yield criteria
in describing granular materials is that the principle of normality
gives a flow rule that predicts unstoppable dilatancy.  Consider a
rough extension of the Coulomb yield criterion into 3D,
$Y=\mu(\text{tr}\mathbf{T})/3+|\mathbf{T_0}|/\sqrt{2}$, which displays the
basic property that yield occurs when a certain multiple of the
pressure equals the shear stress. Its associated flow rule is
\[\mathbf{\dot{E}}=\lambda\left(\frac{\mathbf{T_0}}{\sqrt{2}|\mathbf{T_0}|}+\frac{\mu}{3}\mathbf{I}\right).\]
The trace of this strain-rate tensor is $\lambda\mu$, implying
that material undergoing plastic flow will never stop dilating.

Roscoe and co-workers \cite{roscoe65} present a different viewpoint on
the issue. In what became known as \emph{Critical State Soil Mechanics},
explained in detail in \cite{wroth}, they argue that normality does
hold, but that in fact the Coulomb yield criterion is not technically
the correct yield function.

\begin{figure*}
\centering{\includegraphics[width=6.5in, clip]{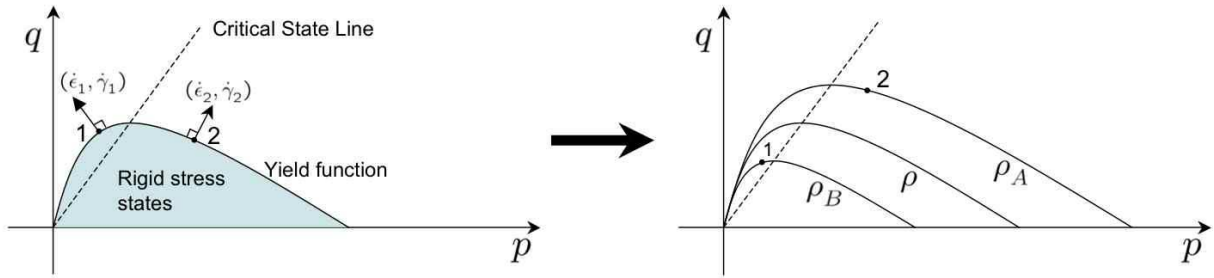} \caption{
    (left) Critical State Theory's spade-shaped yield function for
    material at some density $\rho_s$; (right) Deformations with
    non-zero volumetric strain will cause the material to settle down
    on a new yield function.}\label{critical}} 
\end{figure*}

Backed by results from triaxial stress experiments on soil samples,
Critical State Theory claims that soils have a yield function that
depends on the soil consolidation as measured by the local density
$\rho$. The yield curve for material at a particular density is
defined in terms of two stress tensor invariants: the pressure
$p=-\frac{1}{3}\text{tr}\mathbf{T}$ and the equivalent shear stress
$q=|\mathbf{T_0}|/\sqrt{2}$. Plotted in these variables, the principle
of normality is equivalent to the statement that the strain-rate
vector $(\dot{\epsilon},\dot{\gamma})$ is normal to the yield curve
and pointing outward, where $\dot{\epsilon}=-\text{tr}\mathbf{\dot{E}}$ is
a volumetric strain-rate which determines changes in density, and
$\dot{\gamma}=\frac{\sqrt{2}}{3}|\mathbf{\dot{E}_0}|$ is a shear
strain-rate proportional to the total shear deformation
(volume-conserving part of the deformation).  Figure \ref{critical}
displays the theory's picture of the yield function and how it changes
after material deformation.  Any stress state underneath the yield
curve corrresponds to rigid material. Under normality, material at
point 1 in the initial state will undergo a deformation according to
the vector $(\dot{\epsilon}_1,\dot{\gamma}_1)$. Since
$\dot{\epsilon}_1$ is negative, the material will dilate and settle
down at point 1 on the right on a new yield curve corresponding to
$\rho_{_A}<\rho$. The material at stress state 2 will likewise undergo
compaction and arrive at point 2 on the yield curve corresponding to
$\rho_{_B}>\rho$.

The critical state line is defined as the locus of points for which
normality predicts no volumetric changes during deformation--- note
that wherever a yield curve intersects the critical state line, the
curve becomes parallel to the $p$ axis and thus the corresponding
strain-rate vector has no volumetric component.  The theory reasons
that the critical state line is indeed a straight line of the form
\[q=Mp.\]As flow developes, the stress states throughout the material
will continually move toward the critical state line and once local
volume changes finally stop, all flowing material stress states should
lie on the critical state line. Thus in a steady flow, the critical
state line might falsely appear to \emph{be} the yield function when
in fact it is only a locus of states from a family of yield
functions. So, it is argued then that the reason normality
previously failed to describe granular materials was because it was
applied mistakenly to the critical state line and not to the true
family of yield functions.

  \bibliography{granular}

\begin{thebibliography}{89}
\expandafter\ifx\csname natexlab\endcsname\relax\def\natexlab#1{#1}\fi
\expandafter\ifx\csname bibnamefont\endcsname\relax
  \def\bibnamefont#1{#1}\fi
\expandafter\ifx\csname bibfnamefont\endcsname\relax
  \def\bibfnamefont#1{#1}\fi
\expandafter\ifx\csname citenamefont\endcsname\relax
  \def\citenamefont#1{#1}\fi
\expandafter\ifx\csname url\endcsname\relax
  \def\url#1{\texttt{#1}}\fi
\expandafter\ifx\csname urlprefix\endcsname\relax\def\urlprefix{URL }\fi
\providecommand{\bibinfo}[2]{#2}
\providecommand{\eprint}[2][]{\url{#2}}

\bibitem[{\citenamefont{Nedderman}(1991)}]{nedderman}
\bibinfo{author}{\bibfnamefont{R.~M.} \bibnamefont{Nedderman}},
  \emph{\bibinfo{title}{Statics and Kinematics of Granular Materials}}
  (\bibinfo{publisher}{Nova Science}, \bibinfo{year}{1991}).

\bibitem[{\citenamefont{Sokolovskii}(1965)}]{Sok}
\bibinfo{author}{\bibfnamefont{V.~V.} \bibnamefont{Sokolovskii}},
  \emph{\bibinfo{title}{Statics of Granular Materials}}
  (\bibinfo{publisher}{Pergamon/Oxford}, \bibinfo{year}{1965}).

\bibitem[{\citenamefont{Hill}(1950)}]{hill}
\bibinfo{author}{\bibfnamefont{R.}~\bibnamefont{Hill}},
  \emph{\bibinfo{title}{The Mathematical Theory of Plasticity}}
  (\bibinfo{publisher}{Oxford at the Clarendon Press}, \bibinfo{year}{1950}).

\bibitem[{\citenamefont{Jaeger et~al.}(1996)\citenamefont{Jaeger, Nagel, and
  Behringer}}]{jaeger96}
\bibinfo{author}{\bibfnamefont{H.~M.} \bibnamefont{Jaeger}},
  \bibinfo{author}{\bibfnamefont{S.~R.} \bibnamefont{Nagel}}, \bibnamefont{and}
  \bibinfo{author}{\bibfnamefont{R.~P.} \bibnamefont{Behringer}},
  \bibinfo{journal}{Rev. Mod. Phys.} \textbf{\bibinfo{volume}{68}},
  \bibinfo{pages}{1259} (\bibinfo{year}{1996}).

\bibitem[{\citenamefont{de~Gennes}(1999)}]{degennes99}
\bibinfo{author}{\bibfnamefont{P.~G.} \bibnamefont{de~Gennes}},
  \bibinfo{journal}{Rev. Mod. Phys.} \textbf{\bibinfo{volume}{71}},
  \bibinfo{pages}{S374} (\bibinfo{year}{1999}).

\bibitem[{\citenamefont{Kadanoff}(1999)}]{kadanoff99}
\bibinfo{author}{\bibfnamefont{L.~P.} \bibnamefont{Kadanoff}},
  \bibinfo{journal}{Rev. Mod. Phys.} \textbf{\bibinfo{volume}{71}},
  \bibinfo{pages}{435} (\bibinfo{year}{1999}).

\bibitem[{\citenamefont{Edwards and Grinev}(2001)}]{edwards01}
\bibinfo{author}{\bibfnamefont{S.~F.} \bibnamefont{Edwards}} \bibnamefont{and}
  \bibinfo{author}{\bibfnamefont{D.~V.} \bibnamefont{Grinev}},
  \bibinfo{journal}{Advances in Complex Systems} \textbf{\bibinfo{volume}{4}},
  \bibinfo{pages}{451} (\bibinfo{year}{2001}), \bibinfo{note}{(also reprinted
  in Ref.~\protect\cite{halsey02}).}

\bibitem[{\citenamefont{Halsey and Mehta}(2002)}]{halsey02}
\bibinfo{editor}{\bibfnamefont{T.}~\bibnamefont{Halsey}} \bibnamefont{and}
  \bibinfo{editor}{\bibfnamefont{A.}~\bibnamefont{Mehta}}, eds.,
  \emph{\bibinfo{title}{Challenges in Granular Physics}}
  (\bibinfo{publisher}{World Scientific}, \bibinfo{year}{2002}).

\bibitem[{\citenamefont{Aranson and Tsimring}(2006)}]{aranson06}
\bibinfo{author}{\bibfnamefont{I.~S.} \bibnamefont{Aranson}} \bibnamefont{and}
  \bibinfo{author}{\bibfnamefont{L.~S.} \bibnamefont{Tsimring}},
  \bibinfo{journal}{Rev. Mod. Phys.} \textbf{\bibinfo{volume}{78}},
  \bibinfo{pages}{641} (\bibinfo{year}{2006}).

\bibitem[{\citenamefont{Campbell}(1990)}]{campbell90}
\bibinfo{author}{\bibfnamefont{C.~S.} \bibnamefont{Campbell}},
  \bibinfo{journal}{Ann. Rev. Fluid Mech.} \textbf{\bibinfo{volume}{22}},
  \bibinfo{pages}{57} (\bibinfo{year}{1990}).

\bibitem[{\citenamefont{Balmforth and Provenzale}(2001)}]{alltheories}
\bibinfo{author}{\bibfnamefont{N.}~\bibnamefont{Balmforth}} \bibnamefont{and}
  \bibinfo{author}{\bibfnamefont{A.}~\bibnamefont{Provenzale}},
  \emph{\bibinfo{title}{Geomorph. Fluid Mech.}} (\bibinfo{publisher}{Springer},
  \bibinfo{year}{2001}).

\bibitem[{\citenamefont{Bouchaud et~al.}(1994)\citenamefont{Bouchaud, Cates,
  Prakash, and Edwards}}]{bouchaud94}
\bibinfo{author}{\bibfnamefont{J.-P.} \bibnamefont{Bouchaud}},
  \bibinfo{author}{\bibfnamefont{M.~E.} \bibnamefont{Cates}},
  \bibinfo{author}{\bibfnamefont{J.~R.} \bibnamefont{Prakash}},
  \bibnamefont{and} \bibinfo{author}{\bibfnamefont{S.~F.}
  \bibnamefont{Edwards}}, \bibinfo{journal}{J. Phys. I (France)}
  \textbf{\bibinfo{volume}{4}}, \bibinfo{pages}{1383} (\bibinfo{year}{1994}).

\bibitem[{\citenamefont{Bouchaud et~al.}(1995)\citenamefont{Bouchaud, Cates,
  Prakash, and Edwards}}]{bouchaud95a}
\bibinfo{author}{\bibfnamefont{J.-P.} \bibnamefont{Bouchaud}},
  \bibinfo{author}{\bibfnamefont{M.~E.} \bibnamefont{Cates}},
  \bibinfo{author}{\bibfnamefont{J.~R.} \bibnamefont{Prakash}},
  \bibnamefont{and} \bibinfo{author}{\bibfnamefont{S.~F.}
  \bibnamefont{Edwards}}, \bibinfo{journal}{Phys. Rev. Lett.}
  \textbf{\bibinfo{volume}{74}}, \bibinfo{pages}{1982} (\bibinfo{year}{1995}).

\bibitem[{\citenamefont{Boutreux et~al.}(1998)\citenamefont{Boutreux,
  Rapha\"el, and de~Gennes}}]{boutreux98}
\bibinfo{author}{\bibfnamefont{T.}~\bibnamefont{Boutreux}},
  \bibinfo{author}{\bibfnamefont{E.}~\bibnamefont{Rapha\"el}},
  \bibnamefont{and} \bibinfo{author}{\bibfnamefont{P.-G.}
  \bibnamefont{de~Gennes}}, \bibinfo{journal}{Phys. Rev. E}
  \textbf{\bibinfo{volume}{58}}, \bibinfo{pages}{4692} (\bibinfo{year}{1998}).

\bibitem[{\citenamefont{Boutreux et~al.}(1999)\citenamefont{Boutreux, Makse,
  and de~Gennes}}]{boutreux99}
\bibinfo{author}{\bibfnamefont{T.}~\bibnamefont{Boutreux}},
  \bibinfo{author}{\bibfnamefont{H.~A.} \bibnamefont{Makse}}, \bibnamefont{and}
  \bibinfo{author}{\bibfnamefont{P.-G.} \bibnamefont{de~Gennes}},
  \bibinfo{journal}{Euro. Phys. J. B} \textbf{\bibinfo{volume}{9}},
  \bibinfo{pages}{105} (\bibinfo{year}{1999}).

\bibitem[{\citenamefont{Bagnold}(1954)}]{bagnold54}
\bibinfo{author}{\bibfnamefont{R.~A.} \bibnamefont{Bagnold}},
  \bibinfo{journal}{Proc. Roy. Soc. London Ser. A}
  \textbf{\bibinfo{volume}{225}} (\bibinfo{year}{1954}).

\bibitem[{\citenamefont{Erta{\c s} and Halsey}(2002)}]{ertas02}
\bibinfo{author}{\bibfnamefont{D.}~\bibnamefont{Erta{\c s}}} \bibnamefont{and}
  \bibinfo{author}{\bibfnamefont{T.~C.} \bibnamefont{Halsey}},
  \bibinfo{journal}{Europhys. Lett.} \textbf{\bibinfo{volume}{60}},
  \bibinfo{pages}{931} (\bibinfo{year}{2002}).

\bibitem[{\citenamefont{Savage}(1998)}]{savage98}
\bibinfo{author}{\bibfnamefont{S.~B.} \bibnamefont{Savage}},
  \bibinfo{journal}{J. Fluid Mech.} \textbf{\bibinfo{volume}{377}},
  \bibinfo{pages}{1} (\bibinfo{year}{1998}).

\bibitem[{\citenamefont{Losert et~al.}(2000)\citenamefont{Losert, Bocquet,
  Lubensky, and Gollub}}]{losert00}
\bibinfo{author}{\bibfnamefont{W.}~\bibnamefont{Losert}},
  \bibinfo{author}{\bibfnamefont{L.}~\bibnamefont{Bocquet}},
  \bibinfo{author}{\bibfnamefont{T.~C.} \bibnamefont{Lubensky}},
  \bibnamefont{and} \bibinfo{author}{\bibfnamefont{J.~P.}
  \bibnamefont{Gollub}}, \bibinfo{journal}{Phys. Rev. Lett.}
  \textbf{\bibinfo{volume}{85}}, \bibinfo{pages}{1428} (\bibinfo{year}{2000}).

\bibitem[{\citenamefont{Bocquet et~al.}(2002)\citenamefont{Bocquet, Losert,
  Schalk, Lubensky, and Gollub}}]{bocquet02}
\bibinfo{author}{\bibfnamefont{L.}~\bibnamefont{Bocquet}},
  \bibinfo{author}{\bibfnamefont{W.}~\bibnamefont{Losert}},
  \bibinfo{author}{\bibfnamefont{D.}~\bibnamefont{Schalk}},
  \bibinfo{author}{\bibfnamefont{T.~C.} \bibnamefont{Lubensky}},
  \bibnamefont{and} \bibinfo{author}{\bibfnamefont{J.~P.}
  \bibnamefont{Gollub}}, \bibinfo{journal}{Phys. Rev. E}
  \textbf{\bibinfo{volume}{65}}, \bibinfo{pages}{011307}
  (\bibinfo{year}{2002}).

\bibitem[{\citenamefont{Mills et~al.}(1999)\citenamefont{Mills, Loggia, and
  Tixier}}]{mills99}
\bibinfo{author}{\bibfnamefont{P.}~\bibnamefont{Mills}},
  \bibinfo{author}{\bibfnamefont{D.}~\bibnamefont{Loggia}}, \bibnamefont{and}
  \bibinfo{author}{\bibfnamefont{M.}~\bibnamefont{Tixier}},
  \bibinfo{journal}{Europhys. Lett.} \textbf{\bibinfo{volume}{45}},
  \bibinfo{pages}{733} (\bibinfo{year}{1999}).

\bibitem[{\citenamefont{Pouliquen and Gutfraind}(1996)}]{pouliquen96}
\bibinfo{author}{\bibfnamefont{O.}~\bibnamefont{Pouliquen}} \bibnamefont{and}
  \bibinfo{author}{\bibfnamefont{R.}~\bibnamefont{Gutfraind}},
  \bibinfo{journal}{Phys. Rev. E} \textbf{\bibinfo{volume}{53}},
  \bibinfo{pages}{552} (\bibinfo{year}{1996}).

\bibitem[{\citenamefont{Pouliquen et~al.}(2001)\citenamefont{Pouliquen,
  Forterre, and Dizes}}]{pouliquen01}
\bibinfo{author}{\bibfnamefont{O.}~\bibnamefont{Pouliquen}},
  \bibinfo{author}{\bibfnamefont{Y.}~\bibnamefont{Forterre}}, \bibnamefont{and}
  \bibinfo{author}{\bibfnamefont{S.~L.} \bibnamefont{Dizes}},
  \bibinfo{journal}{Advances in Complex Systems} \textbf{\bibinfo{volume}{4}},
  \bibinfo{pages}{441} (\bibinfo{year}{2001}), \bibinfo{note}{(also reprinted
  in Ref.~\protect\cite{halsey02}).}

\bibitem[{\citenamefont{Litwiniszyn}(1958)}]{lit58}
\bibinfo{author}{\bibfnamefont{J.}~\bibnamefont{Litwiniszyn}},
  \bibinfo{journal}{Rheol. Acta} \textbf{\bibinfo{volume}{2/3}},
  \bibinfo{pages}{146} (\bibinfo{year}{1958}).

\bibitem[{\citenamefont{Mullins}(1972)}]{mullins72}
\bibinfo{author}{\bibfnamefont{J.}~\bibnamefont{Mullins}}, \bibinfo{journal}{J.
  Appl. Phys.} \textbf{\bibinfo{volume}{43}}, \bibinfo{pages}{665}
  (\bibinfo{year}{1972}).

\bibitem[{\citenamefont{Nedderman and T\"uz\"un}(1979)}]{nedderman79}
\bibinfo{author}{\bibfnamefont{R.~M.} \bibnamefont{Nedderman}}
  \bibnamefont{and}
  \bibinfo{author}{\bibfnamefont{U.}~\bibnamefont{T\"uz\"un}},
  \bibinfo{journal}{Powder Technology} \textbf{\bibinfo{volume}{22}},
  \bibinfo{pages}{243} (\bibinfo{year}{1979}).

\bibitem[{\citenamefont{Bazant}(2006)}]{bazant06}
\bibinfo{author}{\bibfnamefont{M.~Z.} \bibnamefont{Bazant}},
  \bibinfo{journal}{Mechanics of Materials} \textbf{\bibinfo{volume}{38}},
  \bibinfo{pages}{717} (\bibinfo{year}{2006}).

\bibitem[{\citenamefont{Rycroft
  et~al.}(2006{\natexlab{a}})\citenamefont{Rycroft, Bazant, Grest, and
  Landry}}]{rycroft06a}
\bibinfo{author}{\bibfnamefont{C.~H.} \bibnamefont{Rycroft}},
  \bibinfo{author}{\bibfnamefont{M.~Z.} \bibnamefont{Bazant}},
  \bibinfo{author}{\bibfnamefont{G.~S.} \bibnamefont{Grest}}, \bibnamefont{and}
  \bibinfo{author}{\bibfnamefont{J.~W.} \bibnamefont{Landry}},
  \bibinfo{journal}{Physical Review E} \textbf{\bibinfo{volume}{73}},
  \bibinfo{pages}{051306} (\bibinfo{year}{2006}{\natexlab{a}}).

\bibitem[{\citenamefont{Lema\^itre}(2002{\natexlab{a}})}]{lemaitre02}
\bibinfo{author}{\bibfnamefont{A.}~\bibnamefont{Lema\^itre}},
  \bibinfo{journal}{Phys. Rev. Lett.} \textbf{\bibinfo{volume}{89}},
  \bibinfo{pages}{195503} (\bibinfo{year}{2002}{\natexlab{a}}).

\bibitem[{\citenamefont{Lema\^itre}(2002{\natexlab{b}})}]{lemaitre02c}
\bibinfo{author}{\bibfnamefont{A.}~\bibnamefont{Lema\^itre}},
  \bibinfo{journal}{Phys. Rev. Lett.} \textbf{\bibinfo{volume}{89}},
  \bibinfo{pages}{064303} (\bibinfo{year}{2002}{\natexlab{b}}).

\bibitem[{\citenamefont{Aranson and Tsimring}(2001)}]{aranson01}
\bibinfo{author}{\bibfnamefont{I.~S.} \bibnamefont{Aranson}} \bibnamefont{and}
  \bibinfo{author}{\bibfnamefont{L.~S.} \bibnamefont{Tsimring}},
  \bibinfo{journal}{Phys. Rev. E} \textbf{\bibinfo{volume}{64}},
  \bibinfo{pages}{020301} (\bibinfo{year}{2001}).

\bibitem[{\citenamefont{Aranson and Tsimring}(2002)}]{aranson02}
\bibinfo{author}{\bibfnamefont{I.~S.} \bibnamefont{Aranson}} \bibnamefont{and}
  \bibinfo{author}{\bibfnamefont{L.~S.} \bibnamefont{Tsimring}},
  \bibinfo{journal}{Phys. Rev. E} \textbf{\bibinfo{volume}{65}},
  \bibinfo{pages}{061303} (\bibinfo{year}{2002}).

\bibitem[{\citenamefont{Midi}(2004)}]{midi04}
\bibinfo{author}{\bibfnamefont{G.~D.~R.} \bibnamefont{Midi}},
  \bibinfo{journal}{Euro. Phys. Journ. E.} \textbf{\bibinfo{volume}{14}},
  \bibinfo{pages}{341} (\bibinfo{year}{2004}).

\bibitem[{\citenamefont{Ostoja-Starzewski}(2005)}]{ostoja05}
\bibinfo{author}{\bibfnamefont{M.}~\bibnamefont{Ostoja-Starzewski}},
  \bibinfo{journal}{Int. J. Plasticity} \textbf{\bibinfo{volume}{21}},
  \bibinfo{pages}{1119} (\bibinfo{year}{2005}).

\bibitem[{\citenamefont{Prager and Drucker}(1952)}]{drucker}
\bibinfo{author}{\bibfnamefont{W.}~\bibnamefont{Prager}} \bibnamefont{and}
  \bibinfo{author}{\bibfnamefont{D.~C.} \bibnamefont{Drucker}},
  \bibinfo{journal}{Q. Appl. Mathematics} \textbf{\bibinfo{volume}{10:2}},
  \bibinfo{pages}{157} (\bibinfo{year}{1952}).

\bibitem[{\citenamefont{Horne and Nedderman}(1976)}]{horne}
\bibinfo{author}{\bibfnamefont{R.~M.} \bibnamefont{Horne}} \bibnamefont{and}
  \bibinfo{author}{\bibfnamefont{R.~M.} \bibnamefont{Nedderman}},
  \bibinfo{journal}{Powder Technol.} \textbf{\bibinfo{volume}{14}},
  \bibinfo{pages}{93} (\bibinfo{year}{1976}).

\bibitem[{\citenamefont{Pitman}(1986)}]{pitman}
\bibinfo{author}{\bibfnamefont{E.~B.} \bibnamefont{Pitman}},
  \bibinfo{journal}{Powder Technol.} \textbf{\bibinfo{volume}{47}},
  \bibinfo{pages}{219} (\bibinfo{year}{1986}).

\bibitem[{\citenamefont{Gremaud and Matthews}(2001)}]{gremaud}
\bibinfo{author}{\bibfnamefont{P.~A.} \bibnamefont{Gremaud}} \bibnamefont{and}
  \bibinfo{author}{\bibfnamefont{J.~V.} \bibnamefont{Matthews}},
  \bibinfo{journal}{J. Comput. Phys.} \textbf{\bibinfo{volume}{166}},
  \bibinfo{pages}{63} (\bibinfo{year}{2001}).

\bibitem[{\citenamefont{Schoefield and Wroth}(1968)}]{wroth}
\bibinfo{author}{\bibfnamefont{A.}~\bibnamefont{Schoefield}} \bibnamefont{and}
  \bibinfo{author}{\bibfnamefont{P.}~\bibnamefont{Wroth}},
  \emph{\bibinfo{title}{Critical State Soil Mechanics}}
  (\bibinfo{publisher}{McGraw-Hill}, \bibinfo{year}{1968}).

\bibitem[{\citenamefont{Jenike}(1964)}]{Jen}
\bibinfo{author}{\bibfnamefont{A.}~\bibnamefont{Jenike}},
  \bibinfo{journal}{Journ. Appl. Mech.} \textbf{\bibinfo{volume}{31}},
  \bibinfo{pages}{499} (\bibinfo{year}{1964}).

\bibitem[{\citenamefont{Spencer}(1964)}]{spencer64}
\bibinfo{author}{\bibfnamefont{A.~J.~M.} \bibnamefont{Spencer}},
  \bibinfo{journal}{J. Mech. Physics} \textbf{\bibinfo{volume}{12}},
  \bibinfo{pages}{337–351} (\bibinfo{year}{1964}).

\bibitem[{\citenamefont{Anand and Gu}(2000)}]{anand00}
\bibinfo{author}{\bibfnamefont{L.}~\bibnamefont{Anand}} \bibnamefont{and}
  \bibinfo{author}{\bibfnamefont{C.}~\bibnamefont{Gu}}, \bibinfo{journal}{J.
  Mech. Phys. Solids} \textbf{\bibinfo{volume}{28}}, \bibinfo{pages}{1701}
  (\bibinfo{year}{2000}).

\bibitem[{\citenamefont{Drescher}(1991)}]{Dre}
\bibinfo{author}{\bibfnamefont{A.}~\bibnamefont{Drescher}},
  \emph{\bibinfo{title}{Analytical Methods in Bin-Load Analysis}}
  (\bibinfo{publisher}{Elsevier}, \bibinfo{year}{1991}).

\bibitem[{\citenamefont{Pitman and Schaeffer}(1987)}]{pitman87}
\bibinfo{author}{\bibfnamefont{E.~B.} \bibnamefont{Pitman}} \bibnamefont{and}
  \bibinfo{author}{\bibfnamefont{D.~G.} \bibnamefont{Schaeffer}},
  \bibinfo{journal}{Commun. Pure. Appl. Math.} \textbf{\bibinfo{volume}{40}},
  \bibinfo{pages}{421} (\bibinfo{year}{1987}).

\bibitem[{\citenamefont{Schaeffer}(1987)}]{schaeffer87}
\bibinfo{author}{\bibfnamefont{D.~G.} \bibnamefont{Schaeffer}},
  \bibinfo{journal}{J. Diff. Eq.} \textbf{\bibinfo{volume}{66}},
  \bibinfo{pages}{19} (\bibinfo{year}{1987}).

\bibitem[{\citenamefont{Drescher}(1998)}]{drescher2}
\bibinfo{author}{\bibfnamefont{A.}~\bibnamefont{Drescher}},
  \bibinfo{journal}{Phil. Trans. R. Soc. Lond. A}
  \textbf{\bibinfo{volume}{356}}, \bibinfo{pages}{2649} (\bibinfo{year}{1998}).

\bibitem[{\citenamefont{Jenike}(1961)}]{jenike61}
\bibinfo{author}{\bibfnamefont{A.~W.} \bibnamefont{Jenike}},
  \bibinfo{journal}{Utah Engineering Experiment Station, Bulletin 108,
  University of Utah, Salt Lake City}  (\bibinfo{year}{1961}).

\bibitem[{\citenamefont{Samadani et~al.}(1999)\citenamefont{Samadani, Pradhan,
  and Kudrolli}}]{samadani99}
\bibinfo{author}{\bibfnamefont{A.}~\bibnamefont{Samadani}},
  \bibinfo{author}{\bibfnamefont{A.}~\bibnamefont{Pradhan}}, \bibnamefont{and}
  \bibinfo{author}{\bibfnamefont{A.}~\bibnamefont{Kudrolli}},
  \bibinfo{journal}{Phys. Rev. E} \textbf{\bibinfo{volume}{60}},
  \bibinfo{pages}{7203} (\bibinfo{year}{1999}).

\bibitem[{\citenamefont{Choi et~al.}(2005)\citenamefont{Choi, Kudrolli, and
  Bazant}}]{choi05}
\bibinfo{author}{\bibfnamefont{J.}~\bibnamefont{Choi}},
  \bibinfo{author}{\bibfnamefont{A.}~\bibnamefont{Kudrolli}}, \bibnamefont{and}
  \bibinfo{author}{\bibfnamefont{M.~Z.} \bibnamefont{Bazant}},
  \bibinfo{journal}{J. Phys.: Condensed Matter} \textbf{\bibinfo{volume}{17}},
  \bibinfo{pages}{S2533} (\bibinfo{year}{2005}).

\bibitem[{\citenamefont{Rycroft
  et~al.}(2006{\natexlab{b}})\citenamefont{Rycroft, Grest, Bazant, and
  Landry}}]{rycroft06b}
\bibinfo{author}{\bibfnamefont{C.~H.} \bibnamefont{Rycroft}},
  \bibinfo{author}{\bibfnamefont{G.~S.} \bibnamefont{Grest}},
  \bibinfo{author}{\bibfnamefont{M.~Z.} \bibnamefont{Bazant}},
  \bibnamefont{and} \bibinfo{author}{\bibfnamefont{J.~W.}
  \bibnamefont{Landry}}, \bibinfo{journal}{Physical Review E} p.
  \bibinfo{pages}{to appear} (\bibinfo{year}{2006}{\natexlab{b}}).

\bibitem[{\citenamefont{Jir\'asek and Ba\v{z}ant}(2002)}]{bazantsr}
\bibinfo{author}{\bibfnamefont{M.}~\bibnamefont{Jir\'asek}} \bibnamefont{and}
  \bibinfo{author}{\bibfnamefont{Z.~P.} \bibnamefont{Ba\v{z}ant}},
  \emph{\bibinfo{title}{Inelastic Analysis of Structures}}
  (\bibinfo{publisher}{John Wiley and Sons Ltd.}, \bibinfo{year}{2002}).

\bibitem[{\citenamefont{Hill}(1967)}]{hill2}
\bibinfo{author}{\bibfnamefont{R.}~\bibnamefont{Hill}}, \bibinfo{journal}{J.
  Mech. Phys. Solids} \textbf{\bibinfo{volume}{15}}, \bibinfo{pages}{79}
  (\bibinfo{year}{1967}).

\bibitem[{\citenamefont{T\"uz\"un and Nedderman}(1979)}]{tuzun79}
\bibinfo{author}{\bibfnamefont{U.}~\bibnamefont{T\"uz\"un}} \bibnamefont{and}
  \bibinfo{author}{\bibfnamefont{R.~M.} \bibnamefont{Nedderman}},
  \bibinfo{journal}{Powder Technolgy} \textbf{\bibinfo{volume}{23}},
  \bibinfo{pages}{257} (\bibinfo{year}{1979}).

\bibitem[{\citenamefont{Medina et~al.}(1998)\citenamefont{Medina, Cordova,
  Luna, and Trevino}}]{medina98a}
\bibinfo{author}{\bibfnamefont{A.}~\bibnamefont{Medina}},
  \bibinfo{author}{\bibfnamefont{J.~A.} \bibnamefont{Cordova}},
  \bibinfo{author}{\bibfnamefont{E.}~\bibnamefont{Luna}}, \bibnamefont{and}
  \bibinfo{author}{\bibfnamefont{C.}~\bibnamefont{Trevino}},
  \bibinfo{journal}{Physics Letters A} \textbf{\bibinfo{volume}{220}},
  \bibinfo{pages}{111} (\bibinfo{year}{1998}).

\bibitem[{\citenamefont{Fenistein and van Hecke}(2003)}]{fenistein03}
\bibinfo{author}{\bibfnamefont{D.}~\bibnamefont{Fenistein}} \bibnamefont{and}
  \bibinfo{author}{\bibfnamefont{M.}~\bibnamefont{van Hecke}},
  \bibinfo{journal}{Nature} \textbf{\bibinfo{volume}{425}},
  \bibinfo{pages}{256} (\bibinfo{year}{2003}).

\bibitem[{\citenamefont{Litwiniszyn}(1963)}]{lit63b}
\bibinfo{author}{\bibfnamefont{J.}~\bibnamefont{Litwiniszyn}},
  \bibinfo{journal}{Bull. Acad. Pol. Sci.} \textbf{\bibinfo{volume}{11}},
  \bibinfo{pages}{593} (\bibinfo{year}{1963}).

\bibitem[{\citenamefont{Choi et~al.}(2004)\citenamefont{Choi, Kudrolli,
  Rosales, and Bazant}}]{choi04}
\bibinfo{author}{\bibfnamefont{J.}~\bibnamefont{Choi}},
  \bibinfo{author}{\bibfnamefont{A.}~\bibnamefont{Kudrolli}},
  \bibinfo{author}{\bibfnamefont{R.~R.} \bibnamefont{Rosales}},
  \bibnamefont{and} \bibinfo{author}{\bibfnamefont{M.~Z.}
  \bibnamefont{Bazant}}, \bibinfo{journal}{Phys. Rev. Lett.}
  \textbf{\bibinfo{volume}{92}}, \bibinfo{pages}{174301}
  (\bibinfo{year}{2004}).

\bibitem[{\citenamefont{Risken}(1996)}]{risken}
\bibinfo{author}{\bibfnamefont{H.}~\bibnamefont{Risken}},
  \emph{\bibinfo{title}{The Fokker-Planck Equation}}
  (\bibinfo{publisher}{Springer}, \bibinfo{year}{1996}).

\bibitem[{\citenamefont{Baxter et~al.}(1989)\citenamefont{Baxter, Behringer,
  Fagert, and Johnson}}]{behringer89}
\bibinfo{author}{\bibfnamefont{G.~W.} \bibnamefont{Baxter}},
  \bibinfo{author}{\bibfnamefont{R.~P.} \bibnamefont{Behringer}},
  \bibinfo{author}{\bibfnamefont{T.}~\bibnamefont{Fagert}}, \bibnamefont{and}
  \bibinfo{author}{\bibfnamefont{G.~A.} \bibnamefont{Johnson}},
  \bibinfo{journal}{Phys. Rev. Lett.} \textbf{\bibinfo{volume}{62}},
  \bibinfo{pages}{2825} (\bibinfo{year}{1989}).

\bibitem[{\citenamefont{P\"oschel}(1994)}]{poschel94}
\bibinfo{author}{\bibfnamefont{T.}~\bibnamefont{P\"oschel}},
  \bibinfo{journal}{J. Phys. (France) I} \textbf{\bibinfo{volume}{4}},
  \bibinfo{pages}{499} (\bibinfo{year}{1994}).

\bibitem[{\citenamefont{Mueth}(2003)}]{mueth03}
\bibinfo{author}{\bibfnamefont{D.~M.} \bibnamefont{Mueth}},
  \bibinfo{journal}{Phys. Rev. E.} \textbf{\bibinfo{volume}{67}},
  \bibinfo{pages}{011304} (\bibinfo{year}{2003}).

\bibitem[{\citenamefont{Lois et~al.}(2005)\citenamefont{Lois, Lemaitre, and
  Carlson}}]{lois05}
\bibinfo{author}{\bibfnamefont{G.}~\bibnamefont{Lois}},
  \bibinfo{author}{\bibfnamefont{A.}~\bibnamefont{Lemaitre}}, \bibnamefont{and}
  \bibinfo{author}{\bibfnamefont{J.~M.} \bibnamefont{Carlson}},
  \bibinfo{journal}{Phys. Rev. E.} \textbf{\bibinfo{volume}{72}},
  \bibinfo{pages}{051303} (\bibinfo{year}{2005}).

\bibitem[{\citenamefont{Latzel et~al.}(2003)\citenamefont{Latzel, Luding,
  Herrmann, Howell, and Behringer}}]{latzel03}
\bibinfo{author}{\bibfnamefont{M.}~\bibnamefont{Latzel}},
  \bibinfo{author}{\bibfnamefont{S.}~\bibnamefont{Luding}},
  \bibinfo{author}{\bibfnamefont{H.~J.} \bibnamefont{Herrmann}},
  \bibinfo{author}{\bibfnamefont{D.~W.} \bibnamefont{Howell}},
  \bibnamefont{and}
  \bibinfo{author}{\bibfnamefont{R.}~\bibnamefont{Behringer}},
  \bibinfo{journal}{Euro. Phys. Journ. E.} \textbf{\bibinfo{volume}{11}},
  \bibinfo{pages}{325} (\bibinfo{year}{2003}).

\bibitem[{\citenamefont{da~Cruz}(2004)}]{dacruz_phd}
\bibinfo{author}{\bibfnamefont{F.}~\bibnamefont{da~Cruz}}, Ph.D. thesis,
  \bibinfo{school}{Ecole Nationale des Ponts et Chaussees, Marne a la vall\'ee,
  France} (\bibinfo{year}{2004}).

\bibitem[{\citenamefont{Bocquet et~al.}(2001)\citenamefont{Bocquet, Losert,
  Schalk, Lubensky, and Gollub}}]{bocquet01}
\bibinfo{author}{\bibfnamefont{L.}~\bibnamefont{Bocquet}},
  \bibinfo{author}{\bibfnamefont{W.}~\bibnamefont{Losert}},
  \bibinfo{author}{\bibfnamefont{D.}~\bibnamefont{Schalk}},
  \bibinfo{author}{\bibfnamefont{T.~C.} \bibnamefont{Lubensky}},
  \bibnamefont{and} \bibinfo{author}{\bibfnamefont{J.~P.}
  \bibnamefont{Gollub}}, \bibinfo{journal}{Phys. Rev. E}
  \textbf{\bibinfo{volume}{65}}, \bibinfo{pages}{011307}
  (\bibinfo{year}{2001}).

\bibitem[{\citenamefont{Mueth et~al.}(2000)\citenamefont{Mueth, Debregeas,
  Karczmar, Eng, Nagel, and Jaeger}}]{mueth00}
\bibinfo{author}{\bibfnamefont{D.~E.} \bibnamefont{Mueth}},
  \bibinfo{author}{\bibfnamefont{G.~F.} \bibnamefont{Debregeas}},
  \bibinfo{author}{\bibfnamefont{G.~S.} \bibnamefont{Karczmar}},
  \bibinfo{author}{\bibfnamefont{P.~J.} \bibnamefont{Eng}},
  \bibinfo{author}{\bibfnamefont{S.~R.} \bibnamefont{Nagel}}, \bibnamefont{and}
  \bibinfo{author}{\bibfnamefont{H.~M.} \bibnamefont{Jaeger}},
  \bibinfo{journal}{Nature} \textbf{\bibinfo{volume}{406}},
  \bibinfo{pages}{385} (\bibinfo{year}{2000}).

\bibitem[{\citenamefont{Chambon et~al.}(2003)\citenamefont{Chambon,
  Schmittbuhl, Corfdir, Vilotte, and Roux}}]{chambon03}
\bibinfo{author}{\bibfnamefont{G.}~\bibnamefont{Chambon}},
  \bibinfo{author}{\bibfnamefont{J.}~\bibnamefont{Schmittbuhl}},
  \bibinfo{author}{\bibfnamefont{A.}~\bibnamefont{Corfdir}},
  \bibinfo{author}{\bibfnamefont{J.~P.} \bibnamefont{Vilotte}},
  \bibnamefont{and} \bibinfo{author}{\bibfnamefont{S.}~\bibnamefont{Roux}},
  \bibinfo{journal}{Phys. Rev. E} \textbf{\bibinfo{volume}{68}},
  \bibinfo{pages}{011304} (\bibinfo{year}{2003}).

\bibitem[{\citenamefont{Sch\"ollmnann}(1998)}]{schollmann98}
\bibinfo{author}{\bibfnamefont{S.}~\bibnamefont{Sch\"ollmnann}},
  \bibinfo{journal}{Phys. Rev. E} \textbf{\bibinfo{volume}{59}},
  \bibinfo{pages}{889} (\bibinfo{year}{1998}).

\bibitem[{\citenamefont{Tsai and Gollub}(2004)}]{tsai04}
\bibinfo{author}{\bibfnamefont{J.-C.} \bibnamefont{Tsai}} \bibnamefont{and}
  \bibinfo{author}{\bibfnamefont{J.~P.} \bibnamefont{Gollub}},
  \bibinfo{journal}{Phys. Rev. E} \textbf{\bibinfo{volume}{70}},
  \bibinfo{pages}{031303} (\bibinfo{year}{2004}).

\bibitem[{\citenamefont{Tsai and Gollub}(2005)}]{tsai05}
\bibinfo{author}{\bibfnamefont{J.-C.} \bibnamefont{Tsai}} \bibnamefont{and}
  \bibinfo{author}{\bibfnamefont{J.~P.} \bibnamefont{Gollub}},
  \bibinfo{journal}{Phys. Rev. E} \textbf{\bibinfo{volume}{72}},
  \bibinfo{pages}{051304} (\bibinfo{year}{2005}).

\bibitem[{\citenamefont{Siavoshi et~al.}(2006)\citenamefont{Siavoshi, Orpe, and
  Kudrolli}}]{siavoshi06}
\bibinfo{author}{\bibfnamefont{S.}~\bibnamefont{Siavoshi}},
  \bibinfo{author}{\bibfnamefont{A.~V.} \bibnamefont{Orpe}}, \bibnamefont{and}
  \bibinfo{author}{\bibfnamefont{A.}~\bibnamefont{Kudrolli}},
  \bibinfo{journal}{Phys. Rev. E} \textbf{\bibinfo{volume}{73}},
  \bibinfo{pages}{010301(R)} (\bibinfo{year}{2006}).

\bibitem[{\citenamefont{Thompson and Grest}(1991)}]{thompson91}
\bibinfo{author}{\bibfnamefont{P.~A.} \bibnamefont{Thompson}} \bibnamefont{and}
  \bibinfo{author}{\bibfnamefont{G.~S.} \bibnamefont{Grest}},
  \bibinfo{journal}{Phys Rev Lett} \textbf{\bibinfo{volume}{67}},
  \bibinfo{pages}{1751} (\bibinfo{year}{1991}).

\bibitem[{\citenamefont{Volfson et~al.}(2003)\citenamefont{Volfson, Tsimring,
  and Aranson}}]{volfson03}
\bibinfo{author}{\bibfnamefont{D.}~\bibnamefont{Volfson}},
  \bibinfo{author}{\bibfnamefont{L.~S.} \bibnamefont{Tsimring}},
  \bibnamefont{and} \bibinfo{author}{\bibfnamefont{I.~S.}
  \bibnamefont{Aranson}}, \bibinfo{journal}{Physical Review E}
  \textbf{\bibinfo{volume}{68}}, \bibinfo{pages}{021301}
  (\bibinfo{year}{2003}).

\bibitem[{\citenamefont{Jalali et~al.}(2002)\citenamefont{Jalali, Polashenski,
  Tynj\"al\"a, and Zamankhan}}]{jalali02}
\bibinfo{author}{\bibfnamefont{P.}~\bibnamefont{Jalali}},
  \bibinfo{author}{\bibfnamefont{W.}~\bibnamefont{Polashenski}},
  \bibinfo{author}{\bibfnamefont{T.}~\bibnamefont{Tynj\"al\"a}},
  \bibnamefont{and}
  \bibinfo{author}{\bibfnamefont{P.}~\bibnamefont{Zamankhan}},
  \bibinfo{journal}{Physica D} \textbf{\bibinfo{volume}{162}},
  \bibinfo{pages}{188} (\bibinfo{year}{2002}).

\bibitem[{dac()}]{dacruz05b}
\bibinfo{howpublished}{cond-mat/0503682}.

\bibitem[{\citenamefont{Lemieux and Durian}(2000)}]{lemieux00}
\bibinfo{author}{\bibfnamefont{P.~A.} \bibnamefont{Lemieux}} \bibnamefont{and}
  \bibinfo{author}{\bibfnamefont{D.~J.} \bibnamefont{Durian}},
  \bibinfo{journal}{Phys. Rev. Lett.} \textbf{\bibinfo{volume}{85}},
  \bibinfo{pages}{4273} (\bibinfo{year}{2000}).

\bibitem[{\citenamefont{Silbert et~al.}(2003)\citenamefont{Silbert, Landry, and
  Grest}}]{silbert03}
\bibinfo{author}{\bibfnamefont{L.~E.} \bibnamefont{Silbert}},
  \bibinfo{author}{\bibfnamefont{J.~W.} \bibnamefont{Landry}},
  \bibnamefont{and} \bibinfo{author}{\bibfnamefont{G.~S.} \bibnamefont{Grest}},
  \bibinfo{journal}{Phys. Fluids} \textbf{\bibinfo{volume}{15}},
  \bibinfo{pages}{1} (\bibinfo{year}{2003}).

\bibitem[{\citenamefont{Komatsu et~al.}(2001)\citenamefont{Komatsu, Inagaki,
  Nakagawa, and Nasuno}}]{komatsu01}
\bibinfo{author}{\bibfnamefont{T.~S.} \bibnamefont{Komatsu}},
  \bibinfo{author}{\bibfnamefont{S.}~\bibnamefont{Inagaki}},
  \bibinfo{author}{\bibfnamefont{N.}~\bibnamefont{Nakagawa}}, \bibnamefont{and}
  \bibinfo{author}{\bibfnamefont{S.}~\bibnamefont{Nasuno}},
  \bibinfo{journal}{Phys. Rev. Lett.} \textbf{\bibinfo{volume}{86}},
  \bibinfo{pages}{1757} (\bibinfo{year}{2001}).

\bibitem[{\citenamefont{Silbert et~al.}(2001)\citenamefont{Silbert, Ertas,
  Grest, Halsey, Levine, and Plimpton}}]{silbert01}
\bibinfo{author}{\bibfnamefont{L.~E.} \bibnamefont{Silbert}},
  \bibinfo{author}{\bibfnamefont{D.}~\bibnamefont{Ertas}},
  \bibinfo{author}{\bibfnamefont{G.~S.} \bibnamefont{Grest}},
  \bibinfo{author}{\bibfnamefont{T.~C.} \bibnamefont{Halsey}},
  \bibinfo{author}{\bibfnamefont{D.}~\bibnamefont{Levine}}, \bibnamefont{and}
  \bibinfo{author}{\bibfnamefont{S.~J.} \bibnamefont{Plimpton}},
  \bibinfo{journal}{Phys. Rev. E.} \textbf{\bibinfo{volume}{64}},
  \bibinfo{pages}{051302} (\bibinfo{year}{2001}).

\bibitem[{\citenamefont{Ancey}(2002)}]{ancey02}
\bibinfo{author}{\bibfnamefont{C.}~\bibnamefont{Ancey}},
  \bibinfo{journal}{Phys. Rev. E} \textbf{\bibinfo{volume}{65}},
  \bibinfo{pages}{011304} (\bibinfo{year}{2002}).

\bibitem[{\citenamefont{Berton et~al.}(2003)\citenamefont{Berton, Delannay,
  Richard, Taberlet, and Valance}}]{berton03}
\bibinfo{author}{\bibfnamefont{G.}~\bibnamefont{Berton}},
  \bibinfo{author}{\bibfnamefont{R.}~\bibnamefont{Delannay}},
  \bibinfo{author}{\bibfnamefont{P.}~\bibnamefont{Richard}},
  \bibinfo{author}{\bibfnamefont{N.}~\bibnamefont{Taberlet}}, \bibnamefont{and}
  \bibinfo{author}{\bibfnamefont{A.}~\bibnamefont{Valance}},
  \bibinfo{journal}{Phys. Rev. E} \textbf{\bibinfo{volume}{68}},
  \bibinfo{pages}{051303} (\bibinfo{year}{2003}).

\bibitem[{\citenamefont{Pouliquen}(1999)}]{pouliquen99}
\bibinfo{author}{\bibfnamefont{O.}~\bibnamefont{Pouliquen}},
  \bibinfo{journal}{Phys. Fluids} \textbf{\bibinfo{volume}{11}},
  \bibinfo{pages}{542} (\bibinfo{year}{1999}).

\bibitem[{\citenamefont{Prochnow}(2002)}]{prochnow_thesis}
\bibinfo{author}{\bibfnamefont{M.}~\bibnamefont{Prochnow}}, Ph.D. thesis,
  \bibinfo{school}{Ecole Nationale de Ponts et Chauss\'{e}es, Marne la
  Vall\'{e}e, France} (\bibinfo{year}{2002}).

\bibitem[{\citenamefont{Azanza}(1997)}]{azanza_thesis}
\bibinfo{author}{\bibfnamefont{E.}~\bibnamefont{Azanza}}, Ph.D. thesis,
  \bibinfo{school}{Ecole des Ponts et Chauss\'{e}es, Marne la Vall\'{e}e,
  France} (\bibinfo{year}{1997}).

\bibitem[{\citenamefont{da~Cruz et~al.}(2005)\citenamefont{da~Cruz, Emam,
  Prochnow, Roux, and Chevoir}}]{dacruz05}
\bibinfo{author}{\bibfnamefont{F.}~\bibnamefont{da~Cruz}},
  \bibinfo{author}{\bibfnamefont{S.}~\bibnamefont{Emam}},
  \bibinfo{author}{\bibfnamefont{M.}~\bibnamefont{Prochnow}},
  \bibinfo{author}{\bibfnamefont{J.}~\bibnamefont{Roux}}, \bibnamefont{and}
  \bibinfo{author}{\bibfnamefont{F.}~\bibnamefont{Chevoir}},
  \bibinfo{journal}{Phys. Rev. E.} \textbf{\bibinfo{volume}{72}},
  \bibinfo{pages}{021309} (\bibinfo{year}{2005}).

\bibitem[{\citenamefont{Duran}(2000)}]{duran_book}
\bibinfo{author}{\bibfnamefont{J.}~\bibnamefont{Duran}},
  \emph{\bibinfo{title}{Sands, Powders, and Grains}}
  (\bibinfo{publisher}{Springer-Verlag}, \bibinfo{year}{2000}).

\bibitem[{\citenamefont{Tardos et~al.}(1998)\citenamefont{Tardos, Khan, and
  Schaeffer}}]{tardos98}
\bibinfo{author}{\bibfnamefont{G.~I.} \bibnamefont{Tardos}},
  \bibinfo{author}{\bibfnamefont{M.~I.} \bibnamefont{Khan}}, \bibnamefont{and}
  \bibinfo{author}{\bibfnamefont{D.~G.} \bibnamefont{Schaeffer}},
  \bibinfo{journal}{Phys. Fluids} \textbf{\bibinfo{volume}{10}},
  \bibinfo{pages}{335} (\bibinfo{year}{1998}).

\bibitem[{\citenamefont{Pouliquen}(2004)}]{pouliquen04}
\bibinfo{author}{\bibfnamefont{O.}~\bibnamefont{Pouliquen}},
  \bibinfo{journal}{Phys. Rev. Lett.} \textbf{\bibinfo{volume}{93}},
  \bibinfo{pages}{248001} (\bibinfo{year}{2004}).

\bibitem[{\citenamefont{Roscoe et~al.}(1965)\citenamefont{Roscoe, Schoefield,
  and Thurairajah}}]{roscoe65}
\bibinfo{author}{\bibfnamefont{K.~H.} \bibnamefont{Roscoe}},
  \bibinfo{author}{\bibfnamefont{A.~N.} \bibnamefont{Schoefield}},
  \bibnamefont{and}
  \bibinfo{author}{\bibfnamefont{A.}~\bibnamefont{Thurairajah}},
  \bibinfo{journal}{Geotechnique} \textbf{\bibinfo{volume}{15}},
  \bibinfo{pages}{127} (\bibinfo{year}{1965}).

\end{thebibliography}
\end{document}